\documentclass[aps,showpacs,showkeys,preprint]{revtex4}
\usepackage{epsfig,amssymb,bm,graphics,color}
\usepackage{epstopdf}
\newcommand*{\fs}[1]{#1\!\!\!/}

\newcommand*{\ee}{e^+e^-}

\begin{document} {\normalsize }

\title{Quantum processes in short and intensive electromagnetic fields}

\author{
 Alexander~I.~Titov$^{a}$,   Burkhard~K\"ampfer$^{b,c}$,
 Atsushi~Hosaka$^{d,e}$, and  Hideaki~Takabe$^f$}
 \affiliation{
 $^a$Bogoliubov Laboratory of Theoretical Physics, JINR, Dubna 141980, Russia\\
 $^b$Helmholtz-Zentrum  Dresden-Rossendorf, 01314 Dresden, Germany\\
 $^c$Institut f\"ur Theoretische Physik, TU~Dresden, 01062 Dresden, Germany\\
 $^d$RCNP, 10-1 Mihogaoka Ibaraki, 567-0047 Osaka, Japan\\
 $^e$J-PARC Branch,
 KEK, Tokai, Ibaraki, 319-1106, Japan\\
 $^f$ILE, Yamada-oka, Suita, Osaka 565-0871, Japan}

\begin{abstract}
 This work provides an overview of our recent results
 in studying two most important and
 widely discussed quantum processes:
 electron-positron pairs production off
 a probe photon propagating through
 a polarized short-pulsed electromagnetic (e.g.\ laser)
 wave field or generalized Breit-Wheeler process,
 and a single a photon emission off an electron interacting
 with the laser pules, so-called non-linear Compton scattering.
 We show that the probabilities of particle production
 in both processes are determined
 by interplay of two dynamical effects, where the first one is
 related to the shape and duration of the pulse and the
 second one is non-linear dynamics of the
 interaction of charged fermions with a strong
 electromagnetic field.
 We elaborate suitable expressions for the production
 probabilities and cross sections, convenient for studying
 evolution of the plasma in presence
 of strong electromagnetic fields

 \end{abstract}
\pacs{12.20.Ds, 13.40.-f, 23.20.Nx}
\keywords{Non-linear dynamics,
 multi-photon effects, sub-threshold processes}
\maketitle

\section{Introduction}

 The rapidly progressing laser technology \cite{Tajima}
 offers unprecedented opportunities for investigations
 of quantum systems with intense laser beams~\cite{Piazza}.
 A laser intensity $I_L$ of $\sim 2\times 10^{22}$  W/cm${}^2$ has been already
 achieved~\cite{I-22}. Intensities of the order of
 $I_L \sim 10^{23}...10^{25}$ W/cm$^2$ are envisaged in near future, e.g.\
 at the CLF~\cite{CLF}, ELI~\cite{ELI}, HiPER~\cite{hiper}.
 Further facilities are in planning on construction stage, e.g.
 PEARL laser facility~\cite{sarov} at Sarov/Nizhny Novgorod, Russia.
 The high intensities are provided in short
 pulses on a femtosecond pulse duration
 level~\cite{Piazza,ShortPulse,ShortPulse_2},
 with only a few oscillations of the electromagnetic (e.m.) field
 or even sub-cycle pulses.
 In order to achieve such high intensities
 in the focal spot of the laser beam a crucial technique
 is required.
(The tight connection of high intensity and short pulse duration
 is further emphasized in \cite{Mackenroth-2011}. The attosecond
 regime will become accessible at shorter wavelengths~\cite{atto,I-222}).

 Quantum processes occurring in the interactions  of charge fermions
 in very (infinitely) long e.m. pulse  were investigated in detail
 in the pioneering works of
 Reiss~\cite{Reiss,Reiss71} as well as Narozhny, Nikishov
 and Ritus~\cite{NikishovRitus,NR-64} and some other papers
 (see for example \cite{Ritus-79}).
 We call the such approaches as an infinite pulse approximation (IPA)
 since it refers to a stationary scattering process.
 Many simple and clear expressions for the production probabilities
 and cross sections have been obtain within IPA. It was shown
 that the charged fermion (electron, for instance)
 can interact with $n\ge1$ photon simultaneously ($n$ is an integer number),
 and cases with $n>1$ correspond to the subthreshold, multi-photon
 events. However, since the new laser generation is expected to
 operate with the finite (short, and ultra-short) pulses,
 the question naturally arises whether predictions of IPA
 are valid for the finite pulses or not.

 Indeed, recently it was shown that for the photon production
 off an electron interacting with short laser pulse (Compton scattering)
 in~\cite{Boca-2009,Heinzl-2009,Mackenroth-2011,Seipt-2011,Dinu,Seipt-2012,Krajewska-2012,TitovEPJD},
 and for $\ee$ pair production off a probe photon interacting with
 short e.m. pulses (Breit-Wheeler process)
 in~\cite{TTKH-2012,TitovPRA,Nousch,Krajewska,VillalbaChavez}
 the finite pulse shape and the pulse duration become important.
 That means the treatment of the intense and short laser field as an infinitely
 long wave train is no longer adequate. The theory must
 operate with essentially finite pulse. We call
 such approaches as a finite pulse approximation (FPA).
 Formation of positrons
 from cascade processes in a photon-electron-positron
 plasma~\cite{Fedotov-2010,Elkina-2011} generated by
 photon-laser~\cite{Ilderton-2011},
 electron-laser~\cite{Hu,Ilderton-2010} or laser-laser
 interactions~\cite{Kirk-2009}
 (see \cite{Bulanov-2011,Ruffini} for surveys)
 is an important problem in laser physics.
 The evaluation of corresponding transport equations needs as an input
 the probabilities/cross sections for the production
 energetic photons (e.g., in the non-linear Compton scattering)
 and direct emission of $e^+e^-$ pairs
 (e.g., in the non-linear Breit-Wheeler process).

 Consider first the non-linear Breit-Wheeler process.
 Corresponding linear Breit-Wheeler $\ee$ pair production
 $\gamma' + \gamma \to e^+ + e^-$~\cite{Breit-Wheeler-1934}
 refers to a perturbative QED
 reaction; the generalization to the multi-photon process
 $\gamma' + n \gamma \to e^+ + e^-$
 (nonlinear Breit-Wheeler process) in IPA were done
 in~\cite{Reiss,NR-64,Ritus-79}.
 Attributing theses processes to
 colliding null fields one can imagine another aspect. In the anti-node of
 suitably counter propagating e.m.\ waves an
 oscillating purely electric field can give rise
 to the dynamical Schwinger effect~\cite{Blaschke-2013};
 in the low-frequency limit one recovers
 the famous Schwinger effect~\cite{Schwinger}
 awaiting still its experimental
 verification.
 These kinds of pair creation processes
 are related to highly non-perturbative effects~\cite{Dunne-2009,Hebenstreit-2011}.
 Once pair production is
 seeded in very intense fields further avalanche like particle
 production can set in
 which then could screen the original field or even limit
 the attainable field strength~\cite{Fedotov-2010}.
 One can relate the Breit-Wheeler process to the absorptive
 part of the probe-photon correlator in an external e.m. field;
 in our case the latter being a null field too.
 Later, we focus on colliding null fields in the multi-photon
 regime and consider the  generalized  Breit-Wheeler effect
 for short pulses of e.m.\ wave fields
 ranging from weak to high intensities.
 Phrased differently we analyze $\ee$
 pair production by a probe photon $\gamma'$ traversing a coherent e.m.\ (i.e.\ laser)
 field. We employ
 the four-potential of a circularly polarized laser field in
 the axial gauge $A^\mu=(0,\,\mathbf{A}(\phi))$ with
\begin{eqnarray}
 \mathbf{A}(\phi)=f(\phi) \left( \mathbf{a}_1\cos(\phi+\tilde\phi)+ \mathbf
 {a}_2\sin(\phi+\tilde\phi)\right)~, \label{III1}
 \end{eqnarray}
 where $\phi=k\cdot x$ is invariant phase with four-wave vector
 $k=(\omega, \mathbf{k})$, obeying the null field property $k^2=k\cdot
 k=0$ (a dot between four-vectors indicates the Lorentz scalar
 product) implying $\omega = \vert\mathbf{k}\vert$,
 $ \mathbf{a}_{(1,2)} \equiv \mathbf{a}_{(x,y)}$;
 $|\mathbf{a}_x|^2=|\mathbf{a}_y|^2 = a^2$, $\mathbf{a}_x \mathbf{a}_y=0$;
 transversality means $\mathbf{k} \mathbf{a}_{x,y}=0$ in the present gauge.
 The envelope function $f(\phi)$ with
 $\lim\limits_{\phi\to\pm\infty}f(\phi)=0$ (FPA) accounts for the
 finite pulse length. (IPA would mean $f(\phi) = 1$).
 To define the pulse duration one can use the number $N$ of cycles in a
 pulse, $N=\Delta/\pi=\frac12\tau\omega$, where the dimensionless
 quantity $\Delta$ or the duration of the pulse $\tau$ are further
 useful measures.
 The carrier envelope phase $\tilde\phi$ is particularly important
 if it is varied in a range comparable with the pulse duration
 $\Delta$. In IPA it is
 anyhow irrelevant; in FPA with $\tilde\phi\simeq\Delta$ the cross
 section of the photon emission would be determined by an involved
 interplay of the carrier phase, the pulse duration and pulse shape
 as well as the intensity of e.m. field  as emphasized, e.g., in
 \cite{Mackenroth}) (see also \cite{Krajewska,Hebenstreit}).
 In the beginning, we drop the carrier phase, thus assuming
 $\tilde\phi\ll\Delta$, and concentrate on the dependence of the
 cross sections on the parameters responsible essentially for
 multi-photon effects. Impact of $\tilde\phi$ on the
 differential production rate is discussed in Sects.~3.~7 and 4.~4.
 In present consideration we drop effect of the pulse focusing which,
 however, is more relevant for longer pulses~\cite{NF1996}
 than those covered in the present review where we consider pulses
 with number of oscillations less than ten.

 The interaction of an electron with e.m. field is characterized by the
 dimensionless field intensity $\xi^2$. For simplicity,  let us consider
 case of generalized Compton scattering where the variable $\xi^2$
 can be determined
 through the average value of the manifestly covariant variable
 $\eta=T^{\mu\nu}p_\mu p_\nu/(p\cdot k)^2$~\cite{HeinzlIlderton}
 (cf.\ also~\cite{NikishovRitus}), where $p$ is the four-momentum
 of a target electron, and
 $T^{\mu\nu}$ is the e.m.\ stress-energy tensor
 $T^{\mu\nu}=g_{\alpha\beta} F^{\mu\alpha} F^{\beta\nu}
 + \frac14 g^{\mu\nu} F_{\alpha\beta}F^{\alpha\beta}$ where
 $F_{\mu\nu}=\partial_\mu A_\nu-\partial_\nu A_\mu$ stands
 for the e.m.\ field strength tensor. In the charge's rest frame
 $\eta= T^{00}/\omega^2$, where the
 stress-energy tensor $T^{00}$ is equal to the
energy density of the  e.m.\ field or to
 the pulse intensity $I_L$. 
 In IPA, the quantity $\xi^2$ is determined by
 \begin{eqnarray}
 \xi^2=\frac{e^2}{m^2}\frac{1}{\tau_{IPA}}
\int\limits_{-\tau_{IPA}/2}^{\tau_{IPA}/2} dt\,\,
 \eta
 =\frac{e^2}{m^2\omega^2}\frac{1}{2\pi}\int\limits_{-\pi}^{\pi}
 d\phi\,\, I_L =\frac{e^2a^2}{m^2}~,
 \label{S22}
 \end{eqnarray}
 where the averaging interval is set equal to the duration of one cycle,
 $\tau_{IPA}=2\pi/\omega$ (we use natural units with
 $c=\hbar=1$, $e^2/4\pi = \alpha \approx 1/137.036$). 
 The generalization to
 a finite pulse may be done in a straightforward manner:
 \begin{eqnarray}
 \xi^2_{FPA}=\frac{e^2}{m^2}\frac{1}{\tau_{FPA}}\int\limits_{-\infty}^{\infty} dt
 \,\, \eta
 =\frac{e^2}{m^2\omega^2} \frac{1}{2\pi N} \int\limits_{-\infty}^{\infty} d\phi\,\,
 I_L~.
 \label{S23}
 \end{eqnarray}
 Now, the interval $\tau_{FPA}$ is determined by the
 number $N$ of oscillations in a pulse as $2\pi N/\omega$.
 That is, the quantity $\xi^2$, which defines the
 production probability and the cross section,
 can be expressed through the averaged value of the intensity
 of a finite laser pulse
 \begin{eqnarray}
 \xi^2=\xi^2_{FPA} \frac{N}{N_0}~,
 \label{S24}
 \end{eqnarray}
or
\begin{eqnarray}
  \xi^2=\frac{N}{N_0}\frac{e^2}{\omega^2m^2}\langle I_L  \rangle
  &\simeq& \frac{N}{N_0} \frac{5.62\cdot 10^{-19}}{\omega^2[{\rm eV}^2]}
  \langle I_L \rangle \left[\frac{\rm W}{\rm  cm^2}\right]\nonumber\\
  &\simeq& \frac{N}{N_0}\,{3.66\cdot 10^{-19}}{\lambda^2[{\mu {\rm m}}^2]}
  \langle I_L \rangle \left[\frac{\rm W}{\rm  cm^2}\right]~,
  \label{S244}
\end{eqnarray}
 where
 $\lambda=2\pi/\omega$ is wave lenght of the background field and
 $N_0\, \langle I_L\rangle=(\omega/2\pi)\int_{-\infty}^{\infty}dt\,I_L$.
 Hence, the normalization
 factor $N_0$ is determined as
\begin{eqnarray}
 N_0=\frac{1}{2\pi}\int\limits_{-\infty}^{\infty}
 d\phi\,(f^2(\phi)+ {f'}^2(\phi))
 \label{S25}
\end{eqnarray}
 and has a meaning of renormalized factor for the photon flux in the case of
 finite pulse.
 The factor $N_0$ is described in some detail below in Sect.~II.
 In fact, for the realistic envelope functions
 $N_0\simeq N$ and, therefore, $\xi^2\simeq \xi^2_{FPA}$.
 The generalization to the Breit-Wheeler process can be done strightforward
 by substitution $p\to k'$ and utilizing the center of mass system.

\begin{figure}[ht]
 \begin{center}
 \includegraphics[width=0.4\columnwidth]{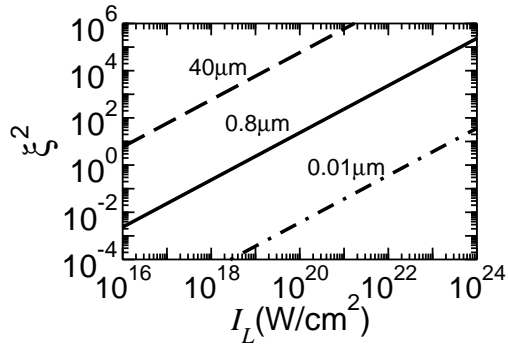}
 \end{center}
 \caption{\small{
 Reduced field intensity $\xi^2$ as a function of averaged
 pulse intensity $I_L$ for different wave lengths.
 Dashed, solid and dot-dashed curves are for $\lambda=$ 40, 0.8,
 and 0.01 $\mu$m, respectively.
 \label{Fig:XI2}}}
 \end{figure}

 For completeness, in Fig.~\ref{Fig:XI2} we exhibit explicit
 dependence of $\xi^2$ on $I_L$
 for different wave lengths for $N=N_0$.
 Dashed, solid and dot-dashed curves are for $\lambda=$ 40, 0.8,
 and 0.01 $\mu$m, respectively, which correspond
 to the infrared, optic, and X-ray scales.

 A second relevant dimensionless variable characterizing both null fields is
\begin{eqnarray}
\zeta=\frac{s_{\rm thr}}{s},
\label{I-zeta}
\end{eqnarray}
 where $s_{\rm thr}=4m^2$ and $s = 2 \omega \omega' (1 - \cos
 \Theta_{\mathbf{k} \mathbf{k'}})$
 (for head-on collision geometry,
 $\Theta_{\mathbf{k k}'}=\pi$); $\omega, \omega'$ and
 $\mathbf{ k},
 \mathbf{ k}'$ are the frequencies and three-wave vectors of the laser
 field and the probe photon, respectively.
 The variable $s_{\rm thr}$ is the square
 of the initial energy at the threshold
 therefore, the variable $\zeta$ is
 a pure kinematic quantity with the meaning that for $\zeta > 1$
 the linear Breit-Wheeler process $\gamma' + \gamma \to e^+ + e^-$
 is sub-threshold, i.e.\ kinematically forbidden. However,
 multi-photon effects enable the non-linear process $\gamma' + n
 \gamma \to e^+ + e^-$ even for $\zeta > 1$ which we refer as
 sub-threshold pair production. The non-linear Breit-Wheeler
 process has been experimentally verified in the experiment E-144
 at SLAC~\cite{SLAC-1997}. There, the minimum number of photons
 involved in one $\ee$ event
 can be estimated by the integer part of $\zeta(1+\xi^2)$, i.e., five.
 To arrive at such an estimate we recall that the reduced strength $\xi$
 is related to the laser intensity $I_L$
 via Eq.~(\ref{S244}),  and therefore,
 at  $\omega'=29$~GeV, $\omega=2.35$~eV,
 and at peak focused laser intensity of $1.3 \times 10^{18} \, {\rm
 W/cm}^2$, one gets $\xi =0.36$ and $\zeta=3.83$. The laser pulses
 contained about thousand cycles in a shot, allowing to neglect the
 details of the pulse shape and duration.

 Some important difference between IPA and FPA is
 that in the first case the variable
 $n = 1, 2,\, \cdots$ is integer, it refers to
 the contribution of the individual harmonics.
 The value $n\omega$ is related to the energy
 of the background field involved into considered
 quantum process. Obviously, this value is a multiple of $\omega$.
 In FPA, the basic subprocess operate with $l$
 {\it background photons},
 where $l$ is a continuous variable.
 The quantity $l\omega$ can be considered as the energy
 partition of the laser beam involved into considered process,
 and it is not a multiple $\omega$.
 Mindful of this fact, without loss of generality, we
 denote the processes with $l>1$ as a generalized multi-photon
 processes, remembering that $l$ is a continuous quantity.

  The Compton process is considered below as a spontaneous emission
  of one photon off an electron in an external e.m.\ wave.
  Evaluation of corresponding transition matrix is close
  to that of case of
  the Breit-Wheeler process because both processes are crossed
  to each other.
  Despite of the similarities
  of these two processes, the physical meaning of
  the dynamical variables and observables is quite different.
  For the sake of completeness, we start our analysis
 from fully differential cross sections which
 are calculated as a  function
 of the frequency of the outgoing photon at fixed scattering angle.
 The main difference to the previous studies
 mentioned above is utilizing a wider class of
 the pulse envelope functions including flat-top envelopes.
 However, the fully
 differential cross section has a complicated structure
 being rapidly oscillating function of
 the energy of the outgoing photons
 $\omega'$ at fixed production angle $\theta'$,
 especially in the kinematically
 forbidden region. It is clear that
 experimental studying the multi-photon dynamics in case of rapidly
 varying cross sections is a challenging task.
 Rather integrated observables may overcome
 this problem.

 But here one has to be careful. The totally
 integrated cross section is not suitable for this aim, because
 in this case the integration starts from the minimum value of
 the energy of the outgoing photon, $\omega'_1$,
 kinetically allowed for the electron - one photon interaction,
 and this region dominates
 in the total cross section, masking the relatively weak effects
 of electron - multi-photon interactions. To highlight the role of the
 multi-photon interaction the lower limit of integration $\omega'$ must be
 shifted relative to $\omega'_1$:  $\omega'>\omega_1'$.
 Such partially integrated cross sections
 are smooth functions of $\omega'$ and allow to study directly the multi-photon
 dynamics. Similarly to the variable $\zeta$ in the Breit-Wheeler process,
 the ratio $\kappa=\omega'/\omega'_1>1$ may be considered as a
 sub-threshold variable in non-linear Compton scattering.

We show below that in the considered quantum processes the
production probability (or cross section) is determined
by the non-trivial interplay of two dynamical effects. The first one is
related to the shape and duration of the pulse, while
the second one is the non-linear dynamics of the
electron (positron) in the
strong
electromagnetic field, independently of the pulse
geometry.
These two effects play quite different roles in two limiting cases.
The pulse shape effects manifest most clearly
in the weak-field regime characterized by
small values of the reduced field intensity $\xi^2$.
The rapid variation
of the e.m. field in very short (and, in particular, in sub-cycle) pulses
enhances strongly few-photon events such that their
probability
may exceed the IPA result by
orders of magnitude.
Non-linear multi-photon dynamics of the strong
electromagnetic field plays
a dominant role at large values of $\xi^2$.
In this case, results of the IPA and the FPA
are close to each other.
In the transition region, i.e.\
at intermediate values $\xi^2 \sim 1$, the
observables are determined
by the interplay of both effects which must be taken
into account simultaneously.
For the quantum processes in IPA we refer the reader
to the review paper~\cite{Ritus-79}.

 This review is based on the methods and results obtained in
 Refs.~\cite{TTKH-2012,TitovPRA,Nousch} and~\cite{TitovEPJD}.
 It is organized as follows.
 In Sect.~2 we discuss the properties of envelope
 functions used below.
 Sect.~3 is devoted to the non-linear Breit-Wheeler
 process for different pulse shapes, pulse durations and e.m. field
 intensities deriving  the basic expressions  for
 the probability of $\ee$ creation in FPA. We successively analyze cases
 of (i) small pulse
 duration with number of oscillations $N=2...10$ at different pulse intensities,
 (ii) the case of large field intensity where
 the pulse shape becomes unessential, and
 (iii) sub-cycle pulses with $N<1$, where the pulse structure
 is particularly important.
 Special attention is paid
 to the impact of the carrier phase.
 In Sect.~4 we discuss several aspects of non-linear Compton scattering
 for short and sub-cycle pulses.
 Our conclusions are presented in Sect.~5.
 In Appendix, for completeness and
 easy reference, we present some details of a derivation of
 the $\ee$ production probability for very high intensities, $\xi^2\gg1$.

\section{Envelope functions}

 Below, we are going to analyze dependence of observables on
 the shape of $f(\phi)$ in Eq.~(\ref{III1}) for two types of envelopes:
 the one-parameter hyperbolic secant~(hs) shape and the two-parameter
 symmetrized Fermi~(sF) shape widely used for parametrization of the
 nuclear density~\cite{Luk}:
 \begin{eqnarray}
 f_{\rm hs}(\phi)=\frac{1}{\cosh\frac{\phi}{\Delta}}~,\qquad
 f_{\rm sF}(\phi)=\frac{\cosh\frac{\Delta}{b} +1}
 {\cosh\frac{\Delta}{b} +\cosh{\frac{\phi}{b}}}~.
 \label{E1}
 \end{eqnarray}
 These two shapes cover
 a variety of relevant envelopes discussed in literature
 (for details see \cite{TitovPRA}).
 The parameter $b$ in the sF shape describes the ramping time
 in the neighborhood of $\phi \sim \Delta$. Small values of ratio $b/\Delta$
 cause a flat-top shaping. At $b/\Delta\to0$, the sF shape
 becomes a rectangular pulse~\cite{Boca-2009}.
 In the following, we choose the ratio
 $b/\Delta$ as the second independent
 parameter for the sF envelope function.
 The parameter $\Delta$ characterizes the
 pulse duration $2\Delta$ with $\Delta=\pi N$,
 where $N$ has a meaning
 of a "number of oscillations" \ in the pulse.
 Certainly, such a definition is rather conditional
 and is especially meaningful for the flat-top envelope
 with small values of $b/\Delta$. In the case of the
 hs envelope shape, the number of oscillations
 with small amplitudes may exceed $N$. Nevertheless,
 for convenience we call $N$ as a "number of oscillations
 in a pulse" \ for given $f(\phi)$, relying on its relation to
 the shape parameter $\Delta$.
 It was shown that
 the properties of the two-parameter sF shape for
 large values of $b/\Delta\simeq0.3\dots0.5$ are close to that of
 the one-parameter hs shape. Therefore, as mentioned above,
 in order to stress the difference between one- and
 two-parameter (flat-top) envelopes we focus our
 consideration on the choice of $b/\Delta=0.15$
 throughout present paper.

\begin{figure}[ht]
 \begin{center}
 \includegraphics[width=0.35\columnwidth]{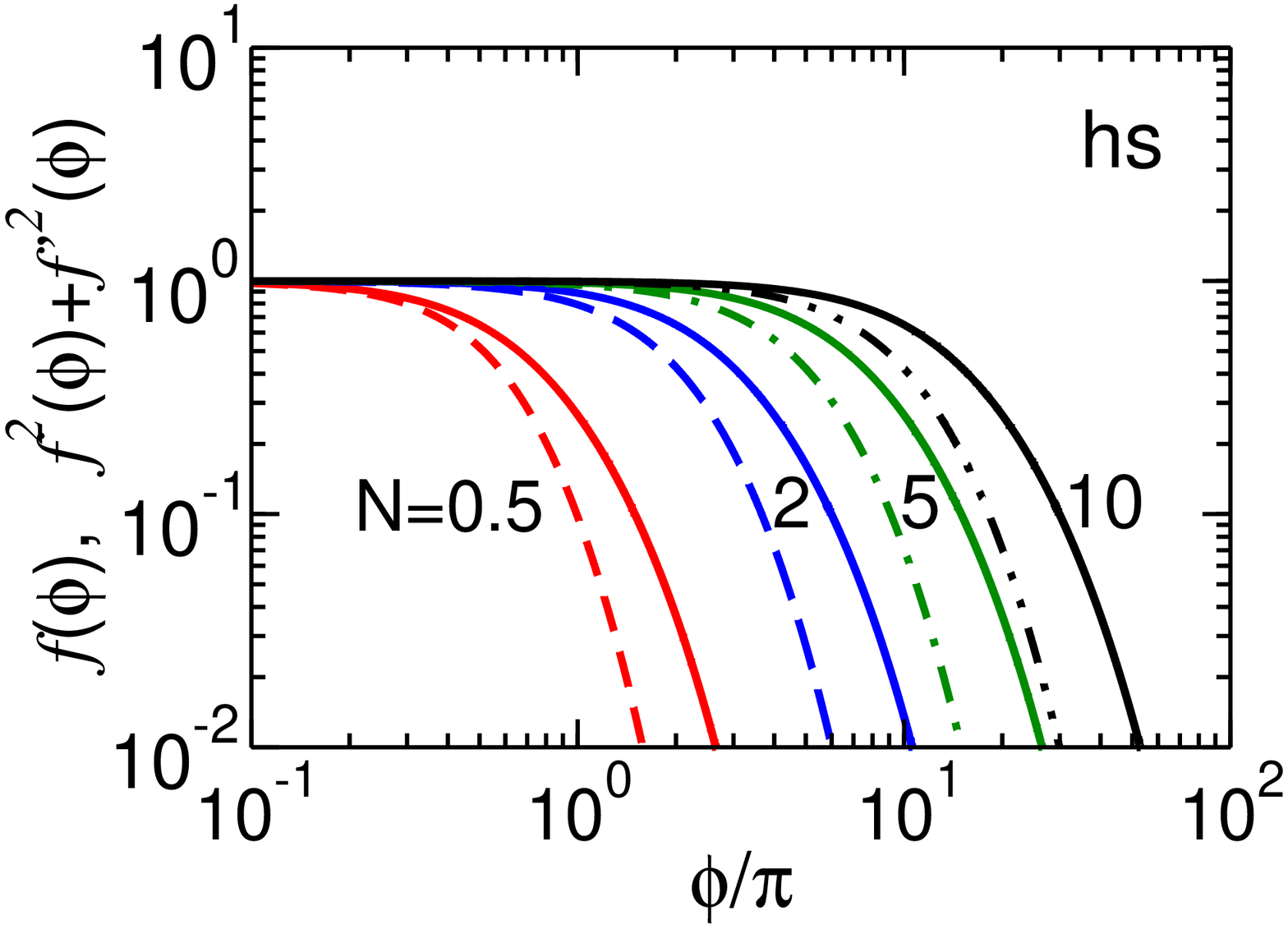}\qquad\qquad
 \includegraphics[width=0.35\columnwidth]{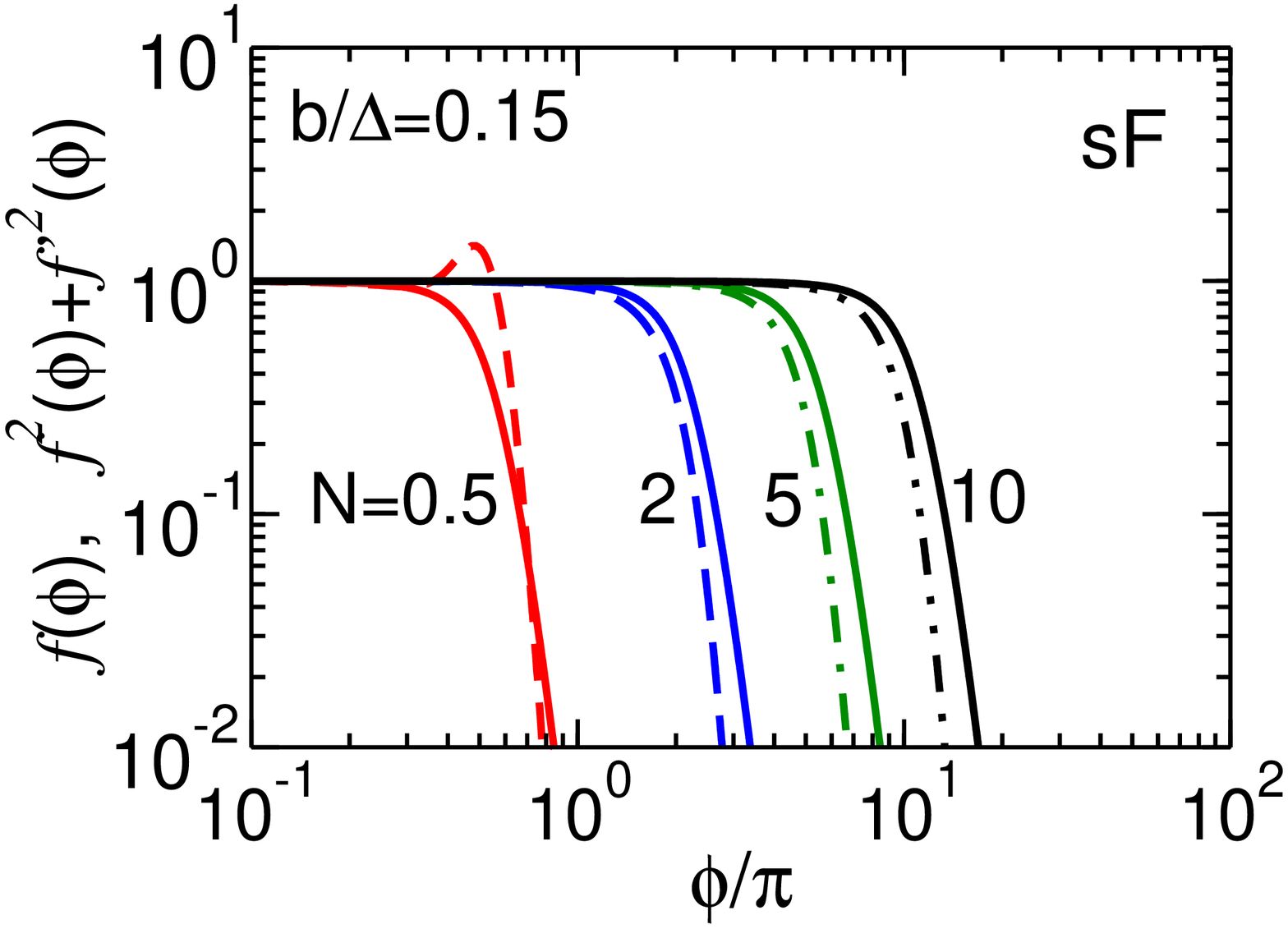}
 \end{center}
 \caption{\small{
 The envelope functions $f(\phi)$
 and the integrand  $f^2(\phi)+{f'}^2(\phi)$ in Eq.~(\ref{S25})
 as the functions of the invariant phase $\phi=kx$.
 The thick solid curves
 labeled by $N$ are for $f(\phi)$.
 The dashed, long-dashed, dot-dashed and dot-dot-dashed curves are for
 $f^2(\phi)+{f'}^2(\phi)$ with $N=0.5$, 2, 5 and 10, respectively.
 Left and right panels are for hs and  sF envelope  shapes,
 respectively. 
 \label{Fig:01}}}
 \end{figure}

 The envelope shape $f(\phi)$ and the integrand  $f^2(\phi)+ {f'}^2(\phi)$
 (which is proportional to the square of the e.m. field strength)
 in Eq.~(\ref{S25}) as functions of the invariant phase for hs and sF shapes
 are  shown in  Fig.~\ref{Fig:01} in left and right
 panels, respectively.
 The numbers in the plot indicate the number of
 oscillations in a pulse $N$. The thick solid curves
 labeled by $N$ are for $f(\phi)$.
 The dashed, long-dashed, dot-dashed and dot-dot-dashed curves are for
 $f^2(\phi)+{f'}^2(\phi)$ with $N=0.5$, 2, 5 and 10, respectively.
 For the smooth  hs shape the integrand is also a smooth
 function (cf.\ Fig.~\ref{Fig:01}, left panel).
 For the flat-top sF envelope shape and  $N \ge 2$
 both, $f(\phi)$ and the integrand $f^2(\phi)+{f'}^2(\phi)$
 are smooth functions of the invariant phase
 which is more compact as compared with the hs shape with
 the same value of $N$. At $N=0.5$ and $\phi\sim\Delta$ the integrand
(see dashed curve in the right panel) displays some
overshoot resulting locally in the height
 $h=1/4 + (\frac{\Delta}{b}/4\Delta)^2\simeq 1.37$.
 Increasing $\Delta$ (or $b/\Delta$) leads
 to a vanishing of this overshoot.

 For the hs envelope, the normalization factor in Eq.~(\ref{S25}) has the form
 \begin{eqnarray}
 N^{\rm hs}_0=\frac{\Delta}{\pi}\left(1+  \frac{1}{3\Delta^2}
 \right)~,
 \label{EE5}
 \end{eqnarray}
while for the sF shape the normalization factor reads
\begin{eqnarray}
N^{\rm sF}_0=\frac{\Delta}{\pi}\left( F_1\left(t \right)
 +  F_2\left(  t \right) \frac{b}{\Delta}\right),\,\,\,
 t =\frac{  1 + \cosh\frac{\Delta} {b} }  {\sinh\frac{\Delta} {b}}~,
\label{EE6}
\end{eqnarray}
where
\begin{eqnarray}
 F_1(t)&=&\frac{(t^2+1)(-t^4 +10 t^2 -1)}{16t}~,\nonumber\\
 F_2(t)&=&\frac {3t^{10}-  35t^8+  90t^6-  90t^4+
 35t^2-3}{24(t^2-1)^3}~.
 \label{EE7}
\end{eqnarray}
In the limit  $\frac{b}{\Delta}\to 0 $,
\begin{eqnarray}
 N_0^{\rm sF} = \frac{\Delta}{\pi} +
 {\cal O}\left(\exp[-\frac{\Delta}{b}] \right) \simeq  \frac{\Delta}{\pi} .
 \label{EE8}
\end{eqnarray}

\begin{figure}[ht]
\begin{center}
\includegraphics[width=0.35\columnwidth]{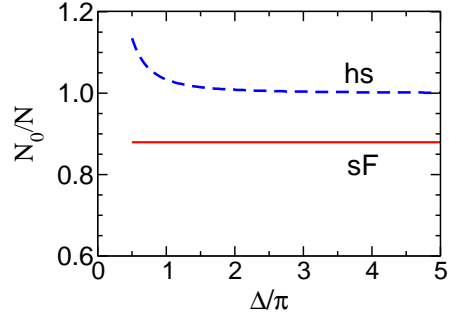}
\end{center}
 \caption{\small{
 The scaled normalization factor $N_0/N$
 as a function of the number of oscillations in the pulse,
 $N={\Delta}/{\pi}$, for hs and sF shapes,
 shown by the dashed blue and
 solid red curves, respectively.
 \label{Fig:02}}}
 \end{figure}

 The normalization factor $N_0$
 scaled by $N=\Delta/\pi$ as a function of $N$
 for hs and sF shapes is exhibited in Fig.~\ref{Fig:02},
 shown by the dashed and
 solid curves, respectively.

 For the hs shape, $N_0\simeq N$ at $N\ge1$
 and slightly increases for the sub-cycle envelopes with $N<1$
 (cf. Eq.~(\ref{EE5})).
 In case of a flat-top envelope, the ratio $N_0/N$ is independent of
 $\Delta$, according to Eq.~(\ref{EE7}).
 The contribution of ${f'}^2$
 in~(\ref{S25}) is weak and varies
 from 0.2\%
 to 3.8\%  for $b/\Delta=0.01$ and 0.2, respectively.
 In the limit $b/\Delta\to 0$ it vanishes and
 $N_0\to N$ and therefore, the overshoot in the integrand does
 not affect the integral in Eq.~(\ref{S25}).
 But taking into account that very small values of $b/\Delta$
 seems to be not realistic, we restrict our actual calculations
 to the finite value $b/\Delta=0.15$,
 where the overshoot in $f^2(\phi)+ {f'}^2(\phi)$ is minor.

 For the sake of completeness,
 we present also the behavior of
 e.m.\ potential $\mathbf{A}$
 and the electric field strength
 ${\mathbf {E}}= - \partial {\mathbf{A}} / \partial t$,
 where ${\mathbf{A}}$ is
 given by Eqs.~(\ref{III1}) and (\ref{E1}) as functions of
 the invariant phase $\phi$.
 The e.m. potential and strength
 for the one- and two-parameter envelope functions read
\begin{eqnarray}
 {A_x}&=&a\,f(\phi)\cos\phi~,\qquad
 {A_y} = a\,f(\phi)\sin\phi~,
 \label{EEA}\\
 {E_x}&=&\omega\, {A_x}
 \left[
 -(\ln f(\phi))' + \tan\phi
 \right]~,\label{EEx}\\
 {E_y}&=&\omega\, {A_y}
 \left[
 -(\ln f(\phi))' - \cot\phi
 \right]~,
 \label{EEy}
\end{eqnarray}
with $a=|~{\mathbf a}_1|=|~{\mathbf a}_2|$ and
\begin{eqnarray}
 -(\ln f(\phi))'=
 \left\{
\begin{array}{ll}
 \frac{1}{\Delta}\tanh\frac{\phi}{\Delta},&{\rm hs},\\
 \frac{1}{b}\frac{\sinh{\frac{\phi}{b}}}{\cosh\frac{\Delta}{b}+\cosh\frac{\phi}{b}},&
 {\rm sF}.
\end{array}
 \right.
 \label{EE2}
\end{eqnarray}

\begin{figure}[ht]
 \begin{center}
\includegraphics[width=0.35\columnwidth]{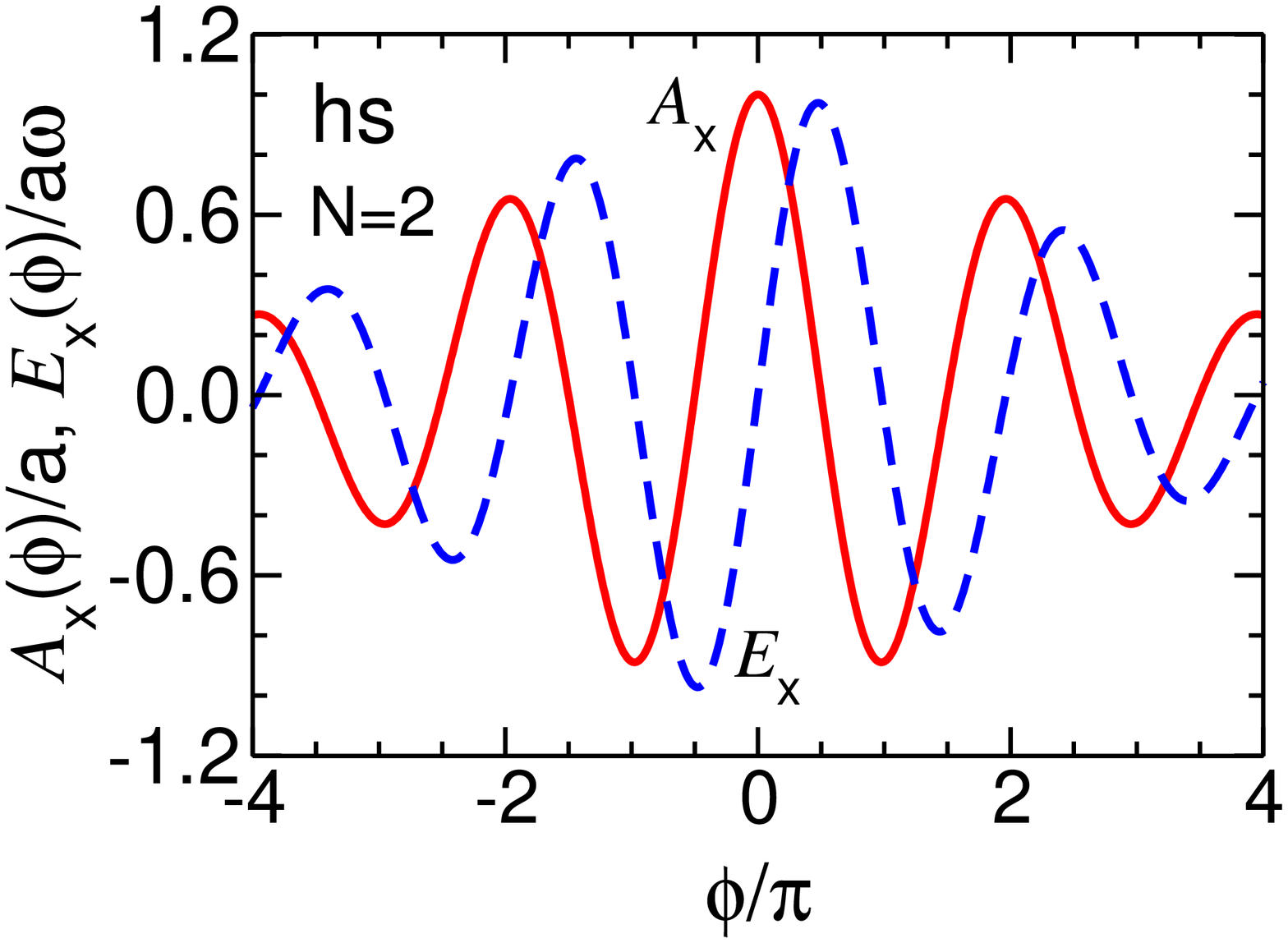}\qquad\qquad
\includegraphics[width=0.35\columnwidth]{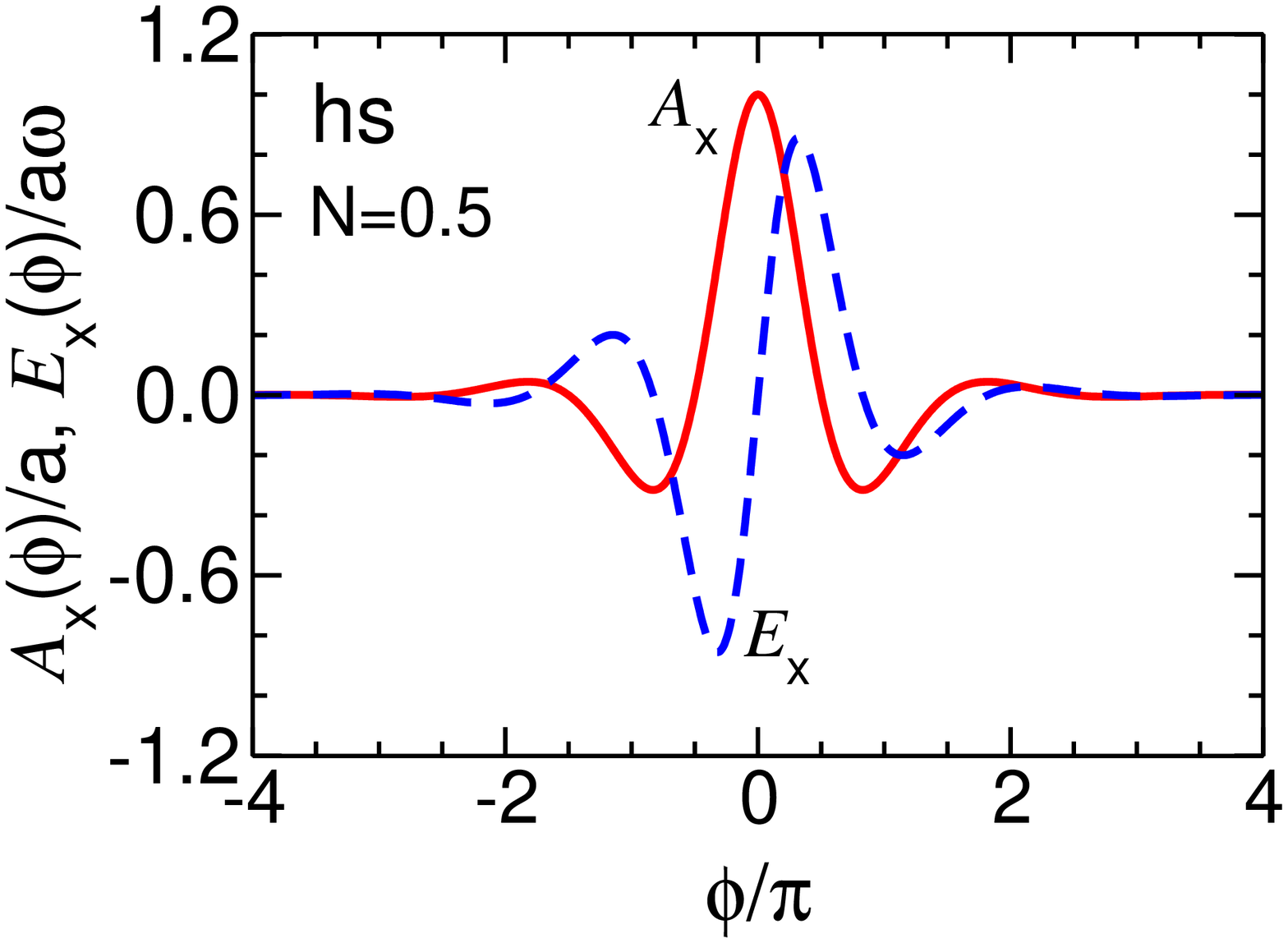}\\
\includegraphics[width=0.35\columnwidth]{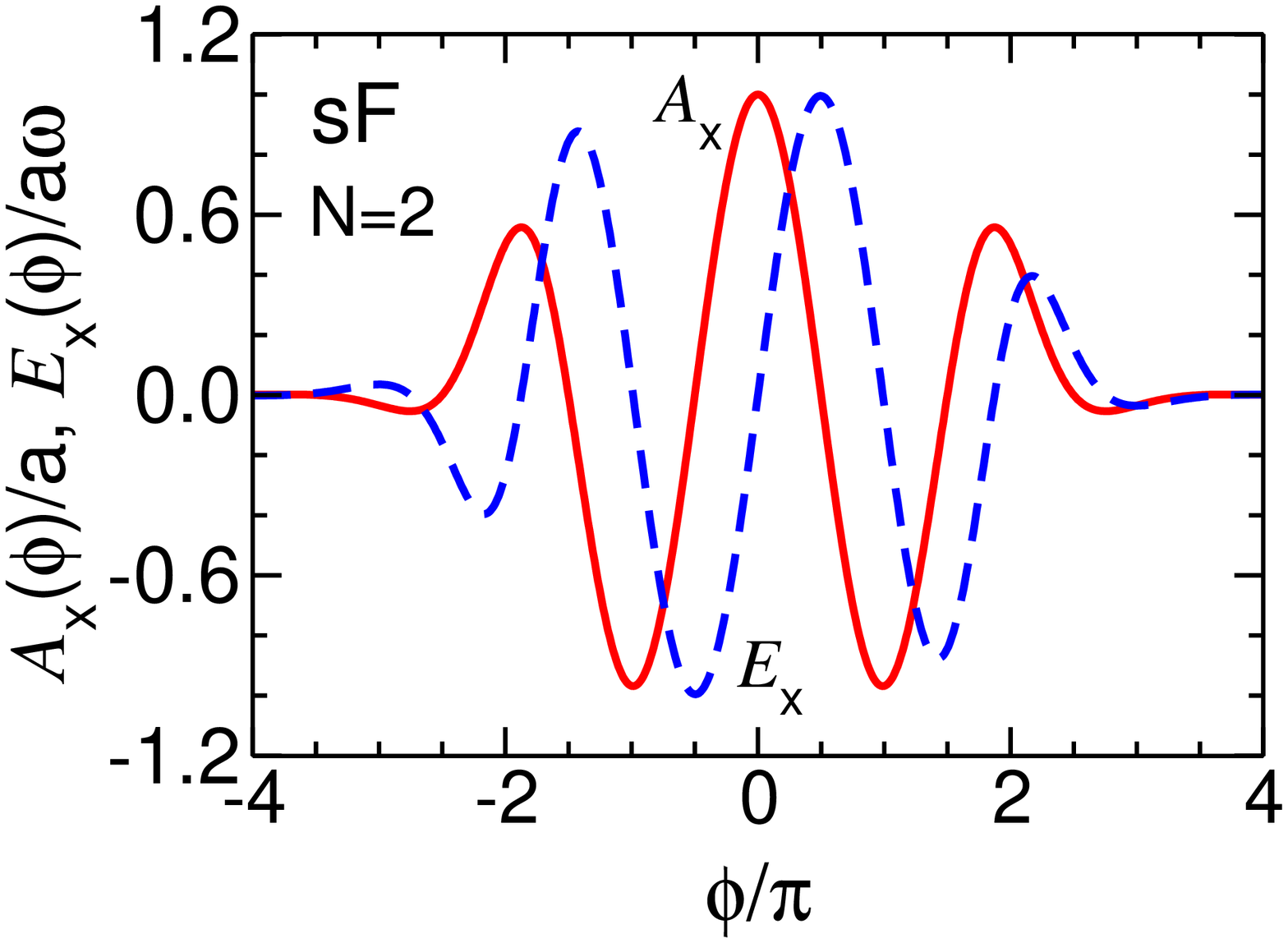}\qquad\qquad
\includegraphics[width=0.35\columnwidth]{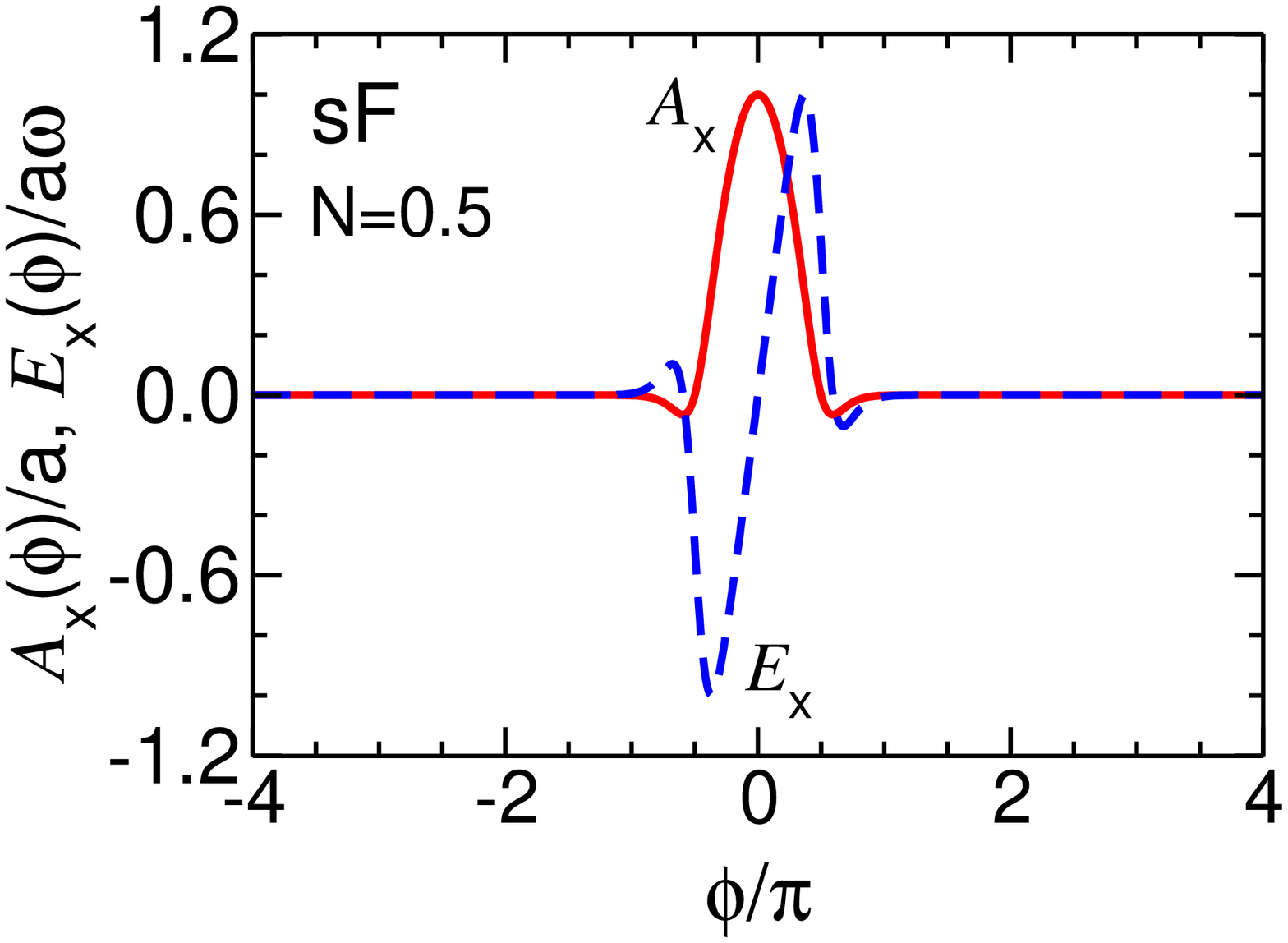}\\
\includegraphics[width=0.35\columnwidth]{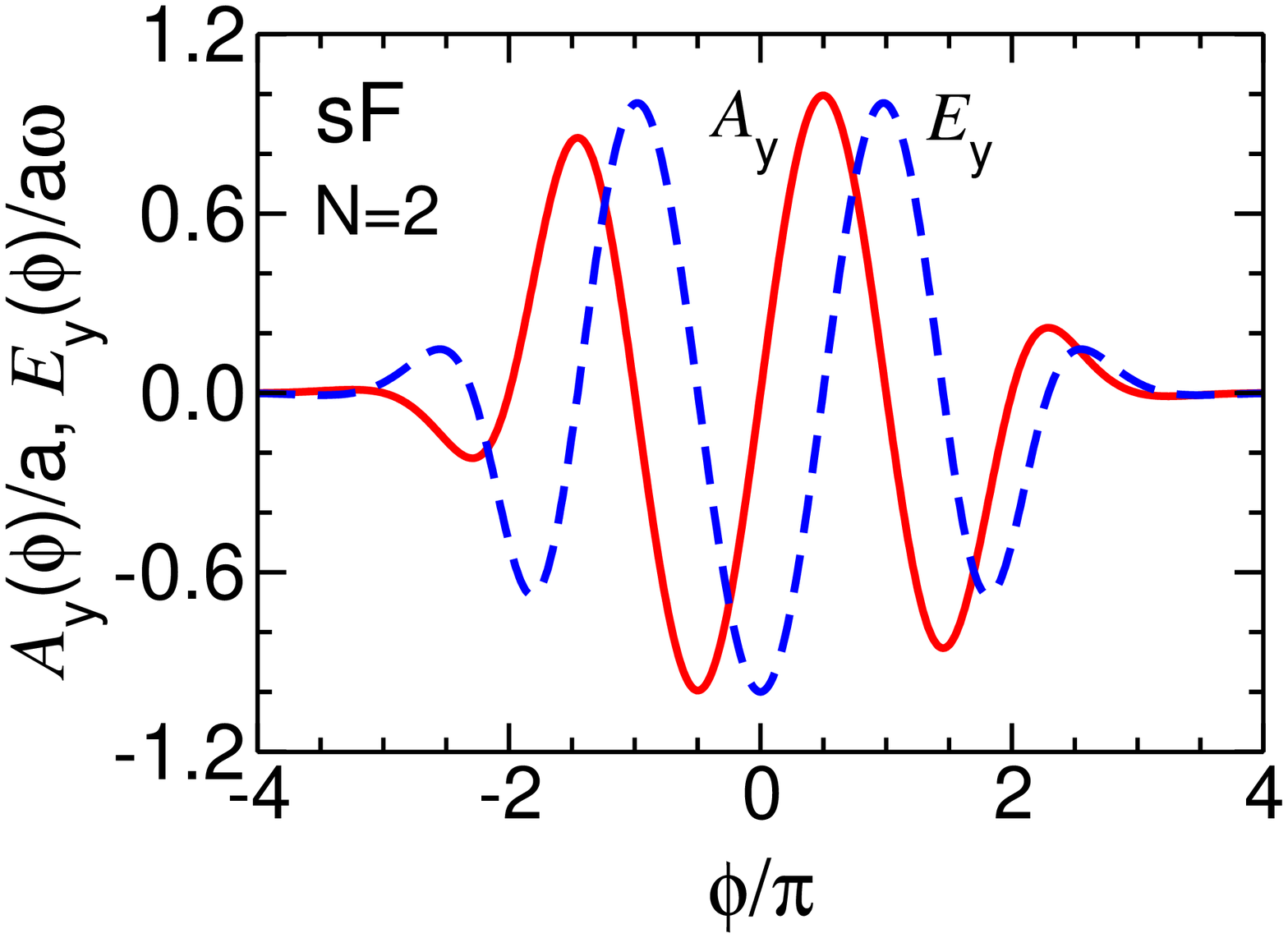}\qquad\qquad
\includegraphics[width=0.35\columnwidth]{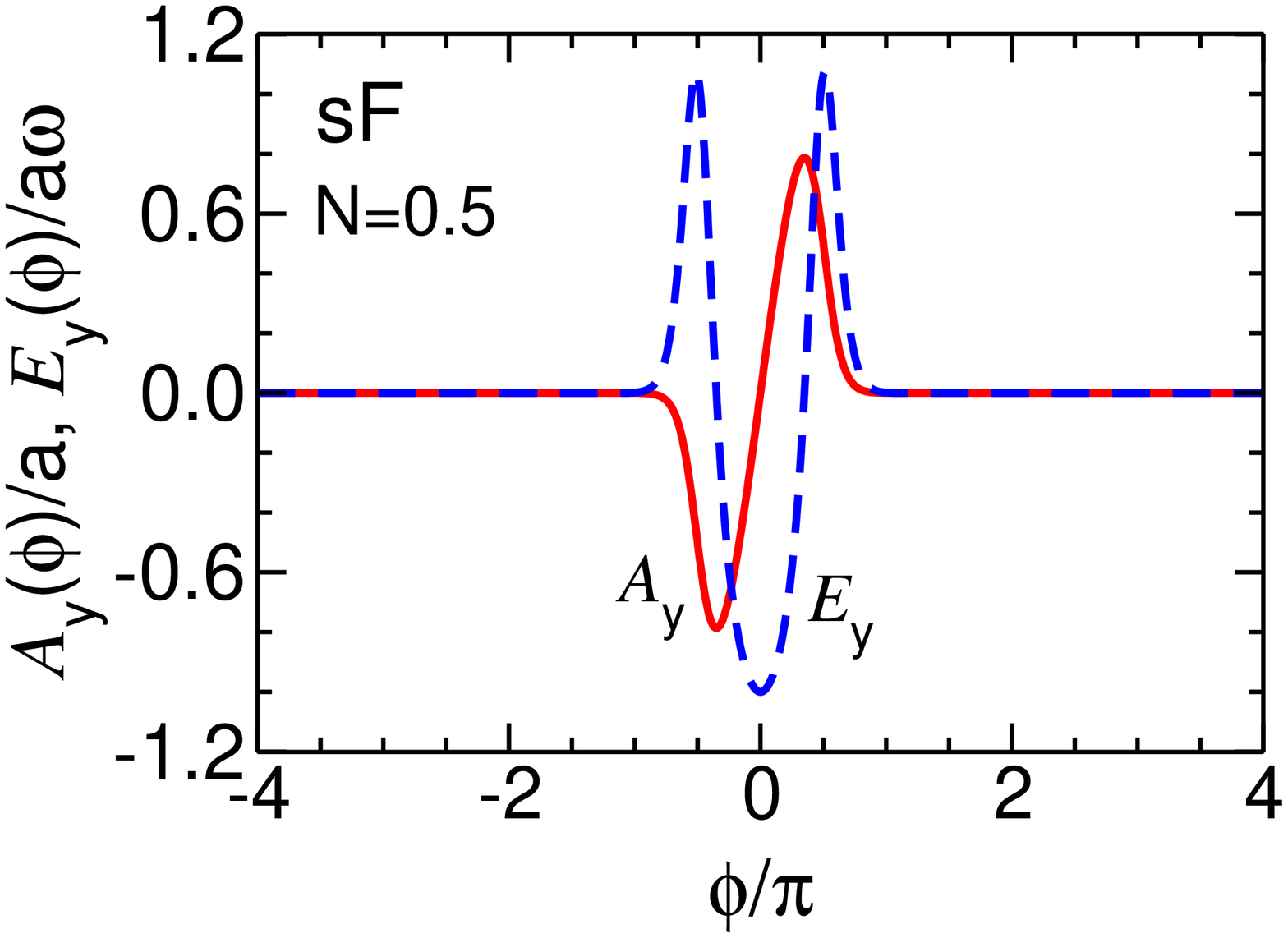}
\end{center}
 \caption{\small{
 The e.m.\ potentials $A/a$ (solid curves) and field
 strengths $E/a\omega$ (dashed curves) as functions of the invariant
 phase $\phi$. The upper and middle panels correspond to the
 hyperbolic secant (hs) and symmetrized Fermi (sF) shapes,
 respectively, for the $x$ components.
 The lower panels correspond to the $y$ components
 for the sF envelope shape.
 The left and right panels are for pulses with $N=2$ and $N=0.5$,
 respectively.
 \label{Fig:03}}}
\end{figure}

 The scaled potentials $A_x/a$ and the scaled
 strengths $E_x/a\omega$ as functions of the invariant
 phase are exhibited by solid and dashed curves,
 respectively, in upper and middle panels in Fig.~\ref{Fig:03}
 for the hs and sF shapes.
 The left and right panels correspond to the pulses with
 $N=2$ and 0.5, respectively.
 The result for the hs shape with $N=0.5$ is close qualitatively  to that
 of~\cite{Mackenroth-2011}.
 One can see that the duration of the pulse increases with
 increasing number of oscillations. The
 flat-top sF shape is more compact compared to the hs shape with
 the same value of the pulse "scale"\ parameter $\Delta$.

 The result for $y$ components is exhibited in Fig.~\ref{Fig:03}, lower
 panels, where we restrict ourselves to the example of the
 flat-top sF envelope shape. For short pulses with $N>2$, the
 contribution of the first terms in Eqs.~(\ref{EEx}) and (\ref{EEy})
 are relatively small and, therefore, the approximate relations
 $A_y\simeq E_x/\omega$ and $E_y\simeq -\omega\,A_x$ are valid.
 Both $A_y$ and $E_y$ are finite. The same is true for the sF shape
 with $N\ge 2$.
 This is illustrated in Fig.~\ref{Fig:03}, lower panel (left), where the
 result for the sF envelope with $N=2$ is shown.
 The approximate relations are valid also
 for sub-cycle pulse with $N=0.5$
 and for the one-parameter hs shape. In case of the flat-top envelope
 for $N=0.5$, the above approximate
 relations are valid for $A_y$ and for the central part of $E_y$
 (cf. Fig.~\ref{Fig:03} lower panel (right)).
 In the border area with $\phi\approx\Delta=\frac{\pi}{2}$,
 the strength $E_y$ has finite narrow peaks with
 height $\tilde h=\left(\frac{\Delta}{b}\right)
 \frac{\sin\Delta}{4\Delta} +{\cal O}(\exp(-\Delta/b))
 \simeq 1.06$. The height of these peaks
 decreases with increasing $\Delta$
 at fixed $b/\Delta$ and
 for $N\ge2$ it becomes negligibly small.
 This "pick-like"\ behavior for the flat-top
 shape can be compared with the
 the popular rectangular pulse~\cite{Boca-2009} where the derivative
 $f'(\phi)=\theta'(\phi-\Delta)=\delta(\phi-\Delta)$ is singular
 at $\phi=\Delta$.
 But such "pick-like" or even singular behavior of $\mathbf{E}$ at
 the border does not affect the transition matrix elements, discussed in the next
 section (cf. Eq.~(\ref{III4})),
 since they are determined  by ${\mathbf A}$ and ${\mathbf A}^2$
 rather than the e.m.\ strength.
 Therefore, our plots and discussions for the e.m.\ strength $E_{x,y}$ have
 an illustrative character since the dynamics of the considered
 process is determined purely by the e.m. potential $A_{x,y}$,
 which is taken, in our approach, as a primary quantity.

\section{The ${\mathbf{\ee}}$ pair production in a finite pulse}

\subsection{General formalism}

 In our consideration of quantum processes,
 we start from $\ee$ pair production
 in the interaction of a probe photon with
 a circularly polarized
 e.m. field described by Eq.~(\ref{III1}).
 Within the Furry picture, the process is diagrammatically
 represented by a one-vertex graph, describing the decay
 of the probe photon with the four-momentum $k'$ into a laser dressed
 $\ee$ pair, where the presence of the background e.m. field
 is included in the Volkov solution of the outgoing
 $e^+$ and $e^-$. (In the weak-field approximation this graph
 turns into the known two two-vertex graphs for the perturbative
 Breit-Wheeler process).
 Utilization of~(\ref{III1}) leads to two significant modifications
 of the transition amplitude in FPA compared to IPA.
 In IPA, the Volkov solutions~\cite{Volkov,LL4} refer to fermions with
 quasi-momenta $q_\mu=p_\mu + k_\mu\frac{\xi^2 m^2}{2(k\cdot p)}$
 and dressed masses $m_*^2=m^2(1+\xi^2)$.
 In FPA, all fermion momenta and masses
 take their vacuum values  $p$ and $m$, respectively,
 whereas the corresponding wave functions are modified
 in accordance with the Volkov solution
 (with more complicated compare to IPA, phase factor).
The finite (in space-time) e.m. potential (\ref{III1}) for FPA
requires the use of Fourier integrals
for invariant amplitudes, instead of Fourier
series which are employed in IPA.
The partial harmonics
become thus continuously in FPA.
The $S$ matrix element is expressed generically as
\begin{eqnarray}
S_{fi}
=\frac{-ie}{\sqrt{2p_02p_0'2\omega'}}
\int\limits_\zeta^\infty dl
\, M_{fi}(l)(2\pi)^4\delta^4(k'+ lk -p-p'),
\label{III3}
\end{eqnarray}
where $k$, $k'$,  $p$ and $p'$ refer to the four-momenta of the
background (laser) field (\ref{III1}),
incoming probe photon, outgoing positron and electron,
respectively, the low limit  $\zeta$ is defined in Eq.~(\ref{I-zeta}).
The transition matrix $ M_{fi}(l)$
consists of four terms
\begin{eqnarray}
\, M_{fi}(l)=\sum\limits_{i=0}^3  M^{(i)}\,C^{(i)}(l)~,
\label{III4}
\end{eqnarray}
where
\begin{eqnarray}
C^{(0)}(l)&=&\frac{1}{2\pi}\int\limits_{-\infty}^{\infty}
d\phi \,{\rm e}^{il\phi -i{\cal P(\phi)}}~,\nonumber\\
C^{(1)}(l)&=&\frac{1}{2\pi}\int\limits_{-\infty}^{\infty}
d\phi f^2(\phi)\,{\rm e}^{il\phi -i{\cal P(\phi)}}~,\nonumber\\
C^{(2)}(l)&=&\frac{1}{2\pi}\int\limits_{-\infty}^{\infty}
d\phi f(\phi)\,\cos\phi\,{\rm e}^{il\phi -i{\cal P(\phi)}}~,\nonumber\\
C^{(3)}(l)&=&\frac{1}{2\pi}\int\limits_{-\infty}^{\infty}
d\phi f(\phi)\,\sin\phi\,{\rm e}^{il\phi -i{\cal P(\phi)}}~,
\label{III5}
\end{eqnarray}
with
\begin{eqnarray}
{\cal P(\phi)}=z\int\limits_{-\infty}^{\phi}\,d\phi'\,\cos(\phi'-\phi_0)f(\phi')
-\xi^2\zeta u\int\limits_{-\infty}^\phi\,d\phi'\,f^2(\phi')~.
\label{III6}
\end{eqnarray}
The quantity $z$ is related to $\xi$, $l$, and
$u\equiv(k'\cdot k)^2/\left(4(k\cdot p)(k\cdot p')\right)$ via
\begin{eqnarray}
z=2l\xi\sqrt{\frac{u}{u_l}\left(1-\frac{u}{u_l}\right)}
\label{III7}
\end{eqnarray}
with $u_l\equiv l/\zeta$.
The phase $\phi_0$ is equal to the azimuthal angle of
the direction of flight of the outgoing electron
in the $\ee$ pair rest frame $\phi_0=\phi_{p'}\equiv\phi_{e}$ and
it is related to the azimuthal angle of
the positron momentum as $\phi_0=\phi_{e^+} + \pi$.
Similarly to IPA,
it can be determined through invariants $\alpha_{1,2}$
as $\cos\phi_0=\alpha_1/z$,  $\sin\phi_0=\alpha_2/z$ with
$\alpha_{1,2}=e\left(a_{1,2}\cdot p/k\cdot p-a_{1,2}\cdot p'/k\cdot p'\right)$.

The transition operators $ M^{(i)}$ in Eq.~(\ref{III4})
have the form
\begin{eqnarray}
M^{(i)}=\bar u_{p'}\,\hat M^{(i)}\,v_p~
\label{B1}
\end{eqnarray}
with
\begin{eqnarray}
 \hat M^{(0)}&=&\fs\varepsilon'~,\quad
 \hat M^{(1)}=-
  \frac{ e^2a^2 \,
 (\varepsilon'\cdot k)\,\fs k}
 {2(k\cdot p)(k\cdot p')}~,\nonumber\\
 \hat M^{(2,3)}&=&\frac{e\fs a_{(1,2)}\fs k\fs
 \varepsilon'}{2(k\cdot p')} - \frac{e\fs \varepsilon'\fs k\fs
 a_{(1,2)}}{2(k\cdot p)}~.
 \label{B2}
\end{eqnarray}
where $u_{p'}$ and $v_{p}$ are the Dirac spinors of the electron and positron,
respectively, and
$\varepsilon'$ is the polarization four-vector of the probe photon.

The integrand of the function $C^{(0)}$ in Eqs.~(\ref{III5}) does not contain the envelope
function $f(\phi)$ and therefore
it is divergent. One can regularize it by using the
prescription of Ref.~\cite{Boca-2009} which leads to
\begin{eqnarray}
C^{(0)}(l)&=&\frac{1}{2\pi l} \int\limits_{-\infty}^{\infty}
d\phi
\left( z\cos(\phi-\phi_0)\,f(\phi)-\xi^2\zeta u\,f^2(\phi)\right)
\,{\rm e}^{il\phi -i{\cal P(\phi)}}\nonumber\\
&&\qquad\qquad\qquad+\,\,\delta(l)\,{\rm e}^{-i{\cal P}(0)}~.
\label{III8}
\end{eqnarray}
This expression contains a singular (last) term which however,
does not contribute because of kinematical
restriction, implying $l > 0$.

The differential probability
of $\ee$ pair production
in terms of the transition matrix $ M_{fi}(l)$ in Eq.~(\ref{III3})
reads
\begin{eqnarray}
{d W}
=\frac{\alpha\zeta^{1/2}}{2\pi N_0 m}
 \,\int\limits_\zeta^{\infty}
 \,dl\,\, |M_{fi}(l)|^2\,\frac{d\mathbf{p}}{2p_0}\,\frac{d\mathbf{p}'}{2p'_{0}}
 \delta^4(k'+ lk -p-p')~.
\label{III9-0}
\end{eqnarray}
It may be represented in conventional form as
a function of $u$ and $\phi_e$
\begin{eqnarray}
\frac{d W}{d\phi_e\,du }
=\frac{\alpha m\zeta^{1/2}}{16\pi N_0}
\,\frac{1}{u^{3/2}\sqrt{u-1}}
 \,\int\limits_\zeta^{\infty}\,dl\ w{(l)}
\label{III9}
\end{eqnarray}
with
\begin{eqnarray}
&&\frac12\,w(l)=
(2u_l+1)|C^{(0)}(l)|^2 +\xi^2(2u-1)(|C^{(2)}(l)|^2 +|C^{(3)}(l)|^2 )
\nonumber\\
&&\,\,\,\,+\,
{\rm Re}\, C^{(0)}(l)\left(
\xi^2 {C^{(1)}(l)} -\frac{2z}{\zeta}(\cos\phi_0 {C^{(2)}(l)}
+\sin\phi_0  {C^{(3)}(l)})
\right)^*~.
\label{III20}
\end{eqnarray}
 The normalization factor $N_0$ is determined by Eq.~(\ref{S25}) and has been discussed
 in the previous section.

It is convenient to express the
$C^{(i)}(l)$ functions defined in Eqs.~(\ref{III5}) and (\ref{III8}) through
the new, basic functions $Y_l$ and $X_l$, which may be considered as an
analog of the Bessel functions in IPA,
\begin{eqnarray}
Y_l(z)&=&\frac{1}{2\pi} {\rm e}^{-il\phi_0}\int\limits_{-\infty}^{\infty}\,
d\phi\,{f}(\phi)
\,{\rm e}^{il\phi-i{\cal P}(\phi)} ~,\nonumber\\
X_l(z)&=&\frac{1}{2\pi}{\rm e}^{-il\phi_0} \int\limits_{-\infty}^{\infty}\,
d\phi\,{f^2}(\phi)
\,{\rm e}^{il\phi-i{\cal P}(\phi)}~.
\label{III24}
\end{eqnarray}
The new
representation of the basic functions $C^{(i)}(l)$ reads
\begin{eqnarray}
C^{(1)}(l)&=&X_l(z)\,{\rm e}^{il\phi_0}~,\nonumber\\
C^{(2)}(l)&=&\frac{1}{2}\left( Y_{l+1}{\rm e}^{i(l+1)\phi_0}
+ Y_{l-1}{\rm e}^{i(l-1)\phi_0}\right)~,\nonumber\\
C^{(3)}(l)&=&\frac{1}{2i}\left( Y_{l+1}{\rm e}^{i(l+1)\phi_0}
- Y_{l-1}{\rm e}^{i(l-1)\phi_0}\right)~,\nonumber\\
C^{(0)}(l)&=&\widetilde Y_l(z){\rm e}^{il\phi_0}, \,\,
\widetilde Y_l(z)=\frac{z}{2l} \left(Y_{l+1}(z) +
Y_{l-1}(z)\right) - \xi^2\frac{u}{u_l}\,X_l(z)~. \label{III25}
\end{eqnarray}
It allows to express $w(l)$ in Eq.~(\ref{III20}) in the form
\begin{eqnarray}
w(l)=
 2 \widetilde Y^2_l(z) +\xi^2(2u-1)
\left(Y^2_{l-1}(z)+ Y^2_{l+1}(z)-2\widetilde
Y_l(z)X^*_l(z)\right)~, \label{III26-0}
\end{eqnarray}
which resembles the expression for the probability in case of IPA
\begin{eqnarray}
w_n=
 2 J^2_n(z) +\xi^2(2u-1)
\left(J^2_{n-1}(z)+ J^2_{n+1}(z)-2J^2_n(z)\right)~, \label{III26IPA}
\end{eqnarray}
with the substitution $\widetilde Y^2_l(z)\to J_n^2(z)$,
$Y^2_{l\pm1}(z)\to J^2_{n\pm1}(z)$, and
$\widetilde Y_l(z)X^*_l(z)\to J^2_n(z)$.

 The differential  probability $dW$ in Eq.~(\ref{III9}), is in fact, the probability per
 unit time (or rate), and it is related to the differential cross section $d\sigma$ as
\begin{eqnarray}
dW=J\cdot \rho_\gamma\,d\sigma=2\cdot \frac{\omega m^2\xi^2}{4\pi\alpha}
\,d\sigma~, \label{IIW-S}
\end{eqnarray}
where $J=2$ and $\rho_\gamma$ are the flux of incoming probe photon
and the density of the background  photons, respectively, $\omega$ is
the frequency of the background photon. Thus, the differential cross
section reads
\begin{eqnarray}
\frac{d \sigma}{d\phi_p\,du }
=\frac{\alpha^2\zeta}{2s_{\rm
 thr}\xi^2 N_0} \,\frac{1}{u^{3/2}\sqrt{u-1}}
 \,\int\limits_\zeta^{\infty}\,dl\ w{(l)}~.
\label{III99}
\end{eqnarray}
 Later, for easy reference and comparison with previous works (cf.~\cite{Ritus-79})
 we present our results for the Breit-Wheeler process in terms of probabilities $dW$
 (production rates) rather than the cross sections $d\sigma$,
 remembering Eq.~(\ref{IIW-S}) connecting these two observables.
\subsection{Short pulses}
In this section we  consider short pulses with the number of
oscillation $N\geq 2$, however, the developed methods
for studying probabilities of $\ee$ pair production
are valid even for
pulses with $N\sim1$. The sub-cycle pulses with $N<1$
will be considered separately, in subsection~F.
Recall that we consider two envelope shapes:
hyperbolic secant (hs) shape and symmetrized Fermi (sF)
shape with $b/\Delta=0.15$.

\begin{figure}[ht]
\includegraphics[width=0.35\columnwidth]{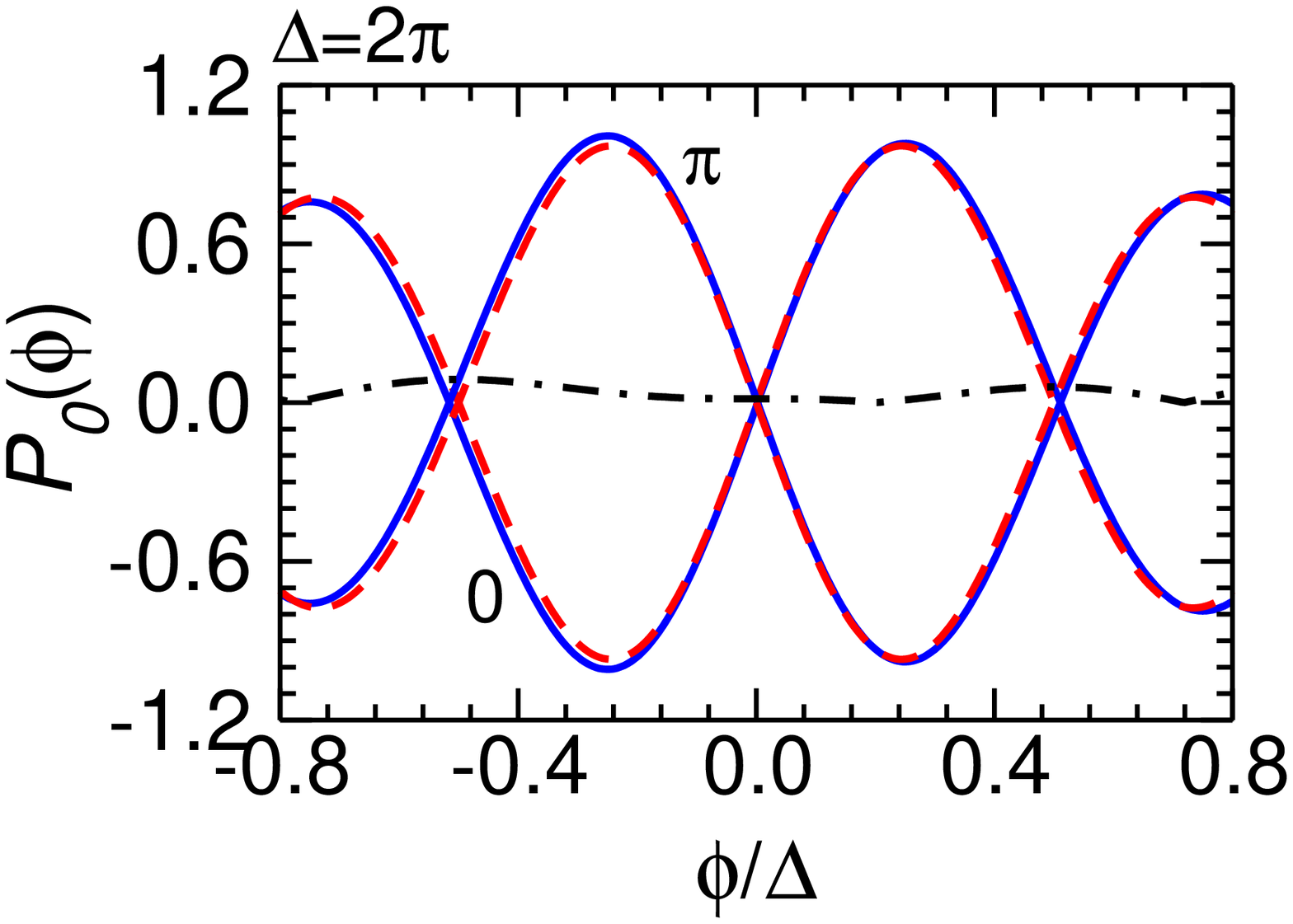}\qquad\qquad
\includegraphics[width=0.35\columnwidth]{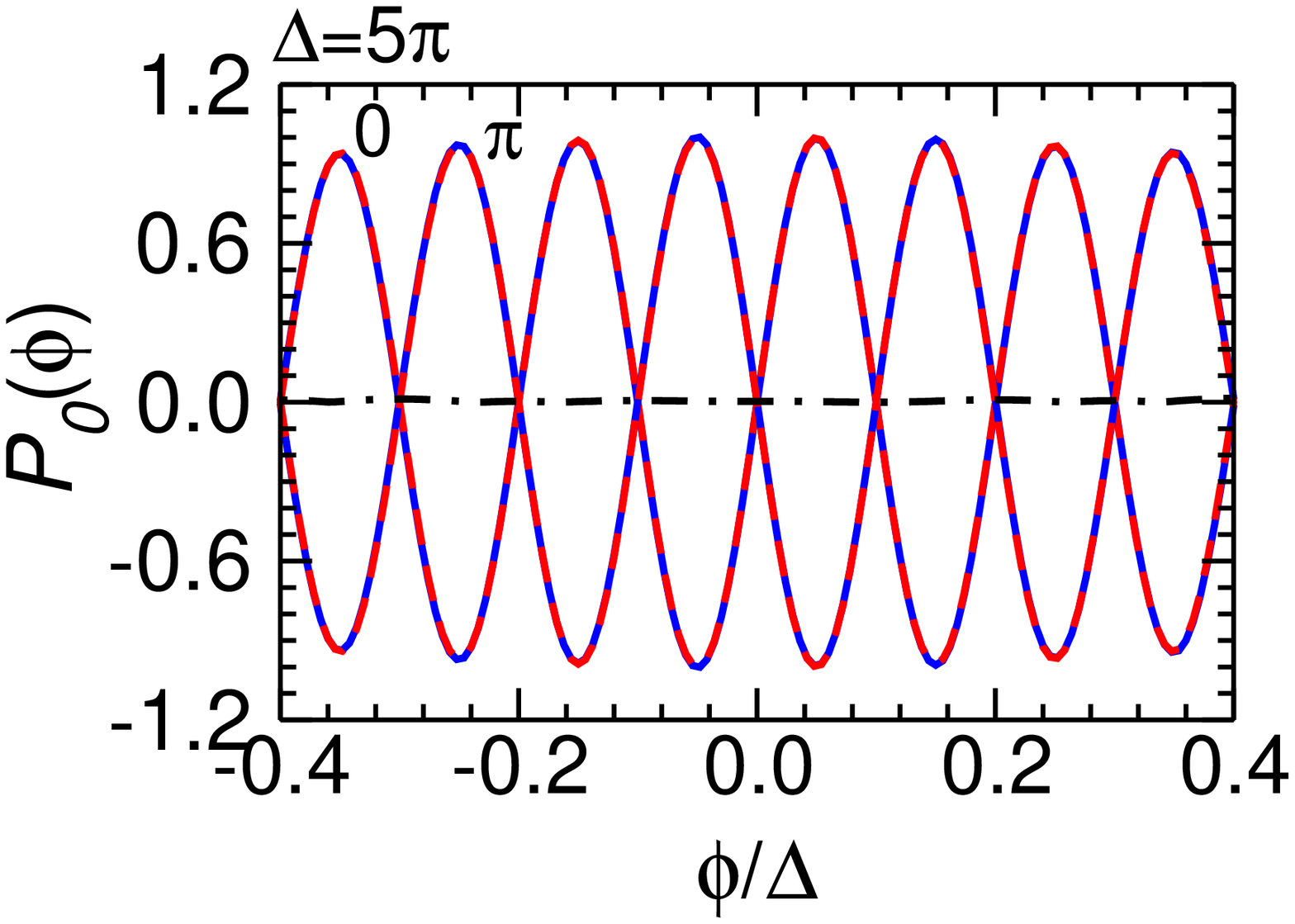}
\caption{\small{ The function ${\cal P}_0(\phi)$
defined in (\ref{III21}) with (solid curves) and without (dashed
curves) the term~(\ref{III22}) for $\Delta=\pi\,N$ with $N=2$ and
5, shown in left and right panels, respectively. The
term~(\ref{III22}) is shown separately by dot-dashed curves.
\label{Fig:8} }}
\end{figure}

As mentioned above,  Eqs.~(\ref{III9}) and~(\ref{III99})
with Eq.~(\ref{III20}) can be
used for numerical estimates of the $\ee$ production
probability or cross section
evaluating five dimensional integral(s) with rapidly oscillating
functions. Technically, such an approach needs long calculation
time for reasonable computational accuracy
which makes it difficult for applications in transport/Monte
Carlo codes. However, a closer inspection of the functions ${\cal
P(\phi)}$ and $Y_l,\,X_l$ shows that the number of integrations
may be reduced and, in some cases, Eq.~(\ref{III26-0}) may be
expressed in an analytical form. Thus, integrating by parts the
function $\cal P(\phi)$ might be rewritten in the following form
\begin{eqnarray}
{\cal P}(\phi)&\equiv&{\cal P}_0(\phi)- \xi^2\zeta
u\int\limits_{-\infty}^\phi d\phi'\,f^2(\phi')~,\nonumber\\
{\cal P}_0(\phi) &=&z\,\left(\sin(\phi-\phi_0)f(\phi) + {\cal
O}\left(\frac{1}{\Delta}\right) \right)\, \label{III21}
\end{eqnarray}
with
\begin{eqnarray}
{\cal O}\left(\frac{1}{\Delta}\right)=
-\frac{1}{\Delta}\int\limits_{-\infty}^\phi\,d\phi'\,
\sin(\phi'-\phi_0)f'(\phi')~. \label{III22}
\end{eqnarray}
The contribution of this term to $\cal P(\phi)$ is sub leading for
the finite pulse size $\Delta=\pi N$ with $N\ge2$. First, because
of the explicit factor ${1}/{\Delta}$, and second because the
derivative $f'(\phi)$ in the integrand reaches its maximum value
at the boundaries of the pulse, where this function is suppressed.
For an illustration, in Fig.~\ref{Fig:8} we present results of a
numerical analysis of ${\cal P}_0(\phi)$ with the hyperbolic
secant envelope function. The solid and dashed curves exhibit
calculations with and without the term~(\ref{III22}), respectively
for $\phi_0=0$ and $\pi$. The left and right panels correspond to
$\Delta=\pi\,N$ with $N=2$ and 5, respectively. The term ($|{\cal
O}({1}/{\Delta})|$) is shown by dot-dashed curves.
One can see, in fact, that this term is rather small and may be
omitted. The second term in expression for ${\cal P}(\phi)$ in Eq.~(\ref{III21})
is a smooth function of $\phi$ and in case of hs shape it can be
given explicitly as $-\xi^2\zeta u \Delta\tanh(\phi/\Delta)$.

Now we are going to discuss separately
the weak-, intermediate- and strong-field regimes.

\subsection{Pair production at small field intensities ($\xi^2\ll1$) }

In case of small $\xi^2\ll1$, implying  $z<1$,
we decompose $l=n +\epsilon$, where $n$ is the integer part of
$l$, yielding
\begin{eqnarray}
Y_l&\simeq&\frac{1}{2\pi}\int\limits_{-\infty}^{\infty}\,d\psi
\,{\rm e}^{il\psi -iz\sin\psi\,f(\psi+\phi_0)} f(\psi+\phi_0)\nonumber\\
&=& \frac{1}{2\pi}\int\limits_{-\infty}^{\infty}\,d\psi
\sum\limits_{m=0}^{\infty} \frac{(iz)^m}{m!}\sin^m\psi \,{\rm
e}^{i(n+\epsilon)\psi} f^{m+1}(\psi+\phi_0)~. \label{B5}
\end{eqnarray}
Similarly, for the function $X_l(z)$ the substitution $f^{m+1}\to
f^{m+2}$ applies. The dominant contribution to the integral
in (\ref{B5}) with
rapidly oscillating integrand comes from the term with $m=n$,
which results in
\begin{eqnarray}
Y_{n+\epsilon}\simeq \frac{z^n}{2^nn!}\,{\rm e}^{-i\epsilon\phi_0}
F^{(n+1)}(\epsilon)~,\qquad X_{n+\epsilon}\simeq
\frac{z^n}{2^nn!}\,{\rm e}^{-i\epsilon\phi_0}
F^{(n+2)}(\epsilon)~, \label{B6}
\end{eqnarray}
where the function $F^{(n)}(\epsilon)$ is the Fourier transform of
the function $f^n(\psi)$.

As an example, let us analyze the $\ee$ production near the
threshold, i.e. $\zeta\sim1$. In this case, the contribution with
$n=1$ is dominant and, therefore, the functions  $Y_{0+\epsilon}$
are crucial, including the first term in (\ref{III26-0}). The
functions $X_{0 +\epsilon}$ are not important because they are
multiplied  by the small $\xi^2$ and may be omitted. Negative
$\epsilon=\zeta-1$ and positive $\epsilon$ correspond to the
above- and sub-threshold pair production, respectively. The
function $Y_{0+\epsilon}$ reads
$Y_{0+\epsilon}=F^{(1)}(\epsilon)\,\exp[-i\phi_0\epsilon]$,
where
the Fourier transforms $F^{(1)}(x)$ for the hs and sF envelope
functions are equal to
\begin{eqnarray}
F_{\rm hs}(x)&=&\frac{\Delta} {  2\cosh  {\frac12\pi\Delta  x}} ~,\nonumber\\
F_{\rm sF}(x)&=&\frac{1+{\exp}\left[{-\frac{\Delta}{b}}\right] }
{1-{\exp}\left[{-\frac{\Delta}{b}}\right] }\,
\frac{ b\,\sin\Delta l } {\sinh \pi bx}~.
%
\label{U5}
\end{eqnarray}
The $\phi_0$ dependence of the
production probability disappears in this case because the latter
one is determined by the quadratic terms of the $Y$ functions.

Consider first the pair production above the threshold. Keeping
the terms with leading power of $\xi^2$ one can express the
production probability as
\begin{eqnarray}
\frac{d W}{du}=\frac{\alpha M_e\zeta^{1/2}}{4N_0} \,\left[
\frac{u}{u_1}\left(1-\frac{u}{u_1}\right) +u  - \frac12 \right]
\frac{\xi^2}{u^{3/2}\sqrt{(u-1)}}\, I_0~, \label{III28}
\end{eqnarray}
where, taking into account that at finite values of $\Delta$,
Fourier transforms for all considered envelopes decrease rapidly
with increasing $\epsilon$, and one can gets
\begin{eqnarray}
I_0 \simeq\int\limits_{1-\zeta}^{1/2} d\epsilon\,
F^{(1)}{}^2(\epsilon) \simeq\int\limits_{-\infty}^{\infty}
d\epsilon\, F^{(1)}{}^2(\epsilon)
=\frac{1}{2\pi}\int\limits_{-\infty}^{\infty}
d\phi\,f^2(\phi)\simeq N_0~. \label{III29}
\end{eqnarray}
Combining these two equations one recovers the IPA
result~\cite{Ritus-79}. Thus, we can conclude that for small field
intensities for a finite pulse duration the probabilities of $\ee$
pair emission above threshold with $\zeta<1$ in IPA and FPA coincide,
independently of the shape of the envelope function.
\begin{figure}[ht]
\begin{center}
\includegraphics[width=0.35\columnwidth]{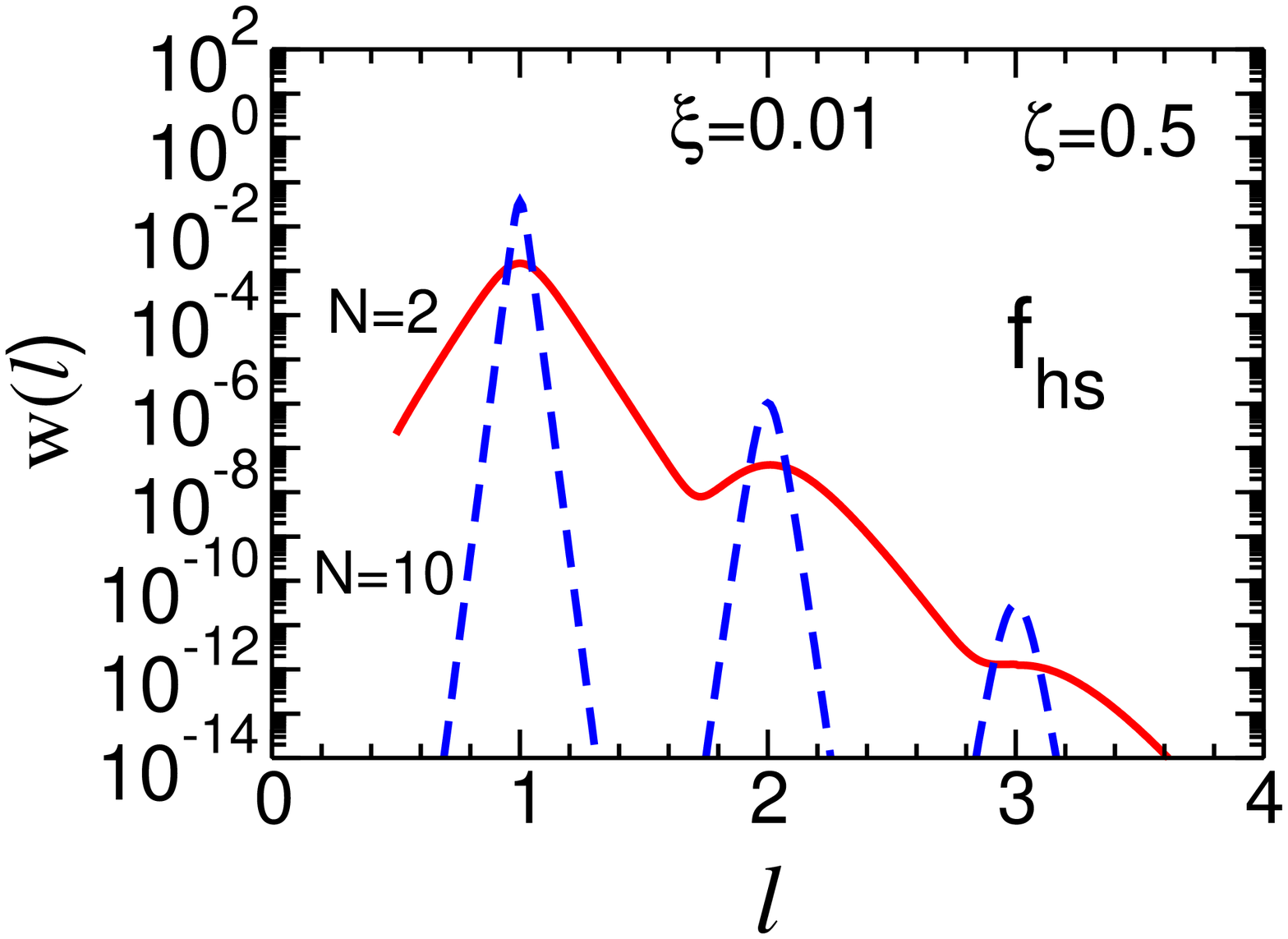}\qquad\qquad
\includegraphics[width=0.35\columnwidth]{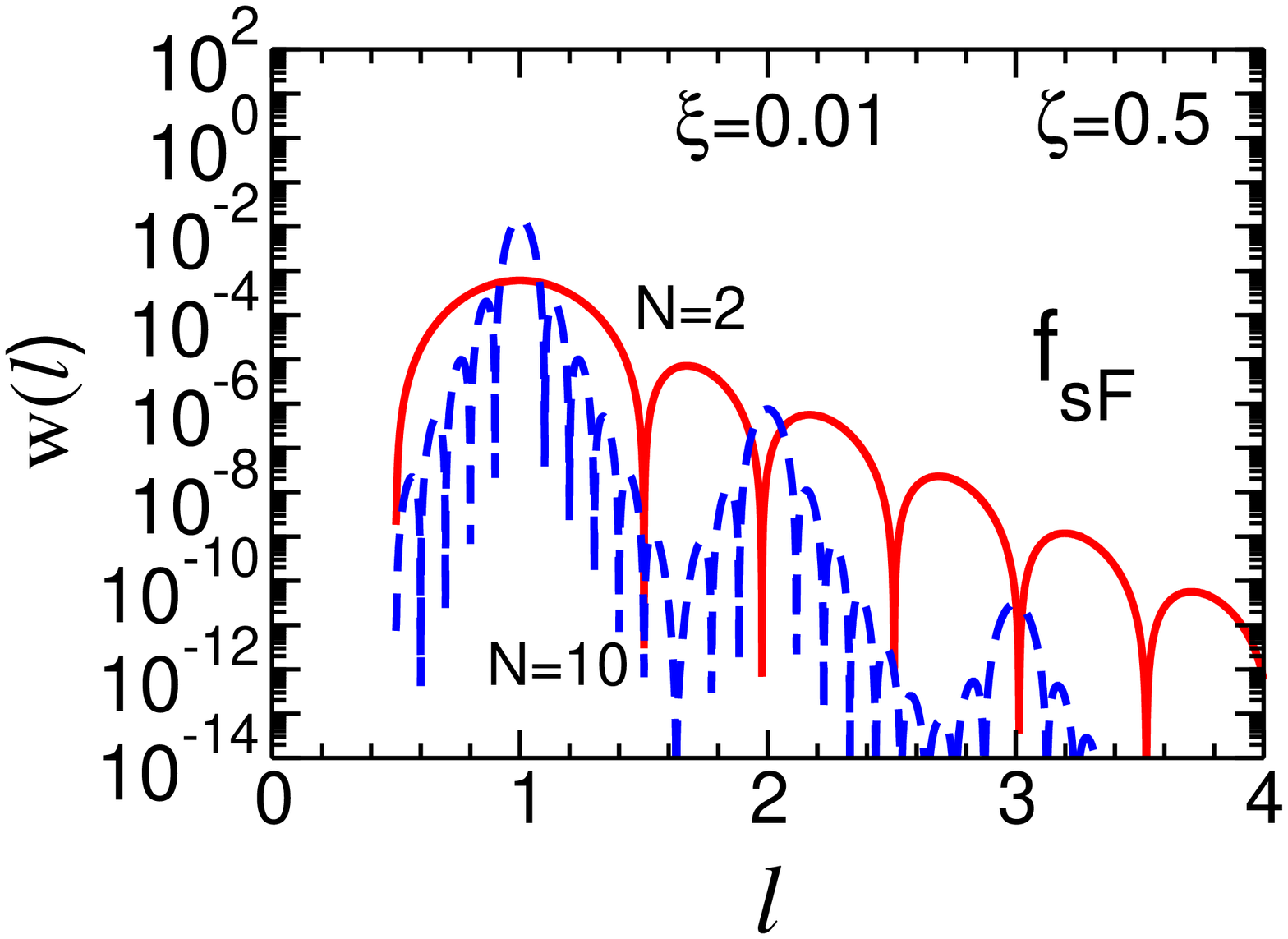}
\end{center}
\caption{\small{The partial probability  $w(l)$
defined in (\ref{III26-0}) as a function of $l$ at $u=1$. The
solid and dashed curves correspond to the parameter $\Delta=\pi N$
with $N=2$ and 10, respectively. Left and right panels exhibit
results for the envelopes with hyperbolic secant and symmetrized
Fermi shapes, respectively. For $\xi^2=10^{-4}$ and $\zeta=0.5$.
\label{Fig:9} }}
\end{figure}
For an illustration, in Fig.~\ref{Fig:9} we show the partial
probability $w(l)$, calculated at $u=1$  for the above-threshold
region with $\xi=10^{-2}$ and $\zeta=0.5$ in a finite region of
$l$ for the envelope size $\Delta=\pi N$ with $N=2$ and 10,
respectively. For the envelope with a hyperbolic secant shape
(left panel) one can see smooth curves with maxima at integer
values of $l$. The widths of bumps decrease with increasing $N$.
However, the integral of $w(l)$ over $l$ in the neighborhood of
the first maximum is independent of $N$ and coincides with the
contribution of the first harmonic in IPA which leads to an
equality of IPA and FPA results. For the symmetrized Fermi shape
(right panel) the situation is different in some sense. The
corresponding Fourier transforms
 $F^{(n)}_{sF}(l)$ in (\ref{B6})
oscillate with $l$. For example, the function $F^{(1)}_{sF}$ goes
to zero at a multiple of $1/N$. This results in an oscillating
structure of $w(l)$. However, the exponential decrease of $w(l)$
with increasing of the integer values of $l$ is the same.

The situation changes when we are slightly below threshold, i.e.
$\zeta>1$. In this case, the function $Y_{0+\epsilon}$ dominates
again and the result for FPA is the same as in (\ref{III28}) but
with the substitution $I_0\to I_1$, with $I_1
\simeq\int\limits_{\zeta-1}^{1} d\epsilon\,
F^{(1)}{}^2(\epsilon)$. In the case of smooth envelope shape
(e.g.~hyperbolic secant) the dominating contribution to this
integral comes from the lower limit and, therefore, $I_1\sim
\left(F_{\rm hs}^{(1)}(\zeta -1)\right)^2$. As a result, the
production probability strongly depends on the duration $\Delta$
of the pulse.
In the case of a flat-top envelope, we have a similar effect, because
$F_{sF}^{(1)}(l)$ in general, decreases exponentially as $\exp(-\pi
b l)$, where $b$ increases with increasing $N$ at fixed
$b/\Delta$.
\begin{figure}[ht]
\begin{center}
\includegraphics[width=0.35\columnwidth]{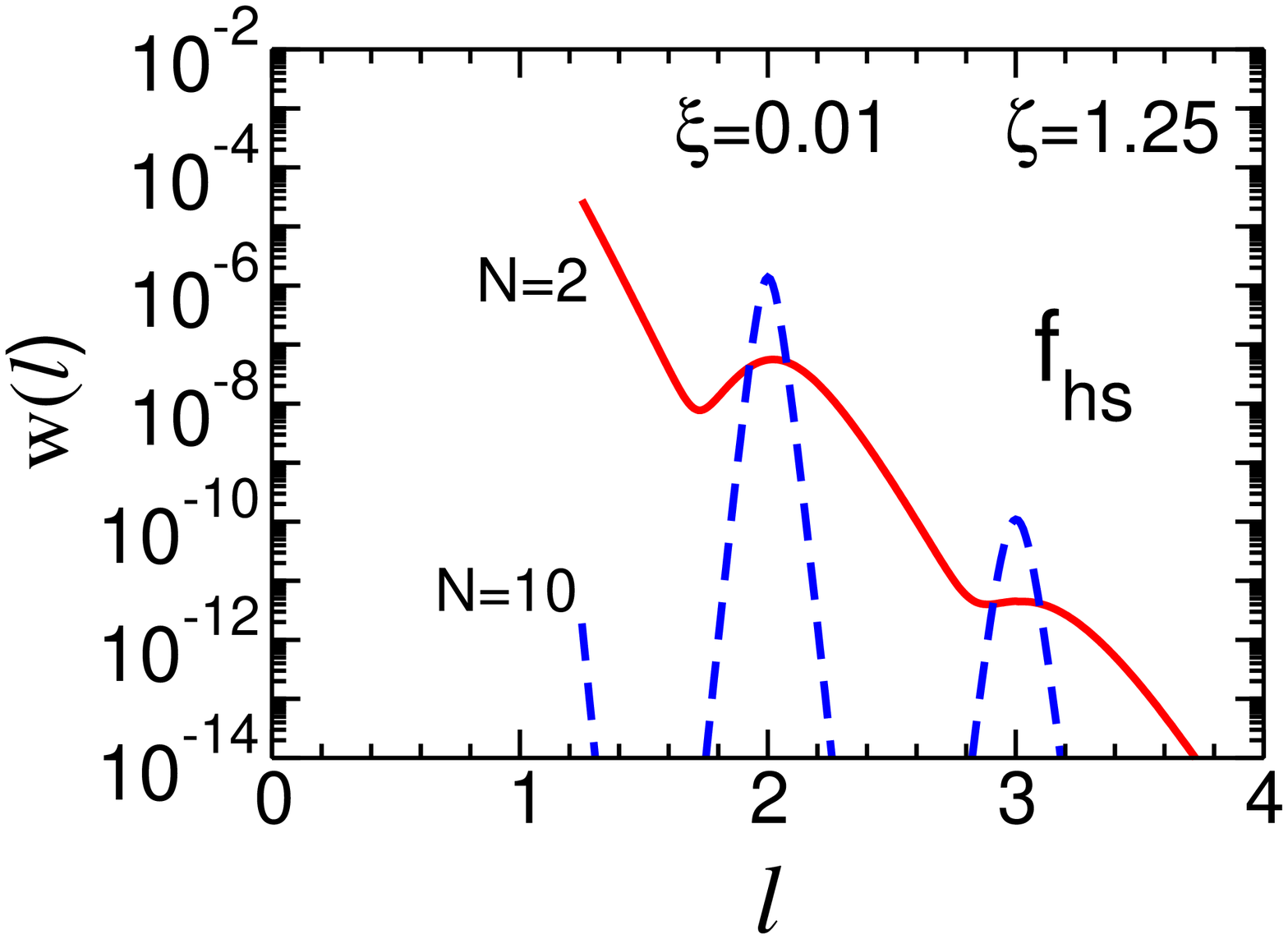}\qquad\qquad
\includegraphics[width=0.35\columnwidth]{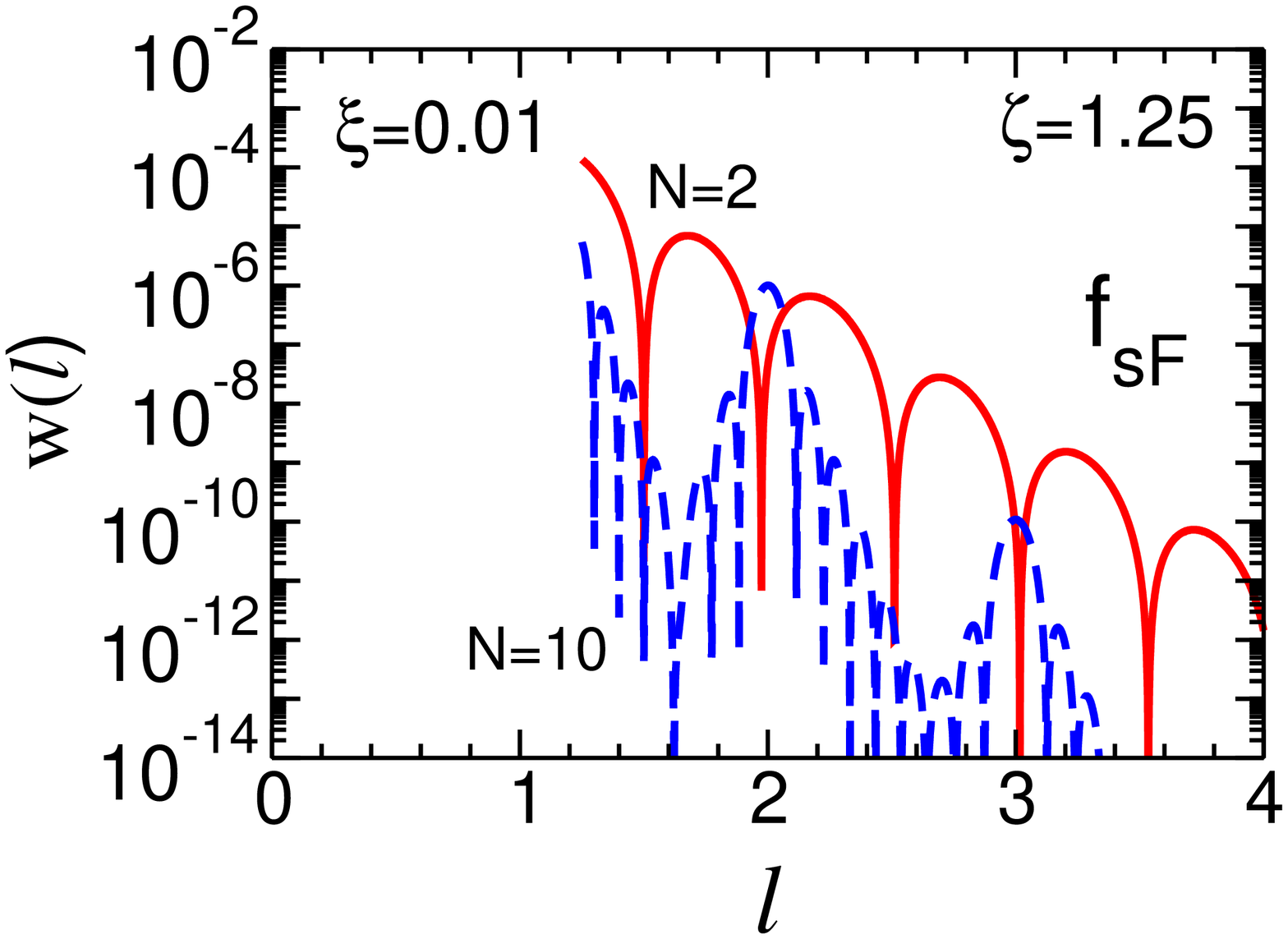}
\end{center}
\caption{\small{The same as in Fig.~\ref{Fig:9} but
for the sub-threshold region at $\zeta=1.25$.
 \label{Fig:10} }}
\end{figure}

In Fig.~\ref{Fig:10} we show the partial probability $w(l)$ in the
sub-threshold region with~$\zeta=1.25$. One can see that for the
hyperbolic secant envelope (left panel) the difference of $w(l)$
at $l\simeq \zeta$ for $N=2$ and $N=10$ is more than several
orders of magnitude, which will be reflected in the total
probability. In the case of the symmetrized Fermi envelope shape,
one also can see a significant enhancement of $w(l)$ for $N=2$
compared to $N=10$. But now, the difference between FPA and IPA is
larger compared to the case of the hyperbolic secant shape.

\begin{figure}[ht]
\begin{center}
\includegraphics[width=0.35\columnwidth]{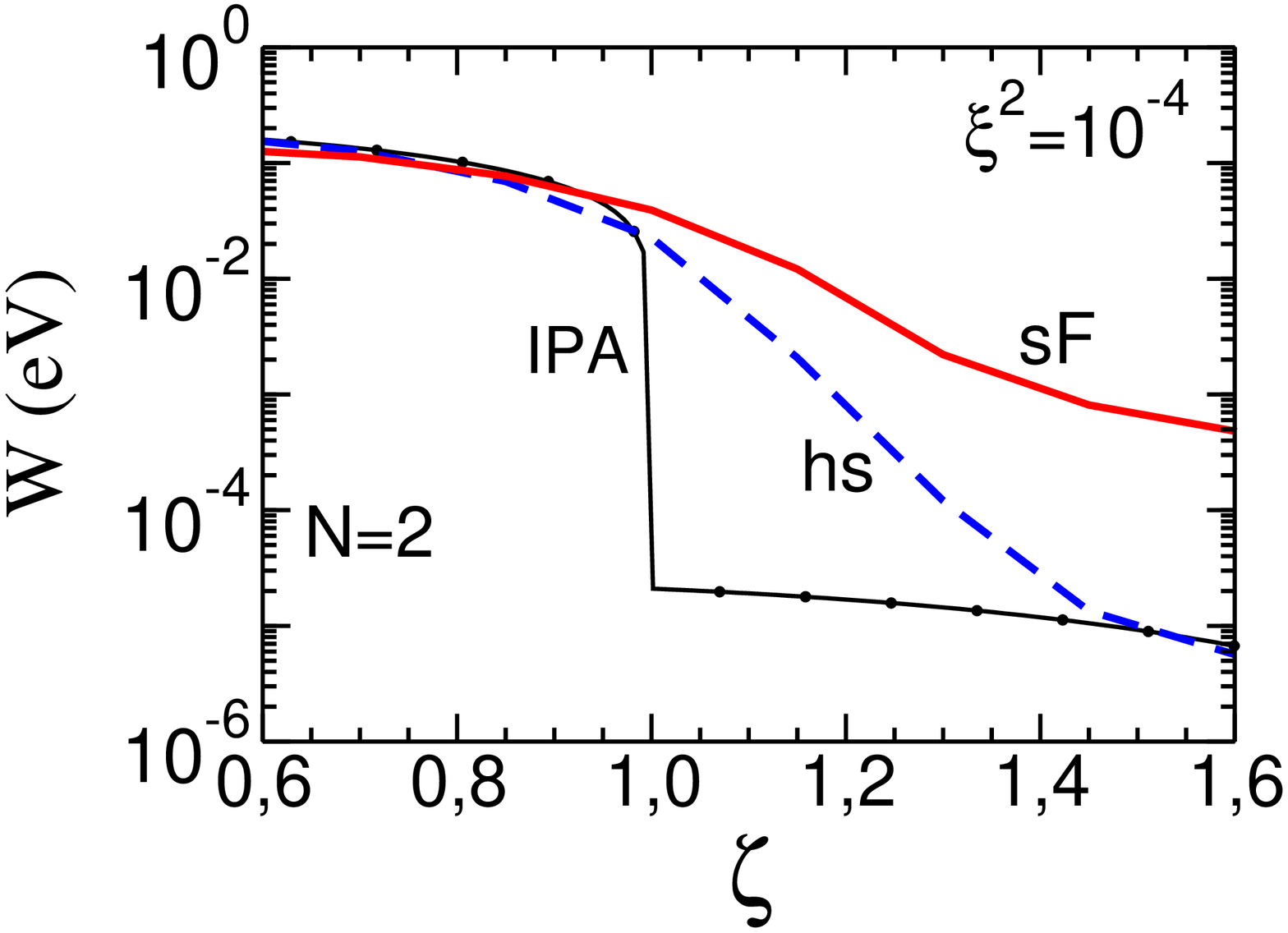}\qquad
\includegraphics[width=0.35\columnwidth]{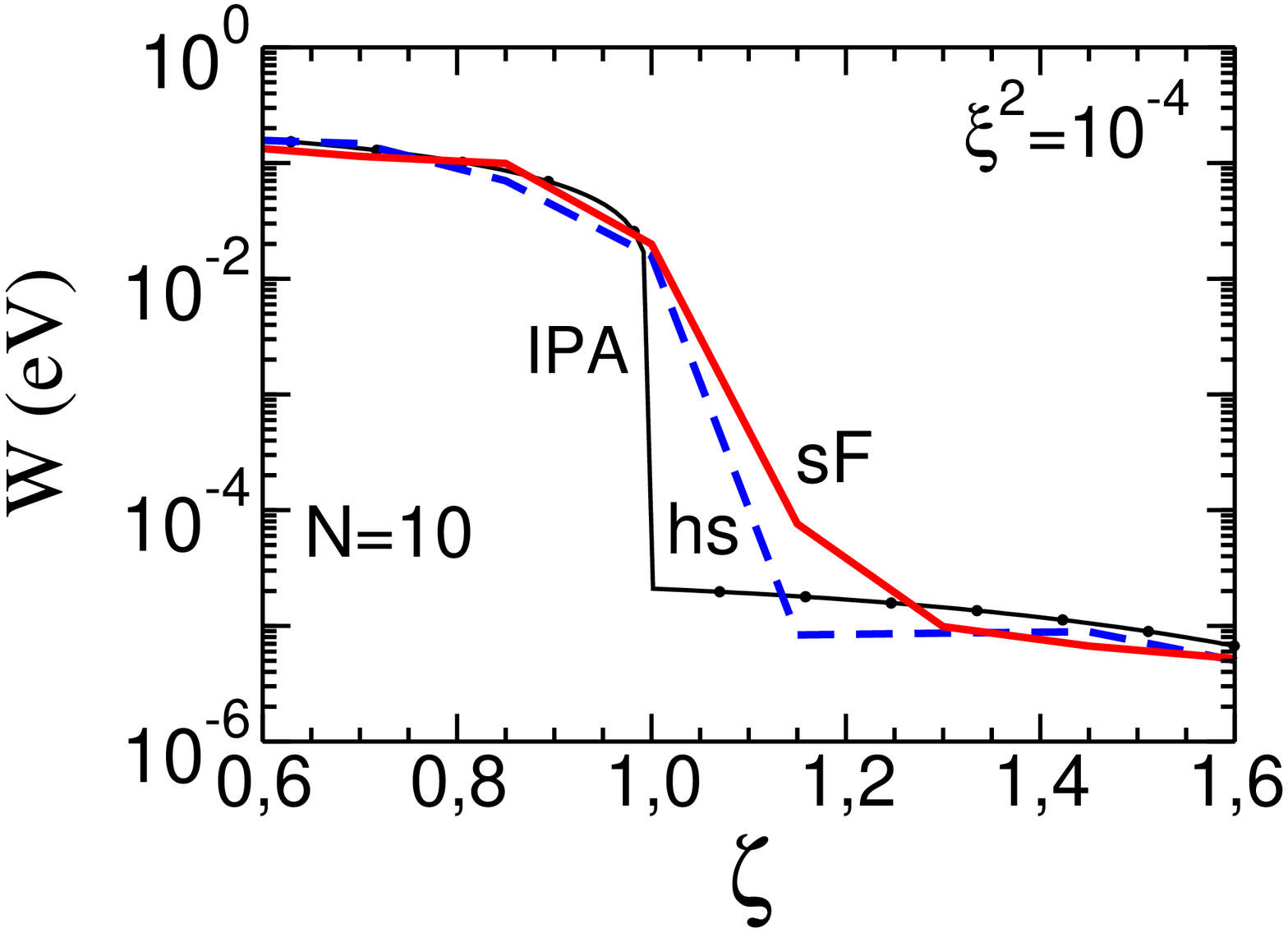}
\end{center}
\caption{\small{The total probability $W$ of the
 $\ee$ pair production as a function of $\zeta$ for short pulses
 with $\Delta=\pi N$ for $N=2$,\ and 10 shown in the left
 and right panels, respectively; $\xi^2=10^{-4}$.
 The dashed and solid curves correspond to the hyperbolic secant
 and symmetrized Fermi envelope shapes, respectively.
 The thin solid curves marked by dots depict the IPA
 result.
 \label{Fig:11} }}
\end{figure}

\begin{figure}[ht]
\begin{center}
\includegraphics[width=0.35\columnwidth]{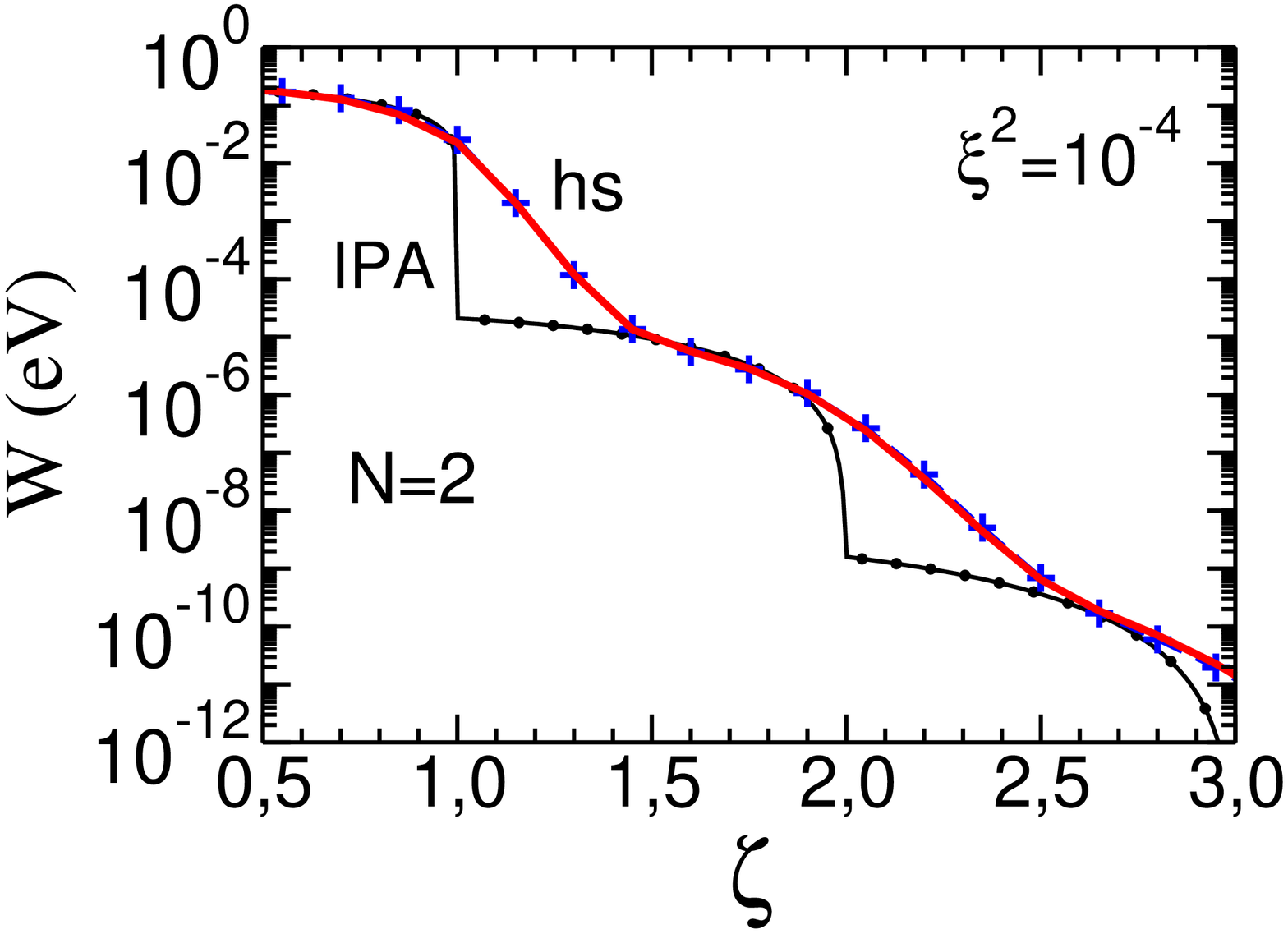}\qquad
\includegraphics[width=0.35\columnwidth]{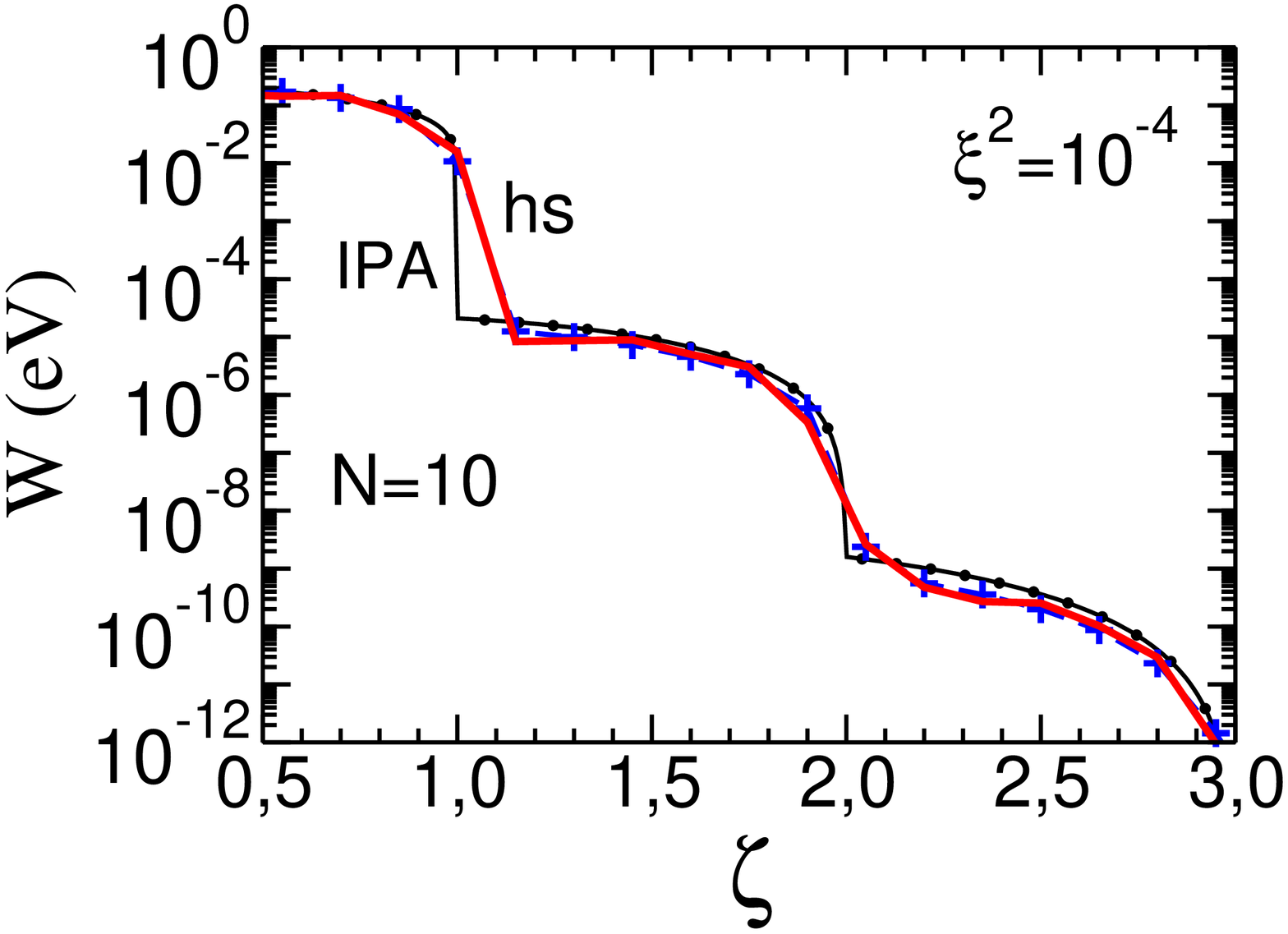}
\end{center}
\caption{\small{The total probability $W$ of the
 $\ee$ pair production as a function of $\zeta$ for the hyperbolic secant
 envelope shape.  The solid curves are for the full calculation,
 while the dashed curves marked by crosses correspond to the
 approximate result with the basic functions taken in the form of
 Eq.~(\ref{B6}). Left and
 right panels correspond to the number of oscillation in a pulse
 $N=2$ and 10, respectively; $\xi^2=10^{-4}$.
 \label{Fig:12_1} }}
\end{figure}

\begin{figure}[ht]
 \begin{center}
\includegraphics[width=0.35\columnwidth]{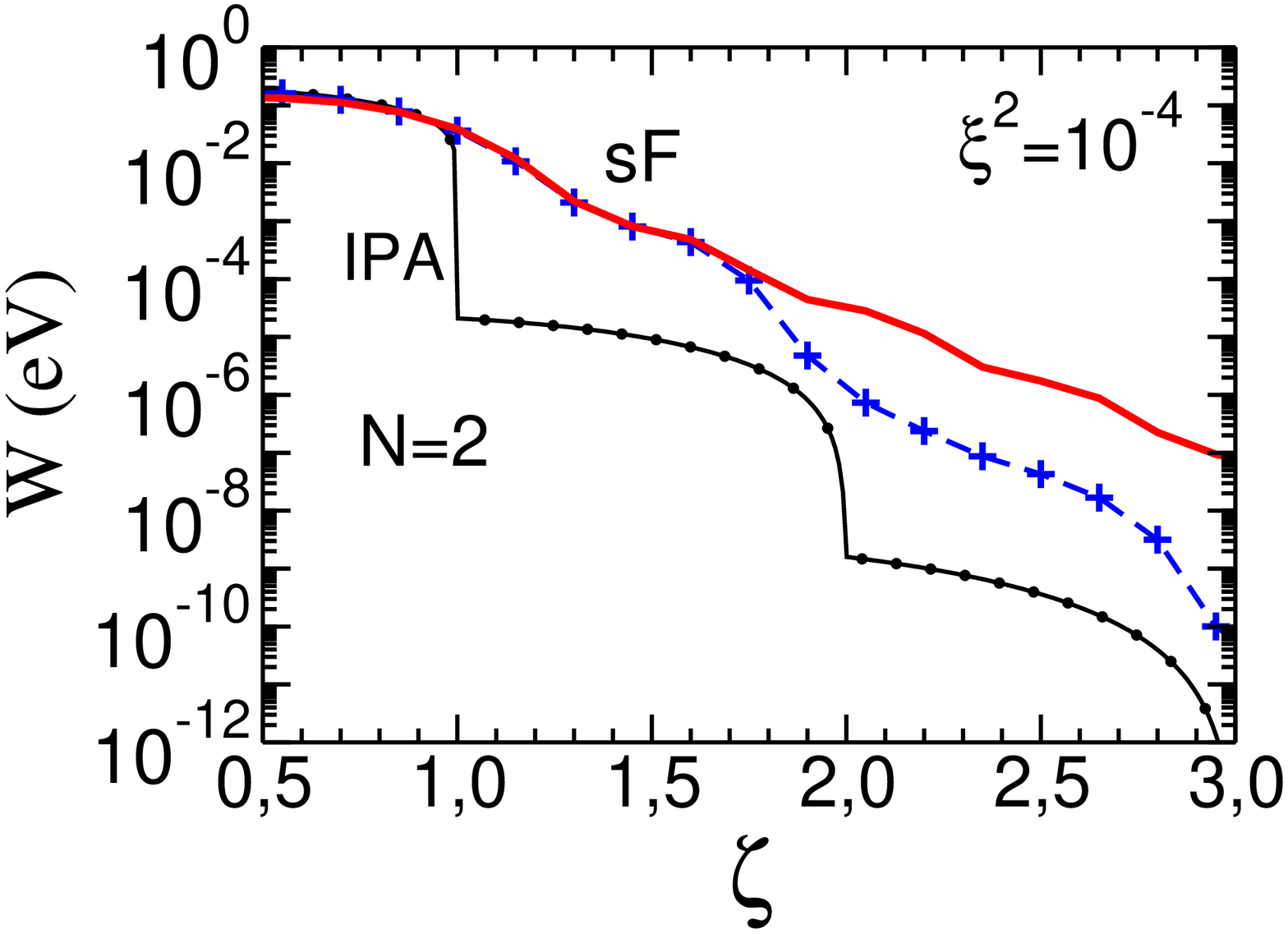}\qquad
\includegraphics[width=0.35\columnwidth]{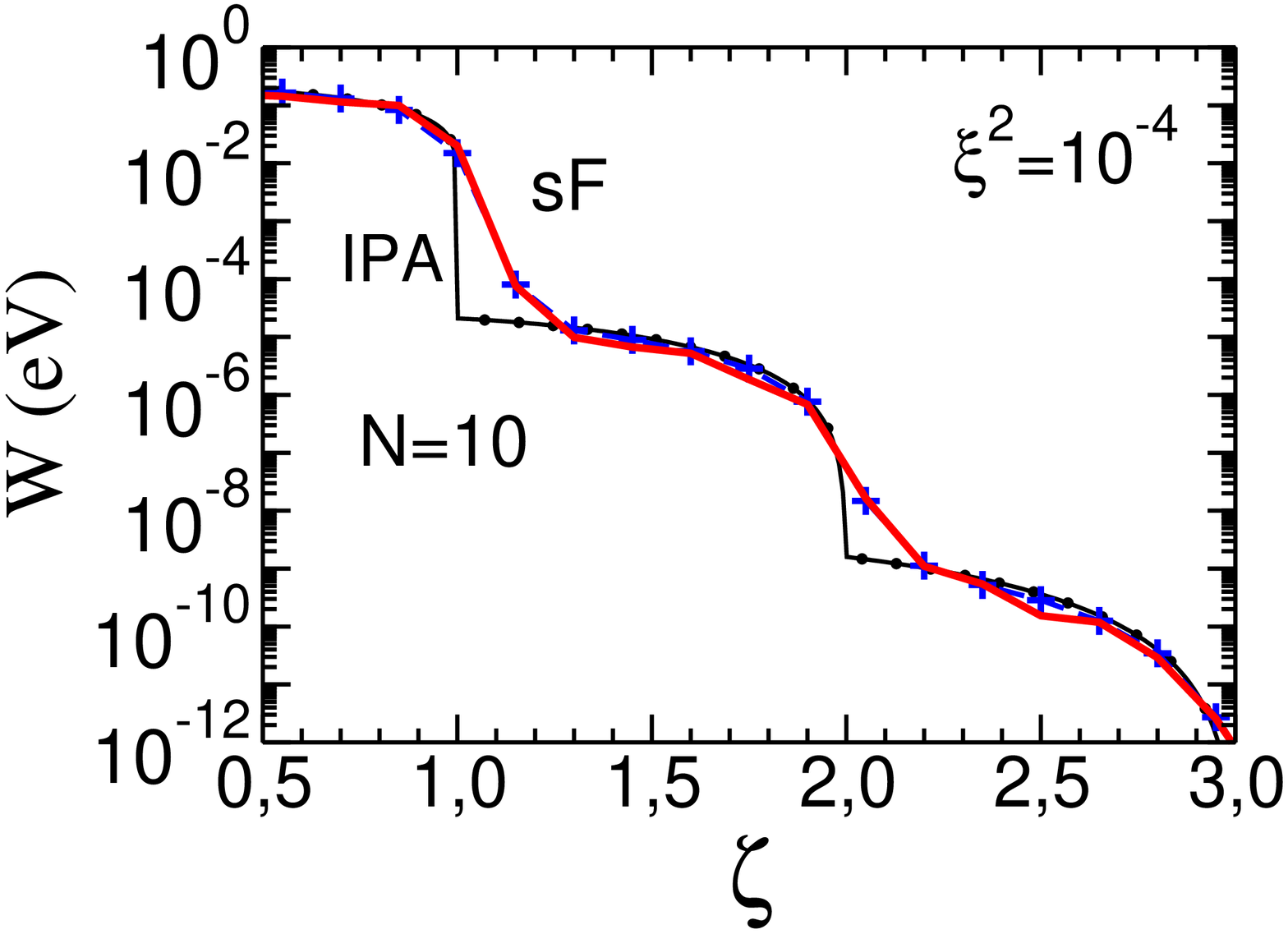}
\end{center}
\caption{\small{The same as in Fig.~\ref{Fig:12_1}
 but for symmetrized Fermi shape.
\label{Fig:12_2} }}
\end{figure}

The total probability $W$ of $\ee$ emission as a function of the
sub-threshold parameter $\zeta$ in the vicinity $\zeta\sim 1$ is
presented in Fig.~\ref{Fig:11}.
 The dashed and solid curves correspond to the hyperbolic secant
 and symmetrized Fermi envelope shapes, respectively.
 The left and right panels correspond to the  short pulses
 with $\Delta=\pi N$ for $N=2$,\ and 10, respectively,
 at $\xi^2=10^{-4}$.
 For comparison, we present also the
 IPA results. In the above-threshold region, results of IPA and FPA
 are equal to each other according to Eqs.~(\ref{III28}) and
(\ref{III29}). However, in the sub-threshold region, where $\zeta$
is close to unity, the probability of FPA considerably exceeds (by more
than two orders of magnitude) the corresponding IPA
result. In the case of the hyperbolic secant envelope function, the
probability increases with decreasing  pulse duration. The results
of FPA and IPA become comparable at $N\geq 10$. Qualitatively,
this result is also valid for the case of the symmetrized Fermi
distribution. However, in this case the enhancement of the
probability in FPA is much greater. This is due to the fact that
the maxima in the partial probability $w(l)$
(cf.~Fig.~\ref{Fig:10}) decreases with increasing $l$ in different
ways for different envelope shapes. In the case of the hyperbolic
secant it decreases as $\exp(-\pi\Delta l)$, whereas in case of
symmetrized Fermi shape it decreases as $\exp(-2\pi b l)$. For the
latter one, at $b/\Delta=0.15$ the slope is much smaller. Such a
strong gain of $\ee$ emission rate is expected for other values of
$\zeta$ when $\zeta$ exceeds an integer number.
This effect is
illustrated in Figs.~\ref{Fig:12_1} and~\ref{Fig:12_2},
where the total $\ee$ production
probability $W$ is presented in a wide region of $\zeta$ at
$\xi^2=10^{-4}$.

 In Fig.~\ref{Fig:12_1} we present the results for the hyperbolic secant
 envelope shape. The solid curves are for the full calculation,
 while the dashed curves marked by crosses correspond to the
 approximate result with the basic functions taken in the form of
 Eq.~(\ref{B6}). Left and
 right panels correspond to the number of oscillation in a pulse
 $N=2$ and 10, respectively.
 One can see that the approximate result is in a very good
 agreement with the full calculation and
 may be used in transport code calculations
 since it is  much easier acceptable.

 Corresponding results for the symmetrized Fermi shape envelopes
 are shown in Fig.~\ref{Fig:12_2}. In the case of short pulse with
 $N=2$, the approximate calculation is valid at $\zeta \lesssim 1.7$.
 However, when $N$ increases, one can find agreement between full and
 approximate results in a wide region of $\zeta$.
 For the flat-top shape with small $b/\Delta$, the probability in
 FPA is larger than the result of IPA near integer values of
 $\zeta$.

 In any case, at large values of $N$ (right panels, $N=10$)
 results of FPA and
 IPA become close to each other, especially for the one-parameter
 envelope shapes. For this case, at least for $\xi=0.1....0.01$, $N\simeq
 10$ can be considered to be near by infinite, when considering the
 overall $\zeta$ dependence.

To summaries this part we would like to note that temporal beam shape
effects for short pulses are strong and even dominant at small
field intensities in the parameter region where the variable $z$
is small, $z < 1$. At finite $z$, the non-linear dynamics of
$\ee$ production at high pulse intensity becomes essential.

\subsection{The case of intermediate field intensity (${\xi^2\sim1}$)}

\begin{figure}[ht]
\includegraphics[width=0.35\columnwidth]{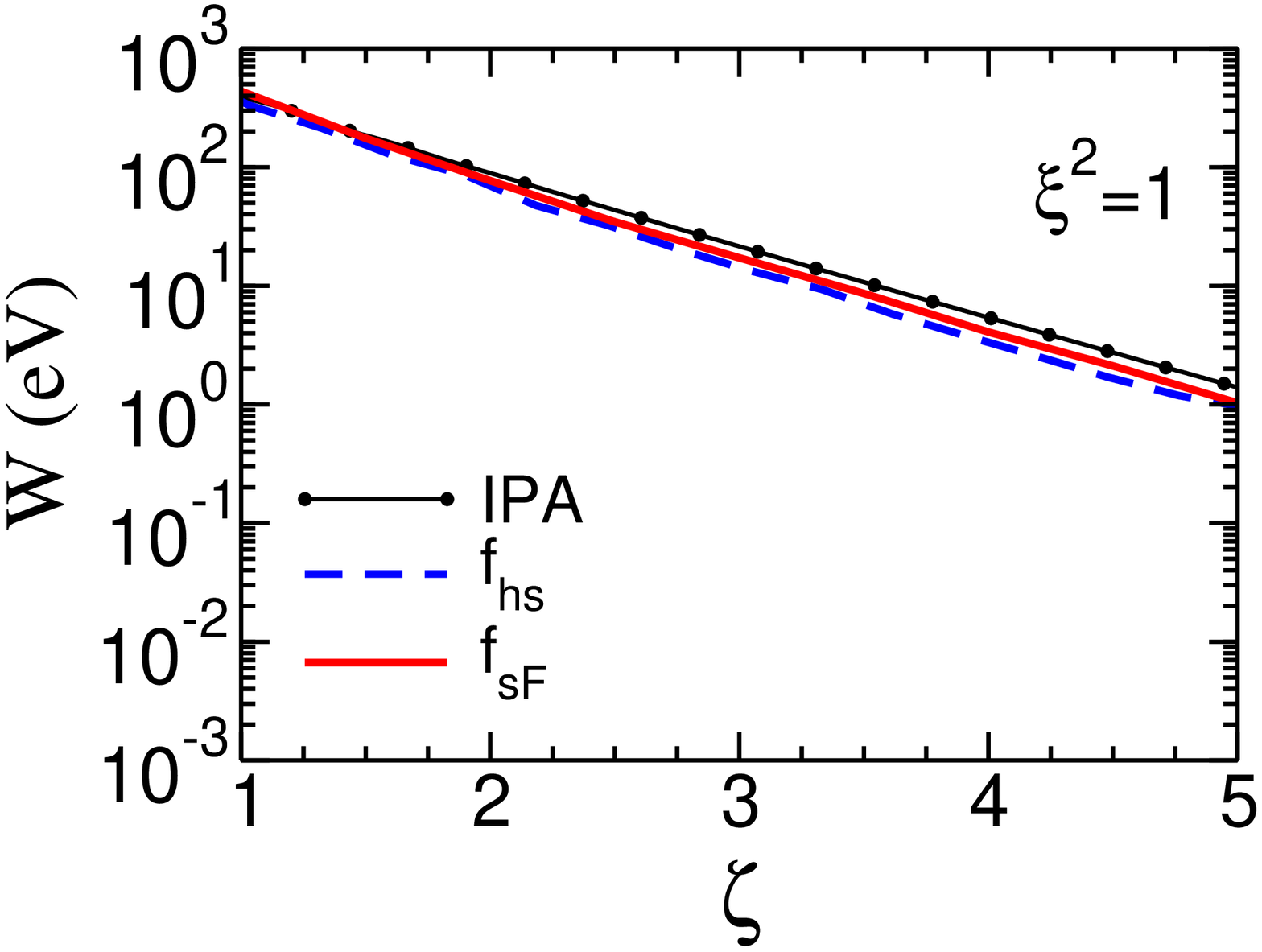}\qquad
\includegraphics[width=0.35\columnwidth]{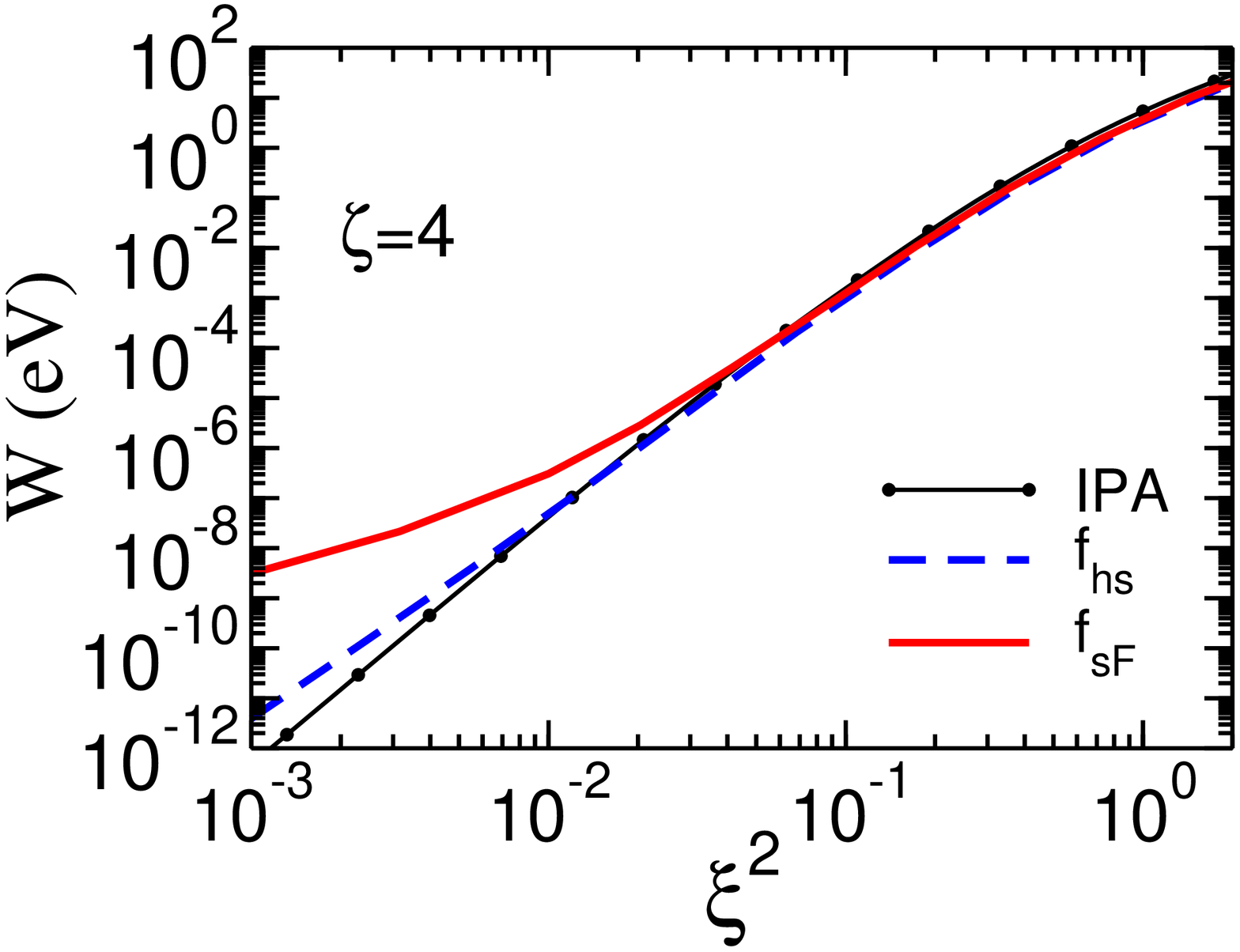}
\caption{\small{The total probability of $\ee$-pair
production for two envelope shapes (dashed and solid curves are
for hyperbolic secant and symmetrized Fermi shapes, respectively).
The thin solid curves marked by dots are the result of IPA. Left
panel: The total probability as a function of $\zeta$ at
$\xi^2=1$. Right panel: The total probability as a function of
$\xi^2$ at $\zeta=4$.
 \label{Fig:13} }}
\end{figure}

At finite values of $z$, $z\gtrapprox 1$, the probability of $\ee$
emission needs to be calculated numerically using
Eqs.~(\ref{III9}), (\ref{III26-0}), and (\ref{III24}). In
Fig.~\ref{Fig:13}, we present the total probability $W$ as a
function of $\zeta$ at fixed $\xi^2=1$ (left panel) and as a
function of $\xi^2$ at fixed $\zeta=4$ (right panel).
The calculations are performed for the hyperbolic secant and
symmetrized Fermi pulse envelope shapes, shown by the dashed and
solid curves, respectively. The duration of the pulse is
$\Delta=\pi N$ with $N=2$. For comparison, we also present IPA
results by the thin solid curves marked by dots. At finite
$\xi^2$, the probability decreases monotonically with increasing
$\zeta$ (left panel), contrary to the step-like decrease typical
for the small $\xi^2\ll1$
(cf. Figs.~\ref{Fig:12_1} and \ref{Fig:12_2} (right panel)).

Concerning the $\xi^2$ dependence (right panel), one can see a
sizeable enhancement of the total probability $W$ at small
values of $\xi^2$ for the flat-top pulse shape compared to the
case of hyperbolic secant and the IPA result. The latter two
results are practically identical to each other. At $\xi^2 \geq 0.1$,
the production probability does not sensitively depend on the
pulse shape, and  FPA and IPA results are close to each other.
This means that at large field intensity the dynamical aspects of
the pair production gain a dominant role in comparison with the
pulse shape and size effects.

Finally, we note that, at finite $\xi^2$, the dependence of the
probability on the azimuthal angle $\phi_{e}$ disappears and the
distribution in the $x-y$ plane becomes isotropic.

\begin{figure}[ht]
\begin{center}
\includegraphics[width=0.35\columnwidth]{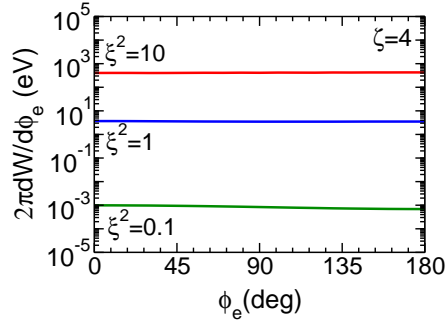}
\end{center}
\caption{\small{The differential probability of
$\ee$ pair production as a function of $\phi_{e'}=\phi_0$ at
$\zeta=4$ and $N=2$ for different values of $\xi^2$.
 \label{Fig:14}}}
\end{figure}
As an example, in Fig.~\ref{Fig:14} we present prediction for
the differential probability of $\ee$-pair
production as a function of $\phi_{e}=\phi_0$ at $\zeta=4$ for the
hyperbolic secant pulse shape with $N=2$ at $\xi^2=0.1$, 1 and 10.
This result reflects the isotropy of the $\ee$ emission and expose
the $\xi^2$ dependence in the considered kinematical region.

\subsection{Pair production at large field intensity $(\xi^2\gg1)$}

At large values of $\xi^2\gg1$,
the basic functions $Y_l$ and $X_l$ in
Eq.~(\ref{III24}) can be expressed in the form of (\ref{U01}):
\begin{eqnarray}
{Y}_l=
\int\limits_{-\infty}^{\infty}dq\,F^{(1)}(q)\,G(l-q)~,\qquad
{X}_l=\int\limits_{-\infty}^{\infty}dq\,F^{(2)}(q)\,G(l-q)~,
\label{H1}
\end{eqnarray}
where $F^{(1)}(q)$ and $F^{(2)}(q)$ are Fourier transforms of
the functions $f(\phi)$ and $f^{2}(\phi)$, respectively,
and $G(l)$ may be written as
\begin{eqnarray}
G(l)=\frac{1}{2\pi}\int\limits_{-\infty}^{\infty}d\phi
\,{\rm e}^{i\left( l\phi -z\sin\phi +\xi^2\zeta u\phi
\right)}~.
\label{H2}
\end{eqnarray}
In deriving this equation we have considered the following facts:
(i) at large $\xi^2$ the probability is isotropic, therefore we put
$\phi_0=0$, (ii) the dominant contribution to the rapidly oscillating
exponent comes from the region $\phi\simeq0$, where the difference of two
large values $l\phi$ and $z\sin\phi$ is minimal, and therefore,
one can decompose the last term in the function
${\cal P}(\phi)$ in (\ref{III21}) around $\phi=0$, and
(iii) replace in exponent $f(\phi)$ by $f(0)=1$.

Equation~(\ref{H2}) represent an asymptotic form of the Bessel
functions $J_{\tilde l}(z)$~\cite{WatsonBook}
with $\tilde l= l + \xi^2\zeta u$ at $\tilde l\gg 1$, $z\gg 1$,
and therefore the following identities
are valid
\begin{eqnarray}
G(\tilde l-1) - G(\tilde l+1)=2G_z'(\tilde l), \qquad
G(\tilde l-1) + G(\tilde l+1)=2\frac{\tilde l}{z} G(\tilde l)~,
\label{H3}
\end{eqnarray}
which allow to express the partial probability $w(\tilde l)$ in
(\ref{III26-0})
as a sum of the diagonal (relative to $\tilde l$) terms: $Y_{\tilde l}^2$,
$Y_{\tilde l}X_{\tilde l}$, $X_{\tilde l}^2$ and $Y^{'2}_{\tilde l}$.
The integral over $\tilde l$
from the diagonal term can be expressed as
\begin{eqnarray}
I_{YY}=\int\limits_{\tilde l_0}^{\infty}d{\tilde l}\,Y_l^2=
\int dq\,dq' F^{(1)}(q)\, F^{(1)}(q')
\int\limits_{\tilde l_0}^{\infty}d\tilde l
G(\tilde l-q)G(\tilde l-q')~,
\label{H4}
\end{eqnarray}
where $\tilde l_0=\zeta(1+\xi^2u$.
Taking into account that for the rapidly oscillating $G$ functions
$G(l-q)G(l-q')\simeq \delta(q-q')G^2(l-q)$ and
$\langle q \rangle\ll\langle l \rangle\ \sim \xi^2$ one gets
\begin{eqnarray}
I_{YY}=
\frac{1}{2\pi}\int\limits_{-\infty}^{\infty}d\phi f^{2}(\phi)
\int\limits_{\tilde l_0}^{\infty}d\tilde l G^2(\tilde l)
=N_{YY}\int\limits_{\tilde l_0}^{\infty}d\tilde l G^2(\tilde l)~.
\label{H44}
\end{eqnarray}
Similar expressions are valid
for the other diagonal terms with own normalization factors.
For  the $X^2_{\tilde l}$ term it is
$N_{XX}=\frac{1}{2\pi}\int\limits_{-\infty}^{\infty}d\phi
f^{4}(\phi)$, and for $Y_{\tilde l}X_{\tilde l}$,  $N_{YX}=\frac{1}{2\pi}\int\limits_{-\infty}^{\infty}d\phi
f^{3}(\phi)$. At large $\xi^2$,
the probability does not depend on the
 envelope shape, because only the central part of
the envelope is important. Therefore, for simplicity,
we choose the flat-top shape
with $N_{YY}=N_{YX}=N_{XX}=N_{0}=\Delta/\pi$ which is valid for any
smooth (at $\phi\simeq 0$) envelopes.

Making a change of the variable $l\to \tilde l= l+\xi^2\zeta u$
the variable $z$ takes the following form
\begin{eqnarray}
z^2=4\xi^2\zeta^2\left(uu_l -u^2\right)
=\frac{4\xi^2l_0^2}{1+\xi^2}\left(uu_{\tilde l} -u^2\right)
\label{H5}
\end{eqnarray}
with $l_0=\zeta(1+\xi^2)$ and $u_{\tilde l}\equiv {\tilde l}/{l_0}$,
that is exactly the same as the variable $z$ in IPA with the substitution
$l \to \tilde l$.
All these transformations
allow to express the total probability in a form similar to the
probability in IPA for large values of $\xi^2$ and a large number of
partial harmonics $n$, replacing the sum over $n$ by an integral
over $n$~\cite{Ritus-79}
\begin{eqnarray}
W&=&\frac12 {\alpha M_e\zeta^{1/2}}
\int\limits_{l_0}^{\infty}d\tilde l
\int\limits_1^{u_{\tilde l}}\frac{du}{u^{3/2}\sqrt{u-1}}
\{J^2_{\tilde l}(z)\nonumber\\
&+&\xi^2(2u-1)[
(\frac{{\tilde l}^2}{z^2} -1  ) {J}^2_{\tilde l}(z) + { J'}^2_{\tilde l}(z)
]\}~.
\label{H6}
\end{eqnarray}

Utilizing Watson's representation~\cite{WatsonBook}
for the Bessel functions at $\tilde l,\,z\gg1$ and
$\tilde l>z$,
$J_{\tilde l}(z)=({{2\pi\tilde l\tanh\alpha}})^{-1/2}
\exp[-\tilde l(\alpha -\tanh\alpha)]$ with
$\cosh\alpha={\tilde l}/{z}$,
and employing a saddle point approximation in the integration in
(\ref{H6}) we find the total
probability of $\ee$ production as (for details see Appendix~A)
\begin{eqnarray}
W=\frac{3}{8}\sqrt{\frac32} \frac{\alpha M_e \xi}{\zeta^{1/2}}
\,d\,
\exp\left[ -\frac{4\zeta}{3\xi}(1-\frac{1}{15\xi^2}) \right],
\,\,d=1+ \frac{\xi}{6\zeta}\left(1+\frac{\xi}{8\zeta} \right)~.
\label{H7}
\end{eqnarray}
This expression resembles the production probability in IPA which
is the consequence of the fact that, at $\xi^2\gg1$
in a short pulse, only the central
part of the envelope at $\phi\simeq 0$ is important.
In case of $\xi/\zeta<<1$, approximating
$d=1+{\cal O}(\xi/\zeta)$, the leading order
term recovers the Ritus result~\cite{Ritus-79}.

\begin{figure}[ht]
\begin{center}
\includegraphics[width=0.35\columnwidth]{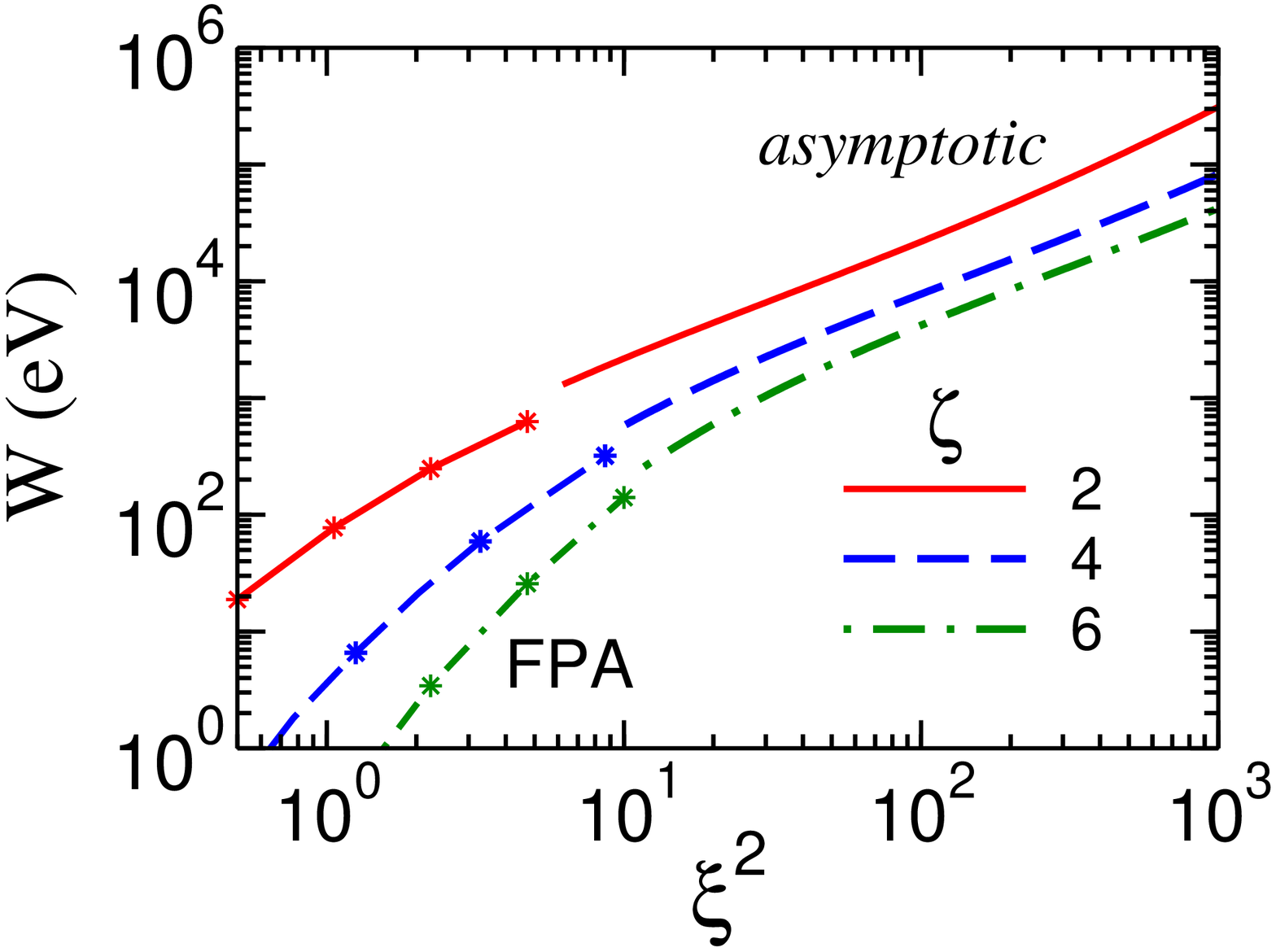}\qquad
\includegraphics[width=0.35\columnwidth]{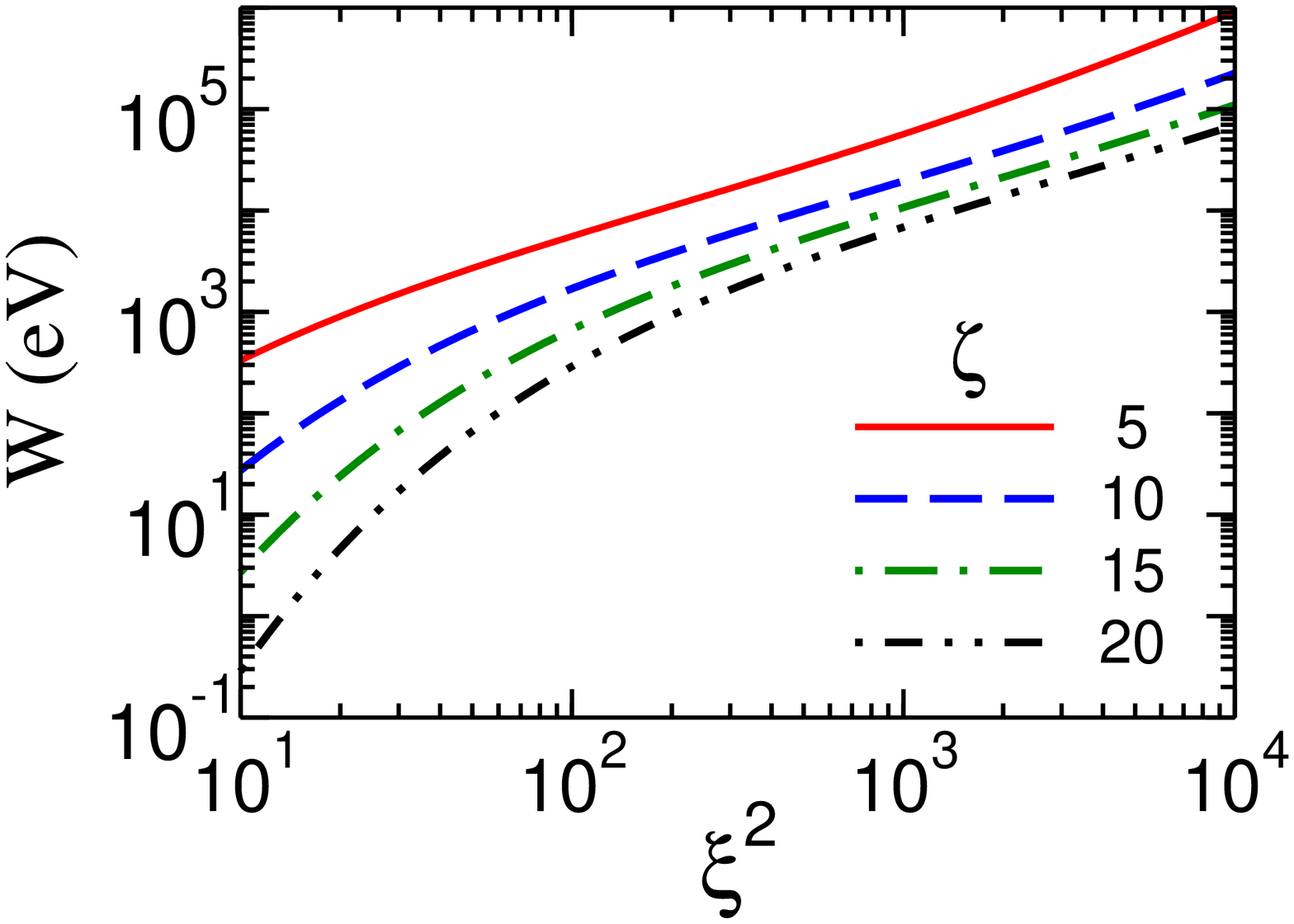}
\end{center}
\caption{\small{The total probability $W$ of the
$\ee$ pair production as a function of $\xi^2$
for various values of $\zeta$.
Left panel: Results of full numerical calculation in FPA
for finite values of $\xi^2\leq 10$
(curves marked by "stars"\ in "FPA"\ sections)
and  the asymptotic probability~(\ref{H7})
for large values of $\xi^2$ (sections labeled by "asymptotic")
at $\zeta=2$, 4 and 6.
Right panel: The asymptotic probability~(\ref{H7})
for various values of
$\zeta$ as indicated in the legend.
\label{Fig:15} }}
\end{figure}

For completeness, in Fig.~\ref{Fig:15} (left panel)
we present FPA results of a full numerical calculation
for finite values of $\xi^2\leq 10$
for the hyperbolic secant envelope shape with $N=2$
(curves are marked by "stars")
and the asymptotic probability calculated by Eq.~(\ref{H7})
at $\zeta=2$, 4 and 6, shown by solid, dashed and dot-dashed
curves, respectively.
The transition region between the two regimes
is in the neighborhood of $\xi^2\simeq 10$.
In the right panel, we show the production probability at
asymptotically large values of $\xi^2$
for $5 \leq \zeta\leq 20$. The exponential factor
in (\ref{H7}) is
most important at relatively low values of $\xi^2\sim 10$
(large  ${\zeta}/{\xi}$).
At extremely large values of $\xi^2$ (small ${\zeta}/{\xi}$ ), the
pre-exponential factor is dominant.

\subsection{Ultra-short pulses}

In this section we consider $\ee$ pair production due to
interaction of the probe photon
with an ultra-short pulse, where the number of cycles less is than one.

\subsubsection{$\ee$ pair production at small field intensity $(\xi^2\ll1)$ }

Consider first the case of small field intensity and a finite sub-threshold
parameter $\zeta$ characterized by the relations $z\ll 1$ or $\xi\zeta\ll1$.

\begin{figure}[ht]
\begin{center}
\includegraphics[width=0.35\columnwidth]{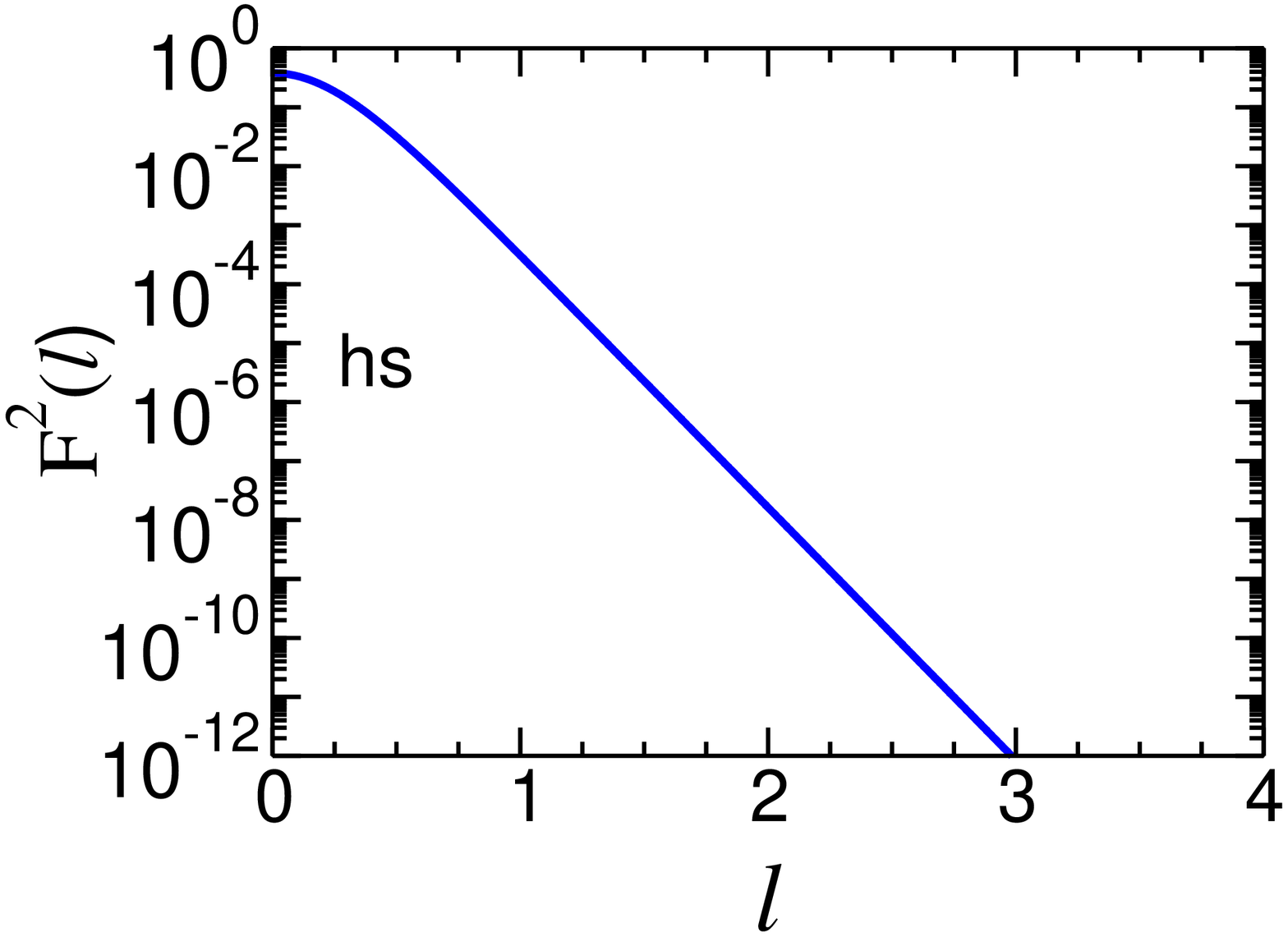}\qquad
\includegraphics[width=0.35\columnwidth]{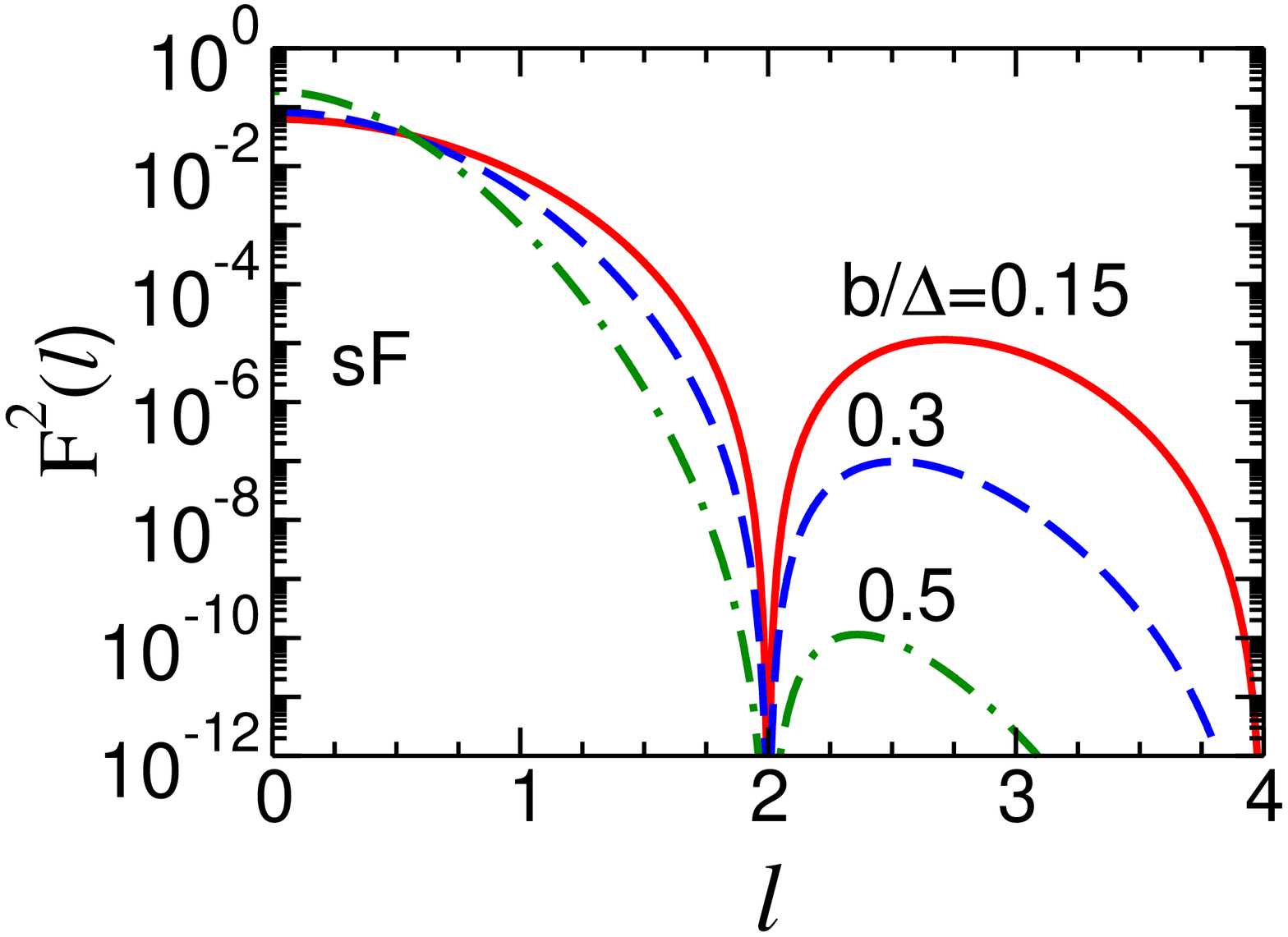}
\end{center}
\caption{\small{Square of the Fourier transforms
of the envelope functions for a sub-cycle pulse
with $N=0.5$. Left panel:
The hyperbolic secant shape.
Right panel: The solid, dashed and dot-dashed curves show
the symmetrized Fermi shape for $b/\Delta=0.15$, 0.3, and 0.5,
respectively.
 \label{Fig:2} }}
\end{figure}

The basic functions $Y_l$ in Eq.~(\ref{III26-0}) can be expressed in this regime as
\begin{eqnarray}
Y_l=\frac{1}{2\pi}\int\limits_{-\infty}^{\infty}d\phi\,{\rm e}^{il\phi}\,f(\phi)\,g(\phi)
\label{U01}
\end{eqnarray}
with
\begin{eqnarray}
g(\phi)\simeq{\rm e}^{-ic}{\rm e}^{-il\xi\cos\phi_0\phi},
\label{U011}
\end{eqnarray}
where $c=z\int_{-\infty}^{0}d\phi' f(\phi')\cos(\phi'-\phi_0)-l\phi_0$
is independent of $\phi$.
As a result one gets
\begin{eqnarray}
|Y_l|\simeq |F(l(1-\xi\cos\phi_0))|\simeq |F(l)|~,
\label{U012}
\end{eqnarray}
where $F(l)$ is the the Fourier transform of the envelope function $f(\phi)$.
Keeping the leading terms in Eq.~(\ref{III26-0}) with
$Y^2_{l-1}\simeq F^2(l-1)$, one can obtain an approximate
expression for the total production probability:
\begin{eqnarray}
W=
{\alpha M_e\zeta^{1/2}\xi^2}
\int\limits_\zeta^{\infty}dl\Phi(l)\,F^2(l-1)~,
\label{U2}
\end{eqnarray}
with
\begin{eqnarray}
\Phi(l)=v\int\limits_0^1 d\cos\theta\,\left(\frac{u}{u_l} -\frac{u^2}{u_l^2} + u -\frac12  \right)~,
\label{U3}
\end{eqnarray}
where $u=1/(1-v^2\cos^2\theta)$;  $\theta$ and $v$ are the polar angle and the velocity of the outgoing electron (positron)
in the $\ee$ c.m.s., respectively: $v=\sqrt{1-\zeta/l}$. An explicit calculation results in
\begin{eqnarray}
\Phi(l)=\frac12\left\{
\left( 1+\frac{\zeta }{l} - \frac{\zeta^2}{2l^2}\right)\log\frac{1+v}{1-v}
-v\left( 1+\frac{\zeta}{l}\right)
\right\}~.
\label{U4}
\end{eqnarray}
The Fourier transforms of the hs and sF envelope functions are given in Eq.~(\ref{U5}),
and  for illustration,
the square of the Fourier transforms for a
sub-cycle pulse with $N=0.5$ are exhibited in Fig.~\ref{Fig:2}.
The left panel corresponds to the hyperbolic secant shape. One can see a
fast monotonic decrease of $F_{\rm hs}$ at large values of  $l$.
The square of the Fourier transform for the symmetrized Fermi shape is shown in
the right panel, where the solid, dashed and dot-dashed curves correspond
to the ratio $b/\Delta=0.15$, 0.3, and 0.5, respectively.
One can see large qualitative and quantitative
differences between the one-parameter and flat-top
symmetrized Fermi shapes, in particularly, at $b/\Delta\leq 0.3$.
In the second case, $F^2$ decreases exponentially as
$\exp\left[-2\pi\Delta\frac{b}{\Delta} \right]$. The slope decreases proportionally
to ${b}/{\Delta}$ (at fixed $\Delta$).
Also, the function oscillates with the half-cycle
$\delta\,l=\pi/\Delta=\pi/0.5\pi=2$.
Contrary to the above one-parameter shapes,
the function $F_{\rm sF}$ has a significant high-$l$ component at
$2\leq l\leq 4$. This strong effect is not seen in the $\phi$ space
(cf. Fig.~\ref{Fig:03}, right top and middle top panels), where all envelope functions look
similar to each other. However, the difference in $l$-space
is very important for the pair production.

\begin{figure}[ht]
\begin{center}
\includegraphics[width=0.35\columnwidth]{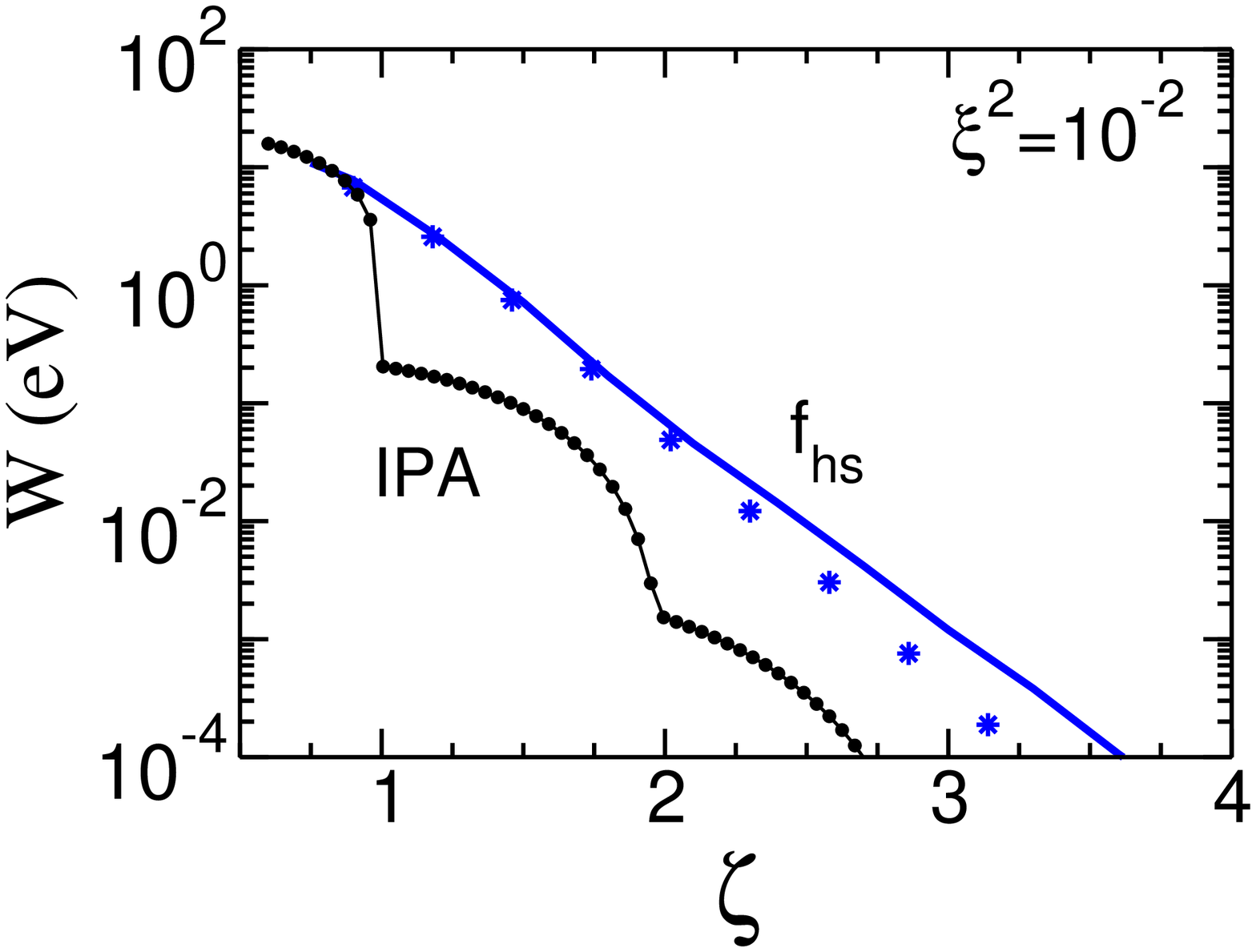}\qquad
\includegraphics[width=0.35\columnwidth]{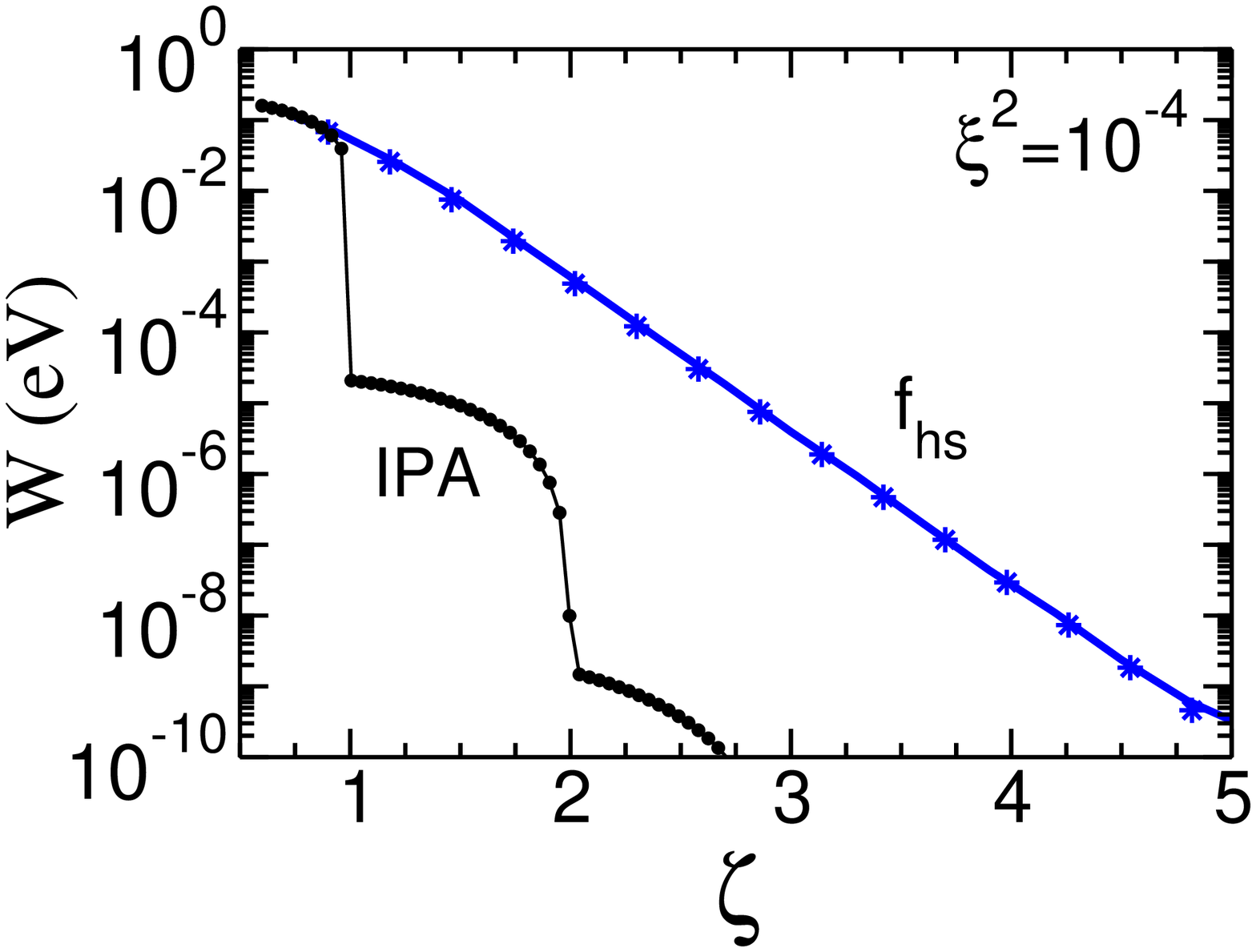}
\end{center}
\caption{\small{
The probability $W$ of $\ee$ production as a function
of the sub-threshold parameter $\zeta$ for one-parameter hs
envelope functions for  an ultra-short pulse with
$N=0.5$. The symbols "star"\
are for the approximation~(\ref{U2}). The thin solid curves marked by
dots correspond to IPA. The left
and right panels are for $\xi^2=10^{-2}$ and  $10^{-4}$, respectively.
 \label{Fig:3}}}
\end{figure}

Our prediction for the total probability of $\ee$ pair production as
a function of the sub-threshold parameter $\zeta$ for the one-parameter
envelope functions for an ultra-short pulse with
$N=0.5$ is shown in Fig.~\ref{Fig:3}.
The solid  curves exhibit result of full numerical
calculations using Eq.~(\ref{III9})
with the hyperbolic secant shape.
The symbols "star" \  display the result obtained by using
the approximation (\ref{U2}).
The thin solid curves marked by
dots correspond to the IPA case. The left
and right panels display results for $\xi^2=10^{-2}$ and  $10^{-4}$, respectively.
One can see the identity of predictions for the ultra-short pulse
and IPA near and above the threshold at $\zeta\lesssim 1$,
and a strong difference
between them below the threshold, i.e. for
$\zeta>1$. Our approximate (analytical)
solution of Eq.~(\ref{U2}) is in a fairly
good agreement with the full numerical calculation.
The function $\Phi(l)$
in Eq.~(\ref{U2}) is rather smooth compared to the Fourier transform
$F(l-1)$, therefore, the dominant contribution to the integral
in  Eq.~(\ref{U2}) comes from the lower limit of $l$, and qualitatively,
the slope of the probability as a function of $\zeta$ is determined
by the scale parameter $\Delta$ of the envelope functions
\begin{eqnarray}
W_{\rm hs}(\zeta)\sim\exp\left[-\pi\Delta\zeta \right]~.
\label{U6}
\end{eqnarray}
Despite of the exponential decrease of the probability $W$
as a function of $\zeta$, one
can see a large difference (several orders of magnitude) between
predictions for the ultra-short pulse
and IPA. In the latter case the probability
decreases much faster with increasing $\zeta$.

\begin{figure}[ht]
\begin{center}
\includegraphics[width=0.35\columnwidth]{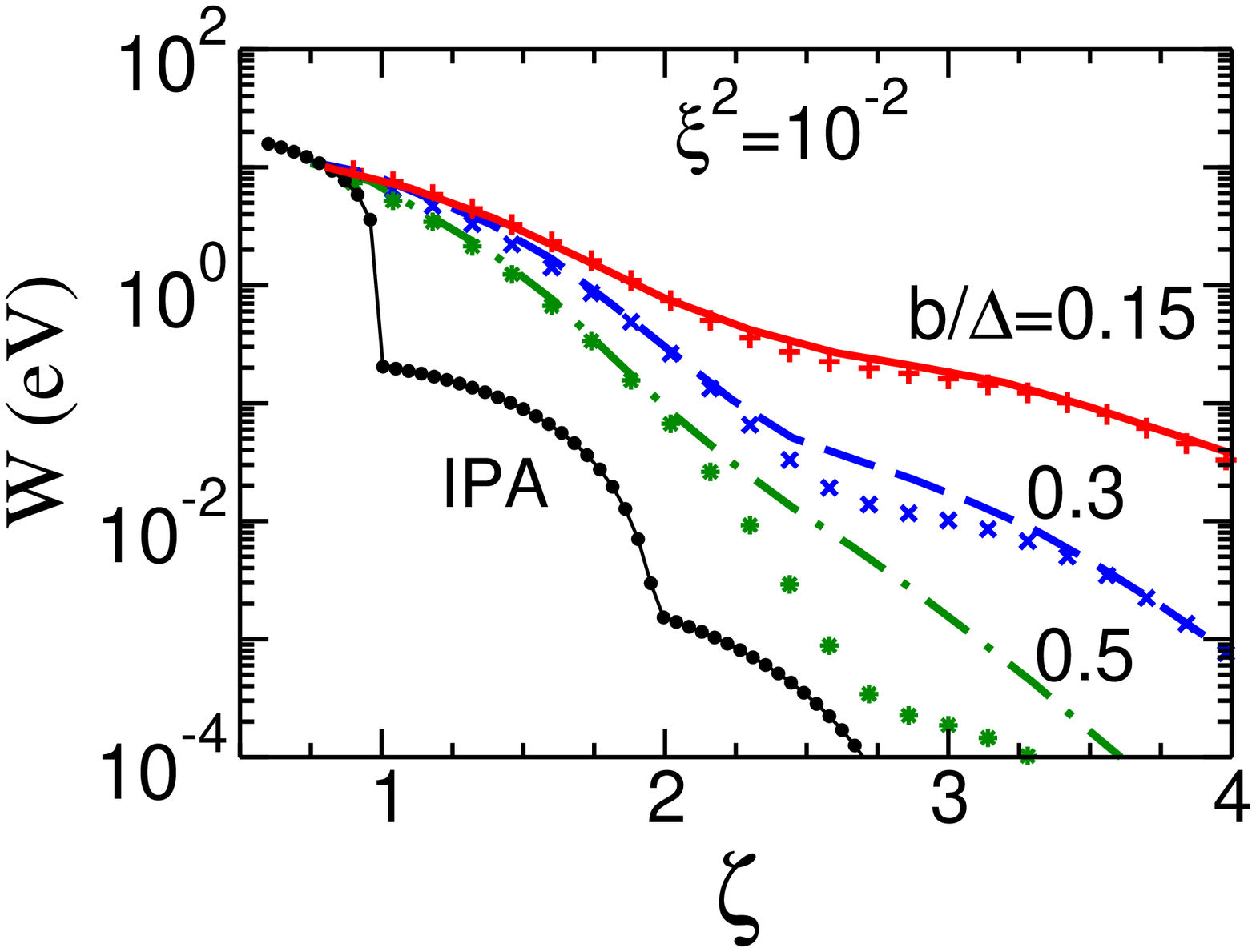}\qquad
\includegraphics[width=0.35\columnwidth]{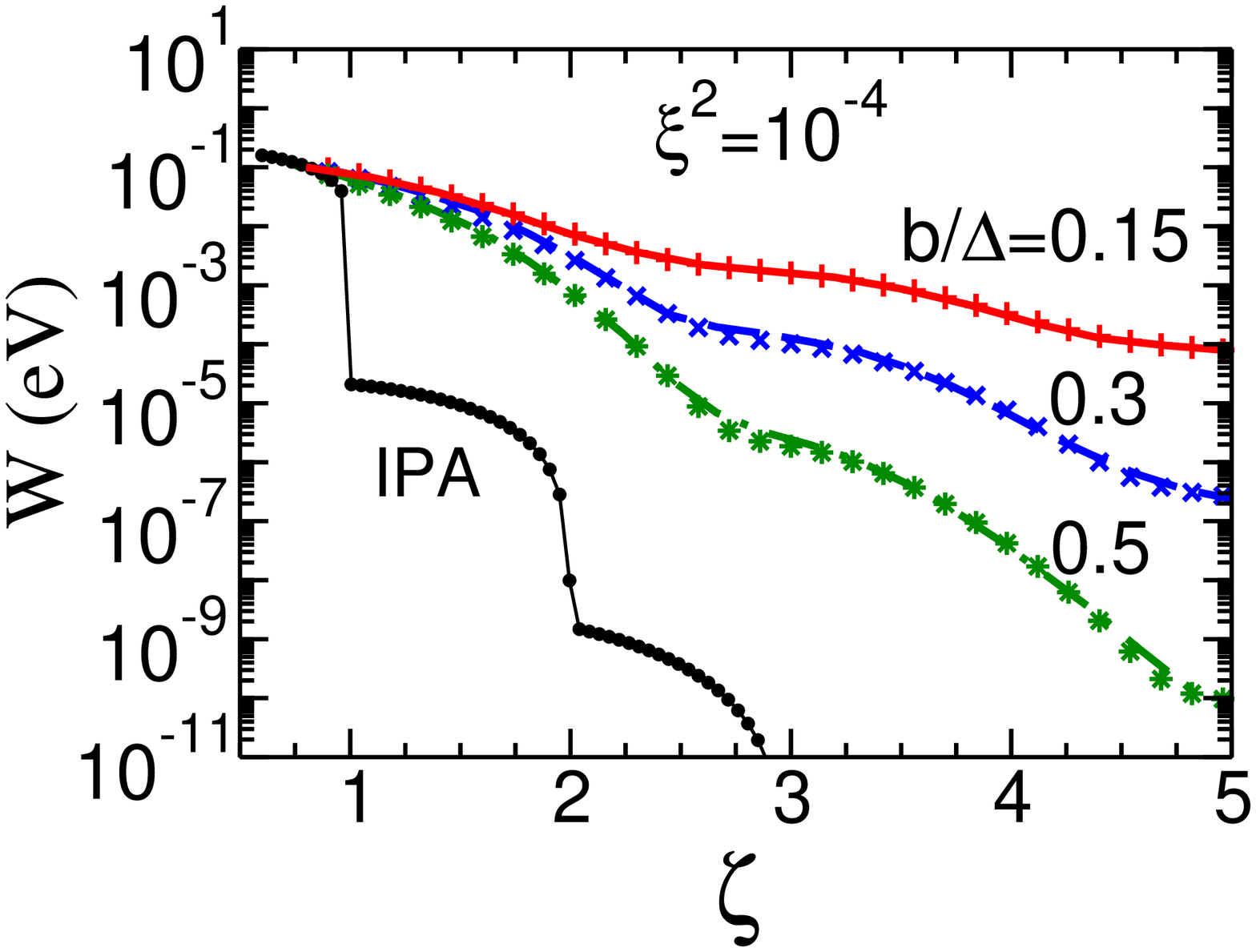}
\end{center}
\caption{\small{
The same as in Fig.~\ref{Fig:3} but for symmetrized Fermi shape envelope.
The solid, dashed and dot-dashed curves are for $b/\Delta=0.15$, 0.3
and 0.5, respectively.
The corresponding approximate solutions are shown by symbols "+"\ , "x"\
and "$\ast$"\ , respectively.
 \label{Fig:4} }}
\end{figure}

Our results for the symmetrized Fermi envelope is presented in Fig.~\ref{Fig:4}.
Now, the shape of the probability is determined
by the two parameters $b$ (or $b/\Delta$)
and $\Delta$
\begin{eqnarray}
W_{\rm sF}(\zeta)\sim\exp\left[-2\pi\Delta\frac{b}{\Delta}\zeta \right]
\sin^2\Delta\zeta~.
\label{U7}
\end{eqnarray}
The first term describes the slope of the probability as a function of
$\zeta$.
The slope is proportional to the "ramping time"\ of
the envelope function, $b$ (or to the ratio $b/\Delta$ at fixed $\Delta$).
The second term, following from the Fourier transform shown in Fig.~\ref{Fig:2},
describes some oscillations with a period inversely proportional
to the duration $\Delta$ of the flat-top envelope and is independent of
the ramping parameter $b$.
Again, one can see a great difference between results
for the ultra-short pulse and IPA on qualitative and quantitative levels.
The probability in IPA has a typical step-like behavior, where each new
step indicates the contribution of the next integer harmonic.
In FPA, the probability decreases monotonically with a slope
determined by the shape of the envelope.
The quantitative difference is rather large and,
as indicated by results shown in Figs.~\ref{Fig:3} and \ref{Fig:4}, can reach
 orders of magnitude depending on the shape of the envelope.

\subsubsection{Intermediate field intensity, anisotropy}

As we have shown above, at small values of $z$, $z\ll1$,
the probability of $\ee$ production
is essentially determined by the pulse shape. The function $g(\phi)$ in
Eq.~(\ref{U01}) is not important and, therefore,
the total probability would be isotropic with respect
to the azimuthal angle $\phi_{e}=\phi_0$
because only the function ${\cal P}(\phi)$ in Eq.~(\ref{III6})
contains a $\phi_0$ dependence.
For finite values of $z$, however,
the function $g(\phi)$ becomes important, and the electron
(positron) azimuthal angle distribution is anisotropic relative
to the direction of the vector $\mathbf{a}_x\equiv\mathbf{a}_1$
in Eq.~(\ref{III1}), at least
for the monotonically rapidly decreasing one-parameter
envelope shapes. The reason of such
anisotropy is the following. At finite values of $z$, the function
${Y}(l)$ in Eq.~(\ref{U01}) is determined by the integral over $d\phi$
with a rapidly oscillating function proportional
to the exponent in
\begin{eqnarray}
{\rm e}^{i
\left[ l\phi
 -z\left(\cos\phi_0\int\limits_{-\infty}^{\phi} d\phi'\, f(\phi')\cos\phi'
 + \sin\phi_0\int\limits_{-\infty}^{\phi} d\phi'\, f(\phi')\sin\phi'
 \right) \right]
 }~.
\label{U8}
\end{eqnarray}
In the case of a fast-decreasing function $f(\phi')$,
the contribution of the term proportional to $\sin\phi_0$ is much smaller
compared to the term proportional to  $\cos\phi_0$, because
the functions $f(\phi')$ and $\sin\phi'$ in the second integral
are in "anti-phase". At finite $z$,
the dominant contribution to the functions $Y_{l}$
comes from the region where
the difference in the exponent is minimal, i.e. $\phi_e=\phi_0 \simeq 0$.
This means that the electrons would be emitted mostly along
the vector $\mathbf{a}_x$
and the positrons in the opposite direction.

We define the anisotropy of the electron emission as
\begin{eqnarray}
{\cal A}=\frac{dW(\phi_e) - dW(\phi_e+\pi)}{dW(\phi_e) + dW(\phi_e+\pi)}~.
\label{U9}
\end{eqnarray}
\begin{figure}[ht]
\includegraphics[width=0.35\columnwidth]{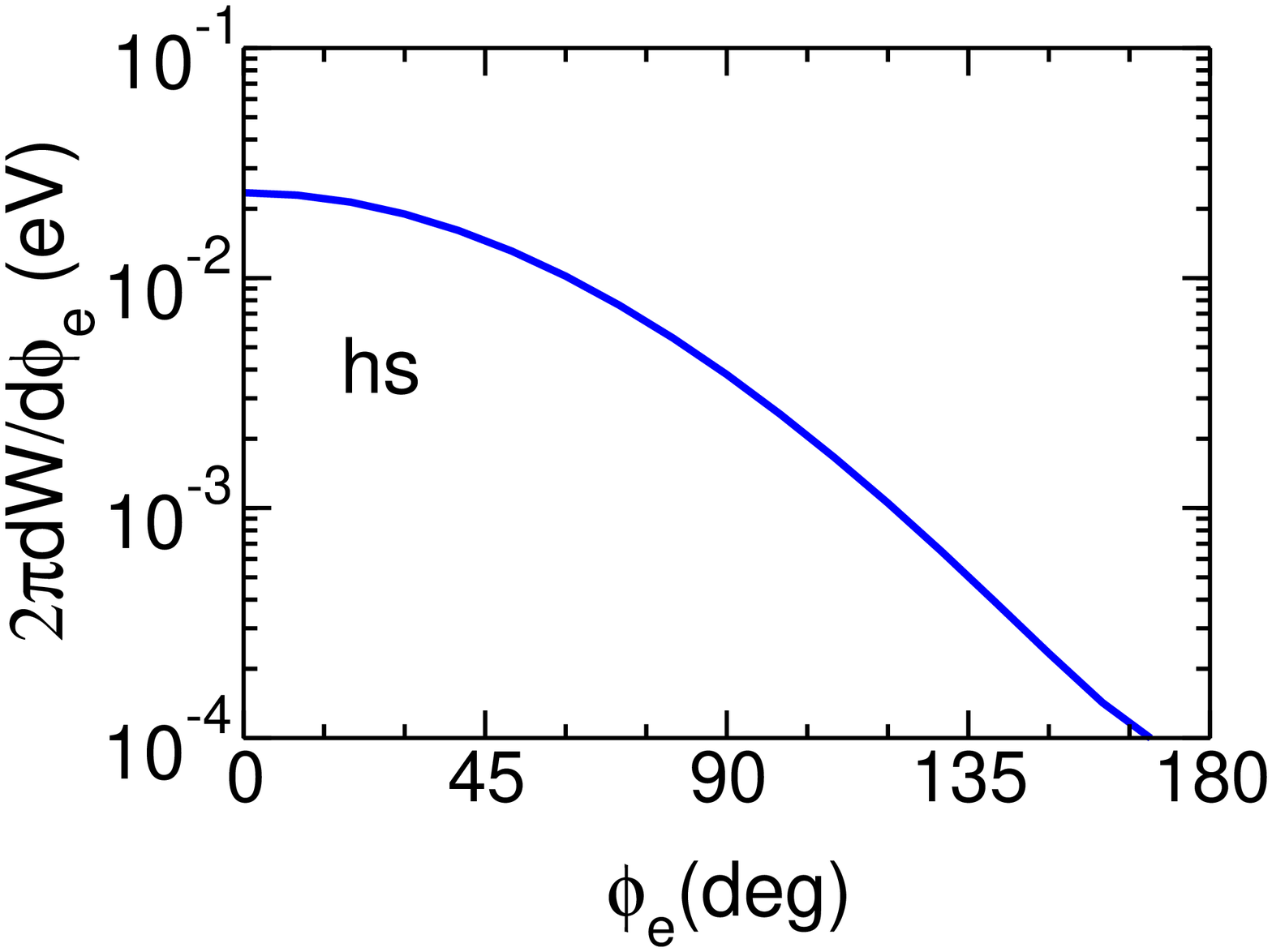}\qquad
\includegraphics[width=0.35\columnwidth]{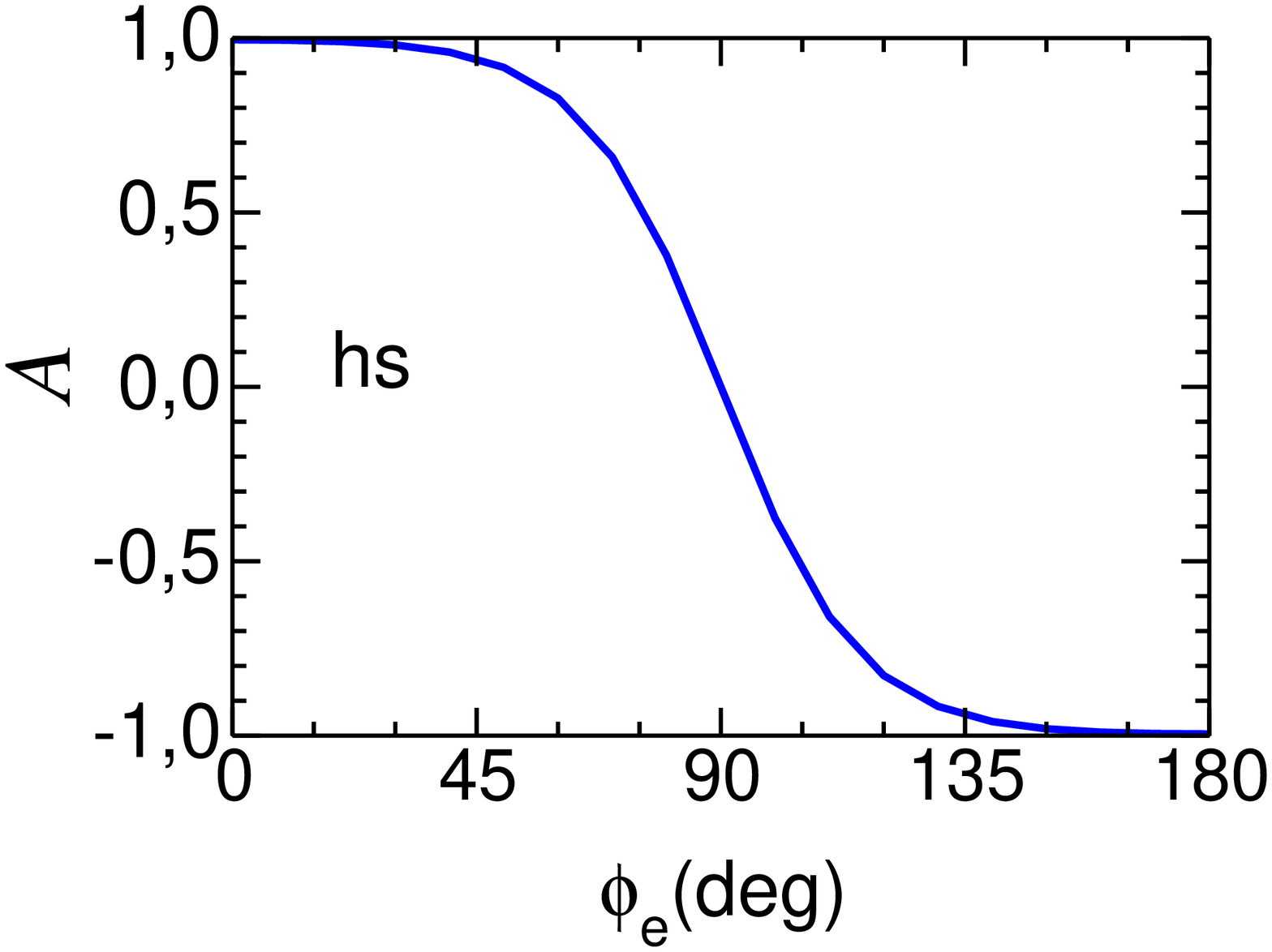}
\caption{\small{
Left panel: The differential production probability as a function of the azimuthal
angle $\phi_e$ of the electron emission.
Right panel: The anisotropy~(\ref{U9}) for  the hyperbolic secant
shapes. For $\xi^2=0.1$ and $ \zeta=4$.
 \label{Fig:5}}}
\end{figure}

The differential probability of the $\ee$ pair emission
and the anisotropy as functions of the azimuthal
angle $\phi_e$ are exhibited in Fig.~\ref{Fig:5}.
The calculations are for the fast-decreasing one-parameter hs
envelope functions for $\Delta=0.5\pi$,
$\zeta=4$ and $\xi^2=0.1$. One can see
a rapidly decreasing probability with $\phi_e$ which leads to the strong
anisotropy  of electron (positron) emission.

In the case the of the symmetrized Fermi distribution with
small $b/\Delta$, the situation changes drastically.
As $b/\Delta\to 0$ the envelope function goes to
the flat-top (step-like)
shape $f_{\rm Fs}(\phi)\to \theta(\Delta^2-\phi^2)$ with $\theta(x)=1$, 0
for $x\ge 0$ or $x< 0$, respectively, and correspondingly
\begin{eqnarray}
Y_l\simeq\frac{1}{2\pi}
\int\limits_{-\Delta}^{\Delta}
d\phi \, {\rm e}^{i\left[\tilde l\phi -z\sin(\phi-\phi_0) \right]}
\label{U99}
\end{eqnarray}
 with $\tilde l= l+ \xi^2\zeta u$. The function $Y_l$
 in the  region $\zeta\leq l< l_{\max}\gg1$
 is alternating, rapidly oscillating with an amplitude that depends
 only on $\xi$, $\zeta$, and $u$.
 It is not sensitive to $\phi_0$. A change in
 $\phi_0$ leads to some phase shift of ${Y}(l)$ in a range of integration,
 leaving $\langle|Y_l|^2\rangle$ to be independent of $\phi_0$
 \begin{figure}[ht]
\includegraphics[width=0.35\columnwidth]{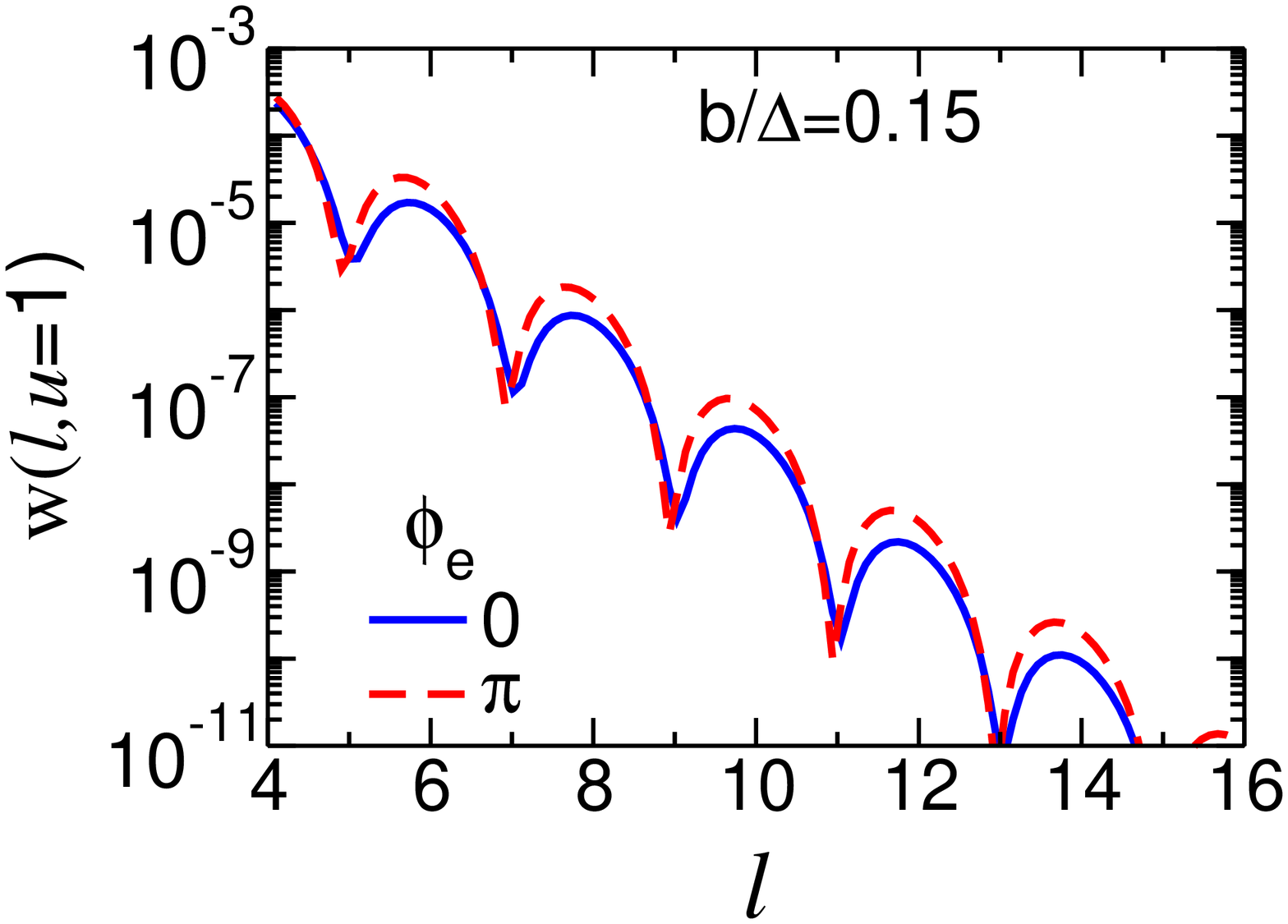}\qquad
\includegraphics[width=0.35\columnwidth]{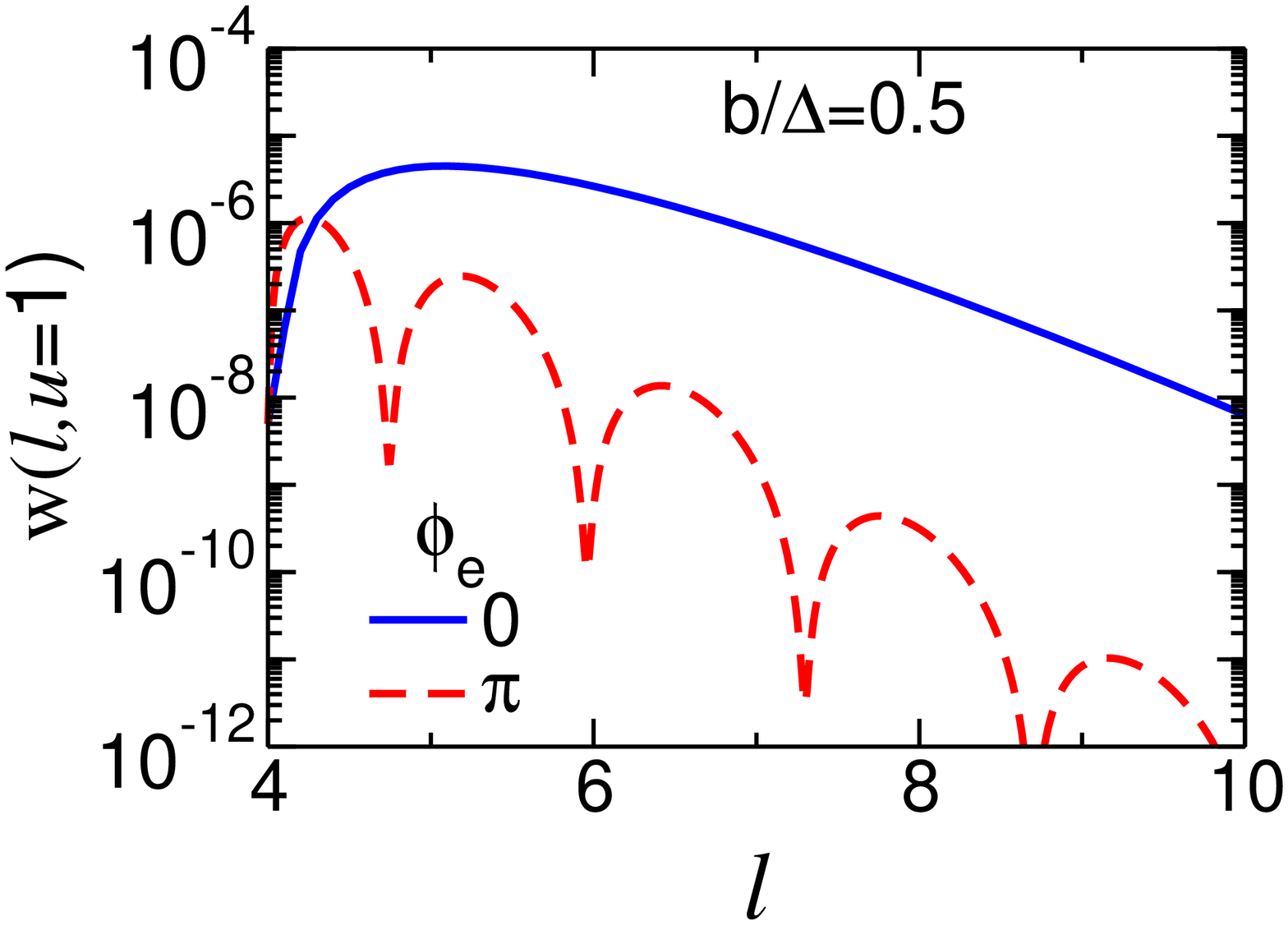}
\caption{\small{
The partial probability $w(l)$ defined in~(\ref{III26-0})
at $\phi_0=0$ and $\pi$ shown by solid and
dashed curves, respectively,
for the symmetrized Fermi envelope shape.
The left panels correspond to small values of
$b/\Delta=$0.15, while the right panel is for $b/\Delta=0.5$.
For $\xi^2=0.1$ and $ \zeta=4$.
 \label{Fig:6}}}
\end{figure}
Therefore, the dependence of the integral of
the partial probability $w(l)\sim |Y_l|^2$ in Eq.~(\ref{III9})
on $\phi_0$ is negligible.
As an example, in the left panel of Fig.~\ref{Fig:6} we present the
partial probability $w(l)$ as a function of $l$, calculated at
$\xi^2=0.1$, $\zeta=4$ and $u=1$ for the small values of
$b/\Delta$ equal to 0.15 at $\phi_0=0$ and $\pi$, shown by
solid and dashed curves, respectively. One can see some small
modification of the frequency of oscillations at $l\sim l_{\rm min}=\zeta$
at two extreme values of $\phi_0$, but the amplitudes of the oscillations
are close to each other.
This situation is quite different from the case of a large
value of $b/\Delta=0.5$ exhibited in the right panel of Fig.~\ref{Fig:6}.
One can see a strong difference in the $l$ dependence of $w(l)$ for $\phi_0=0$
and $\pi$. In the first case, the function
$w(l)$ has only one oscillation in a wide range of $l$ and decreases smoothly
with $l$. In the second case, the probability has a number of oscillations
decreasing rapidly with increasing $l$. As a result, the total probability
in the second case is much smaller.
\begin{figure}[ht]
\includegraphics[width=0.35\columnwidth]{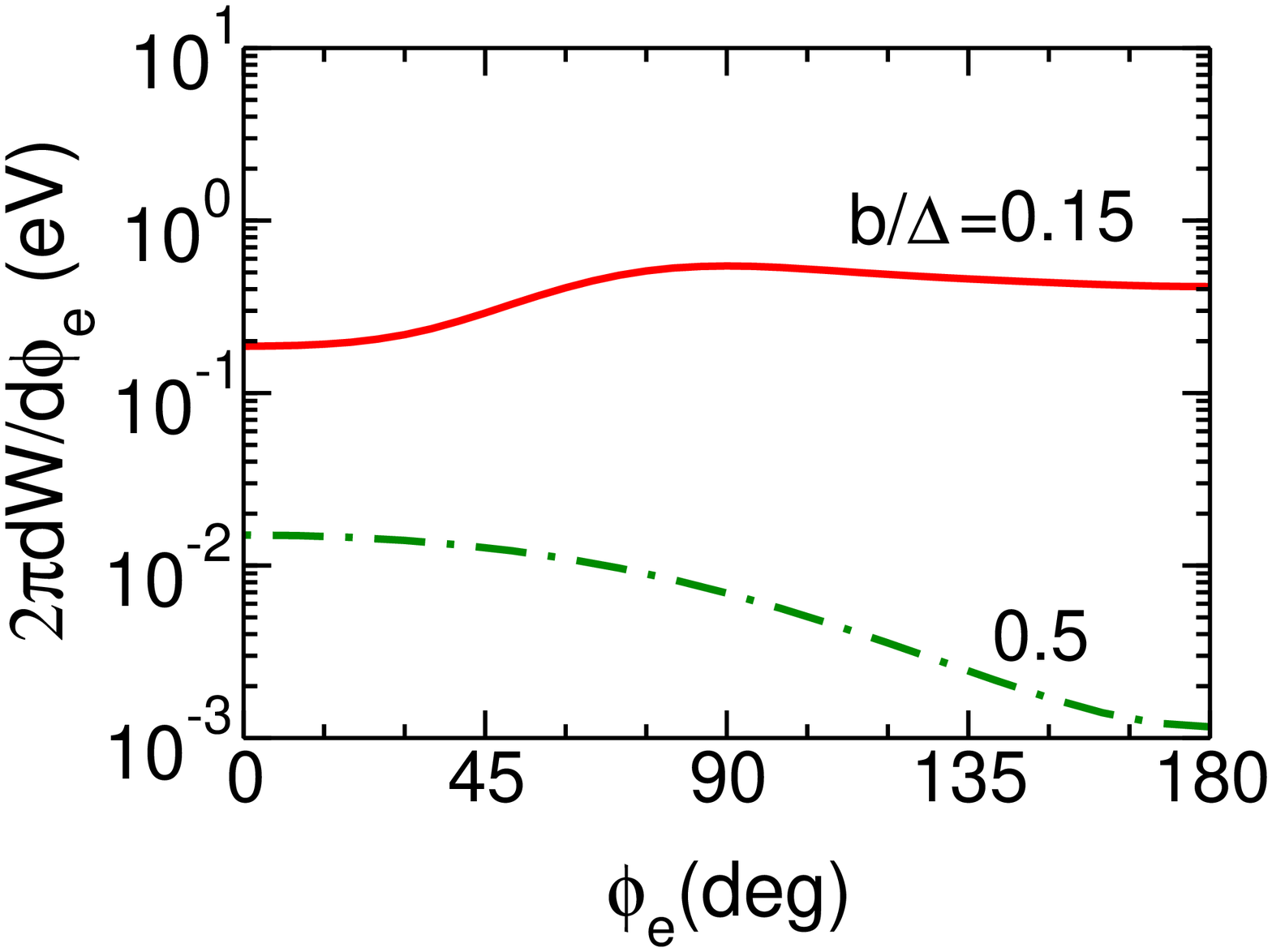}\qquad
\includegraphics[width=0.35\columnwidth]{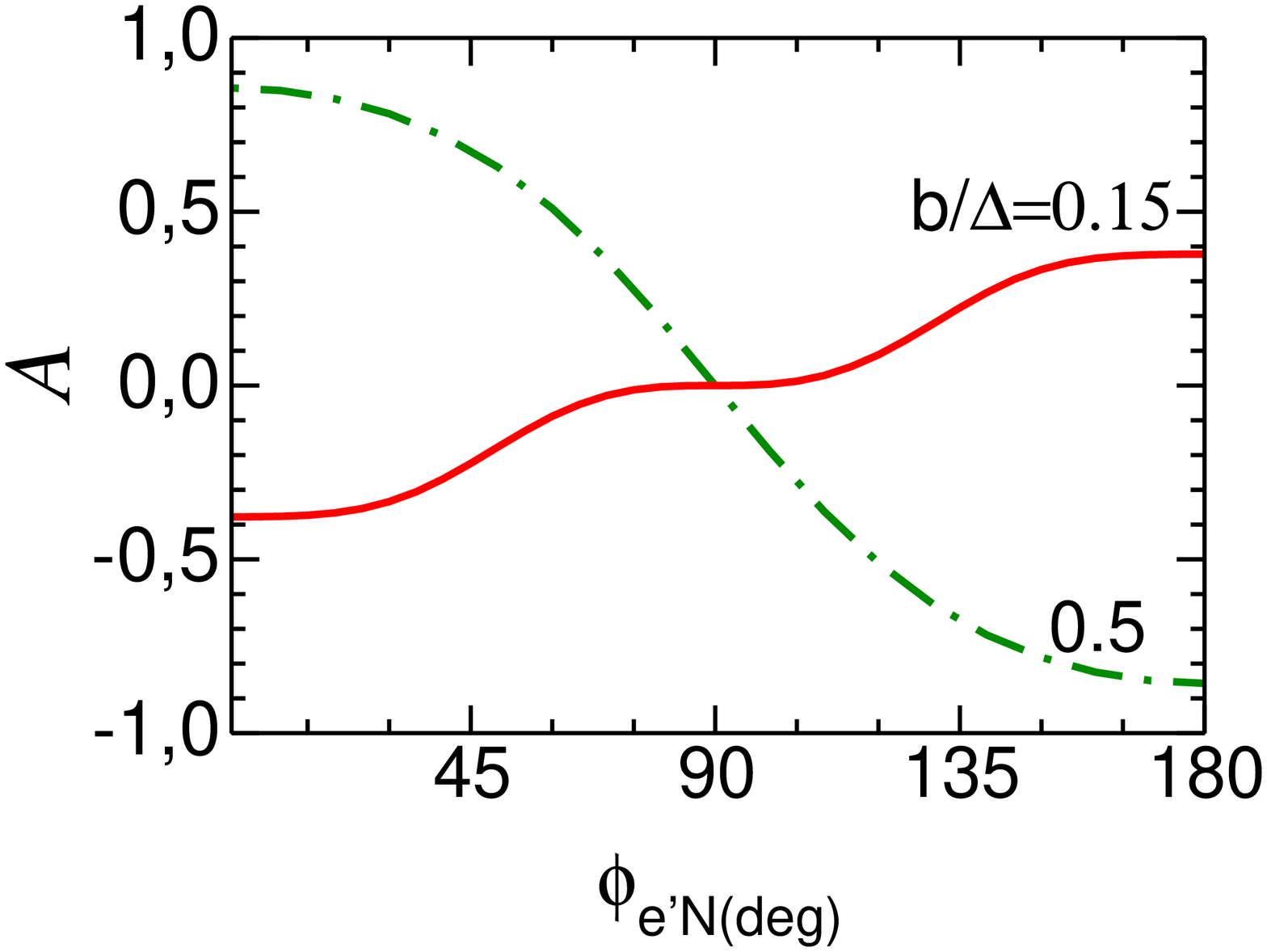}
\caption{\small{
The same as in  Fig.~\ref{Fig:5} but for the symmetrized Fermi shape.
The solid and dashed curves are for
$b/\Delta=0.15$, and 0.5, respectively.
 \label{Fig:7}}}
\end{figure}

In Fig.~\ref{Fig:7} we present our results for
the symmetrized Fermi shape for the production probability (left panel)
and for the anisotropy (right panel) for $b/\Delta=0.15$, and 0.5.
The result for $b/\Delta=0.5$ is similar to that
shown in Fig.~\ref{Fig:5}. However, for smaller values of $b/\Delta$,
the probability is a smooth function of $\phi_e$
which leads to a small absolute value of the anisotropy.

\subsection{Effect of the finite carrier phase}

Consider now the impact of the finite carrier phase $\tilde \phi$ in
the e.m. potential~(\ref{III1}) for the $\ee$ production.
In the case of finite $\tilde\phi$, the functions $C^{(i)}_l$
in the transition matrix~(\ref{III4}) are modified as follows
\begin{eqnarray}
C^{(0)}(l)&=&\frac{1}{2\pi l} \int\limits_{-\infty}^{\infty}
d\phi
\left( z\cos(\phi-\phi_0+\tilde\phi)\,f(\phi)-\xi^2\zeta u\,f^2(\phi)\right)
\,{\rm e}^{il\phi -i{\cal P(\phi)}}~,\nonumber\\
C^{(1)}(l)&=&\frac{1}{2\pi}\int\limits_{-\infty}^{\infty}
d\phi f^2(\phi)\,{\rm e}^{il\phi -i{\cal P(\phi)}}~,\nonumber\\
C^{(2)}(l)&=&\frac{1}{2\pi}\int\limits_{-\infty}^{\infty}
d\phi f(\phi)\,\cos(\phi+\tilde\phi)\,{\rm e}^{il\phi -i{\cal P(\phi)}}~,\nonumber\\
C^{(3)}(l)&=&\frac{1}{2\pi}\int\limits_{-\infty}^{\infty}
d\phi f(\phi)\,\sin(\phi+\tilde\phi)\,{\rm e}^{il\phi -i{\cal P(\phi)}}~,
\label{CP1}
\end{eqnarray}
with
\begin{eqnarray}
{\cal P(\phi)}=z\int\limits_{-\infty}^{\phi}\,d\phi'\,\cos(\phi'-\phi_0+\tilde\phi)f(\phi')
-\xi^2\zeta u\int\limits_{-\infty}^\phi\,d\phi'\,f^2(\phi')~.
\label{CP2}
\end{eqnarray}
Utilizing the new basic functions
\begin{eqnarray}
Y_l(z)&=&\frac{1}{2\pi} {\rm e}^{-il(\phi_0-\tilde\phi)}\int\limits_{-\infty}^{\infty}\,
d\phi\,{f}(\phi)
\,{\rm e}^{il\phi-i{\cal P}(\phi)} ~,\nonumber\\
X_l(z)&=&\frac{1}{2\pi}{\rm e}^{-il(\phi_0-\tilde\phi)} \int\limits_{-\infty}^{\infty}\,
d\phi\,{f^2}(\phi)
\,{\rm e}^{il\phi-i{\cal P}(\phi)}~,
\label{CP3}
\end{eqnarray}
one can obtain the following
representation of the functions $C^{(i)}(l)$
\begin{eqnarray}
C^{(0)}(l)&=&\widetilde Y_l(z){\rm e}^{il(\phi_0-\tilde\phi)}~,\,
\widetilde Y_l(z)=\frac{z}{2l} \left(Y_{l+1}(z) +
Y_{l-1}(z)\right) - \xi^2\frac{u}{u_l}\,X_l(z)~,\nonumber\\
C^{(1)}(l)&=&X_l(z)\,{\rm e}^{il(\phi_0-\tilde\phi)}~,\nonumber\\
C^{(2)}(l)&=&\frac{1}{2}\left( Y_{l+1}{\rm e}^{i(l+1)\phi_0}
+ Y_{l-1}{\rm e}^{i(l-1)\phi_0}\right){\rm e}^{-il\tilde\phi} ~,\nonumber\\
C^{(3)}(l)&=&\frac{1}{2i}\left( Y_{l+1}{\rm e}^{i(l+1)\phi_0}
- Y_{l-1}{\rm e}^{i(l-1)\phi_0}\right){\rm e}^{-il\tilde\phi}~,
\label{CP4}
\end{eqnarray}
 which allows to express the partial probabilities
 $w(l)$ in Eq.~(\ref{III20}) in the form
 of Eq.~(\ref{III26-0}) but with the new basic functions~(\ref{CP3}).
 We recall that $\phi_0$ in above expressions
 is equal to the azimuthal angle $\phi_e$ of
 the outgoing electron momentum in c.m.s..

 It is naturally to expect that the  effect of the finite carrier
 phase essentially appears in the azimuthal angle distribution of
 the outgoing electron
 because the carrier phase is included in the expressions for the basic
 functions (\ref{CP3}) in the combination $\phi_e-\tilde\phi$.

\begin{figure}[ht]
\begin{center}
\includegraphics[width=0.35\columnwidth]{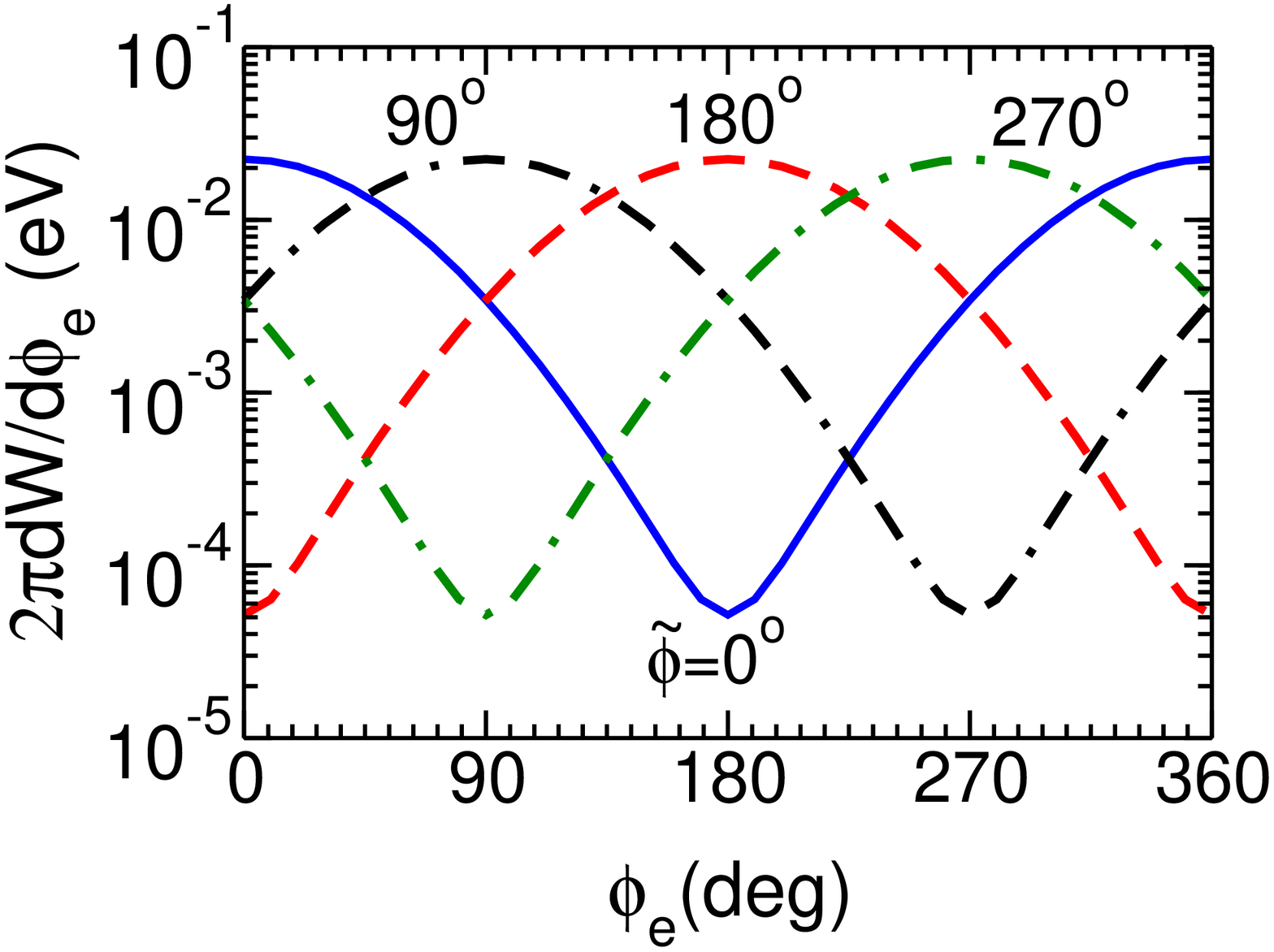}\qquad
\includegraphics[width=0.35\columnwidth]{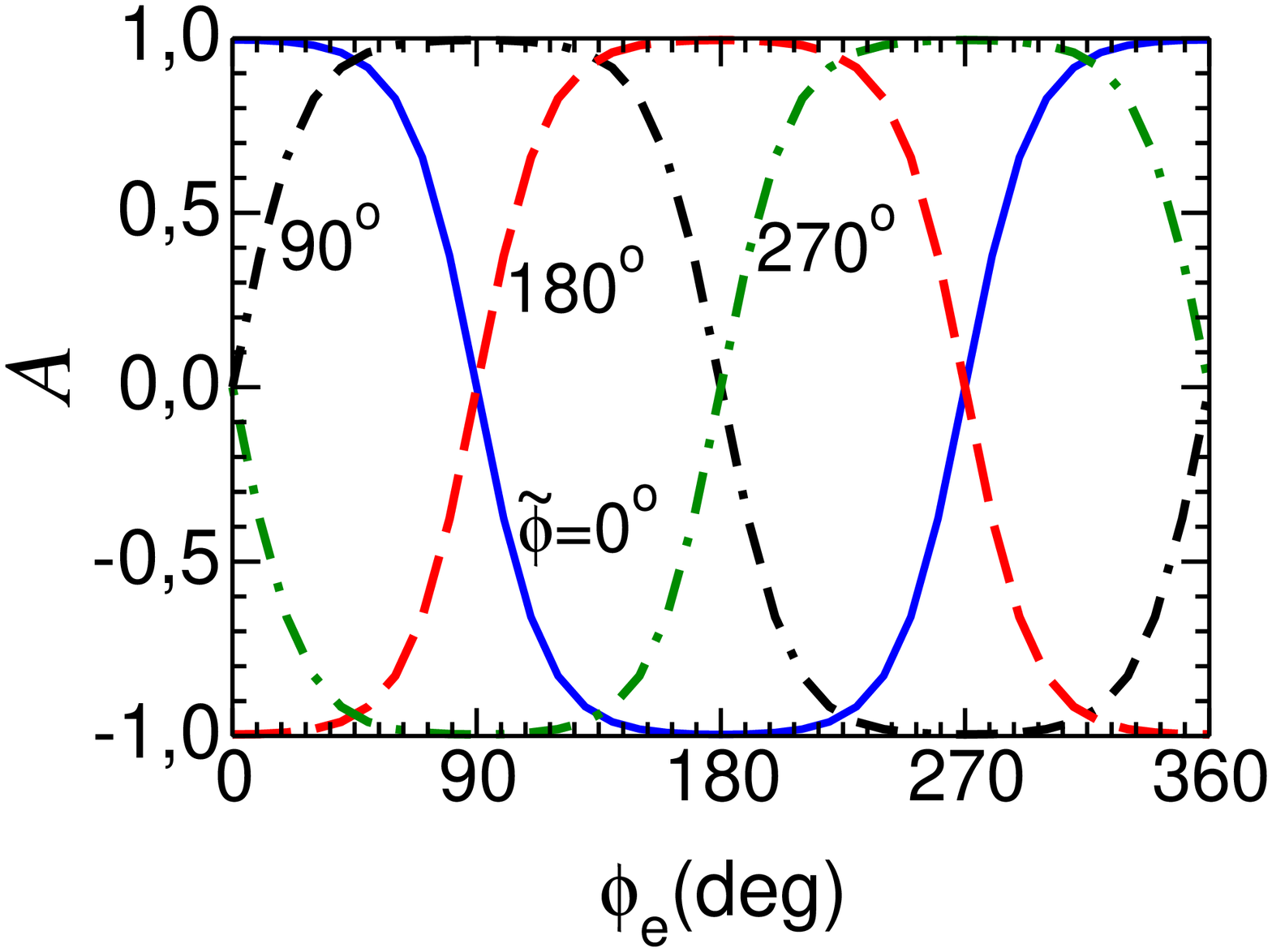}
\end{center}
\caption{\small{
Left panel: The production probability as a function of the azimuthal
angle of the direction of flight of the outgoing electron
$\phi_e$ for different values of the
carrier phase $\tilde\phi$.
The solid, dash-dash-dotted, dashed and dash-dotted
curves are for the carrier phase equal to 0,
90, 180 and 270 degrees, respectively.
Right panel: The anisotropy~(\ref{U9}) for different values of $\tilde\phi$.
For the hyperbolic secant shape with $N=0.5$;
 $\xi^2=0.1$ and $ \zeta=4$.
 \label{Fig:14A}}}
\end{figure}
 As an example, in Fig.~\ref{Fig:14A} (left panel) we show the probability
 of $\ee$ production as a  function of the azimuthal angle
 $\phi_e$ for different values of the carrier phase $\tilde\phi$
 for the sub-cycle pulse with $N=0.5$ for a hyperbolic secant shape
 with $\zeta=4$ and $\xi^2=0.1$.
 One can see a clear bump-like structure of the distribution,
 where the bump position coincides with the corresponding value of
 the carrier phase.
 The reason of such behaviour is the same as an alignment
 of the probability along $\phi_e=0$ for
 $\tilde\phi=0$ described in previous subsection.
 Indeed, now the basic functions $Y_l$ and $X_l$ are determined by the integral
 over $d\phi$
with a rapidly oscillating function proportional to the exponent
\begin{eqnarray}
{\rm e}^{i
\left[ l\phi
 -z\left(\cos(\phi_e-\tilde\phi)\int\limits_{-\infty}^{\phi} d\phi'\, f(\phi')\cos\phi'
 + \sin(\phi_e-\tilde\phi)\int\limits_{-\infty}^{\phi} d\phi'\, f(\phi')\sin\phi'
 \right) \right]
 }~.
\label{UU8}
\end{eqnarray}
 Then, taking into account the inequality
 \begin{eqnarray}
 \int\limits_{-\infty}^{\phi} d\phi'\, f(\phi')\cos\phi'
 \gg
 \int\limits_{-\infty}^{\phi} d\phi'\, f(\phi')\sin\phi'~,
 \label{UU81}
\end{eqnarray}
 which is valid for the sub-cycle pulse with hyperbolic secant shape,
  one can conclude
 that the main contribution to the probability comes from the region
 $\phi_e\simeq\tilde\phi$, which is confirmed by the result of our full
 calculation shown in Fig.~\ref{Fig:14A} (left panel).

  The corresponding anisotropies defined by Eq.~(\ref{U9}) are exhibited in
  Fig.~\ref{Fig:14A} (right panel). One can see a strong dependence
  of the anisotropy on the carrier phase which leads to the
  "bump"\ structure of the differential probabilities shown in the
  left panel.
  The anisotropy takes a maximum value ${\cal A}\simeq 1$
  at $\phi_e=\tilde\phi$ and  $|{\cal A}|<1$ at $\phi_e\neq\tilde\phi$.
  It takes a minimum value ${\cal A}\simeq -1$
  at  $\phi_e-\tilde\phi=\pm\pi$.

\begin{figure}[ht]
\begin{center}
\includegraphics[width=0.35\columnwidth]{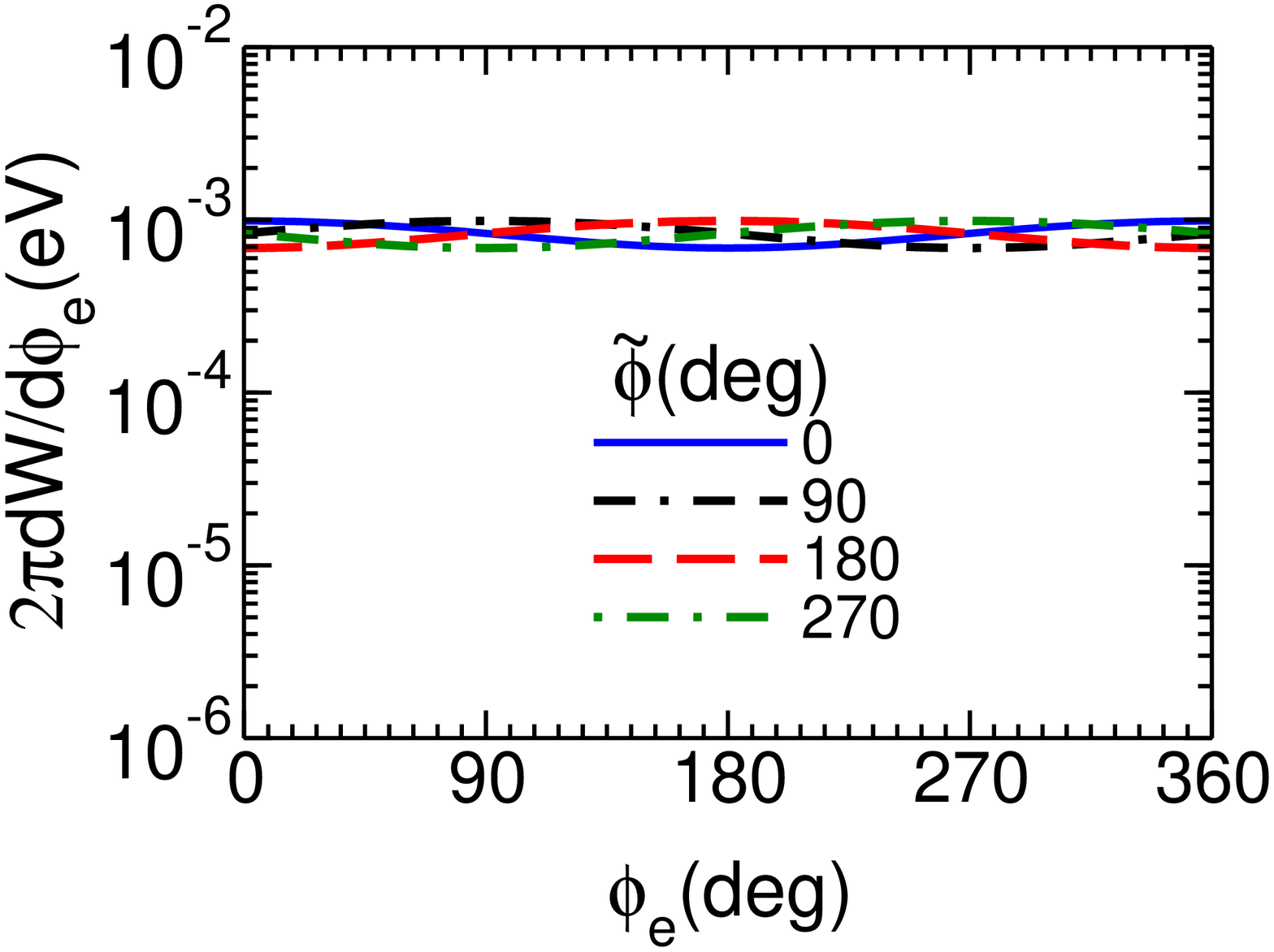}\qquad
\includegraphics[width=0.35\columnwidth]{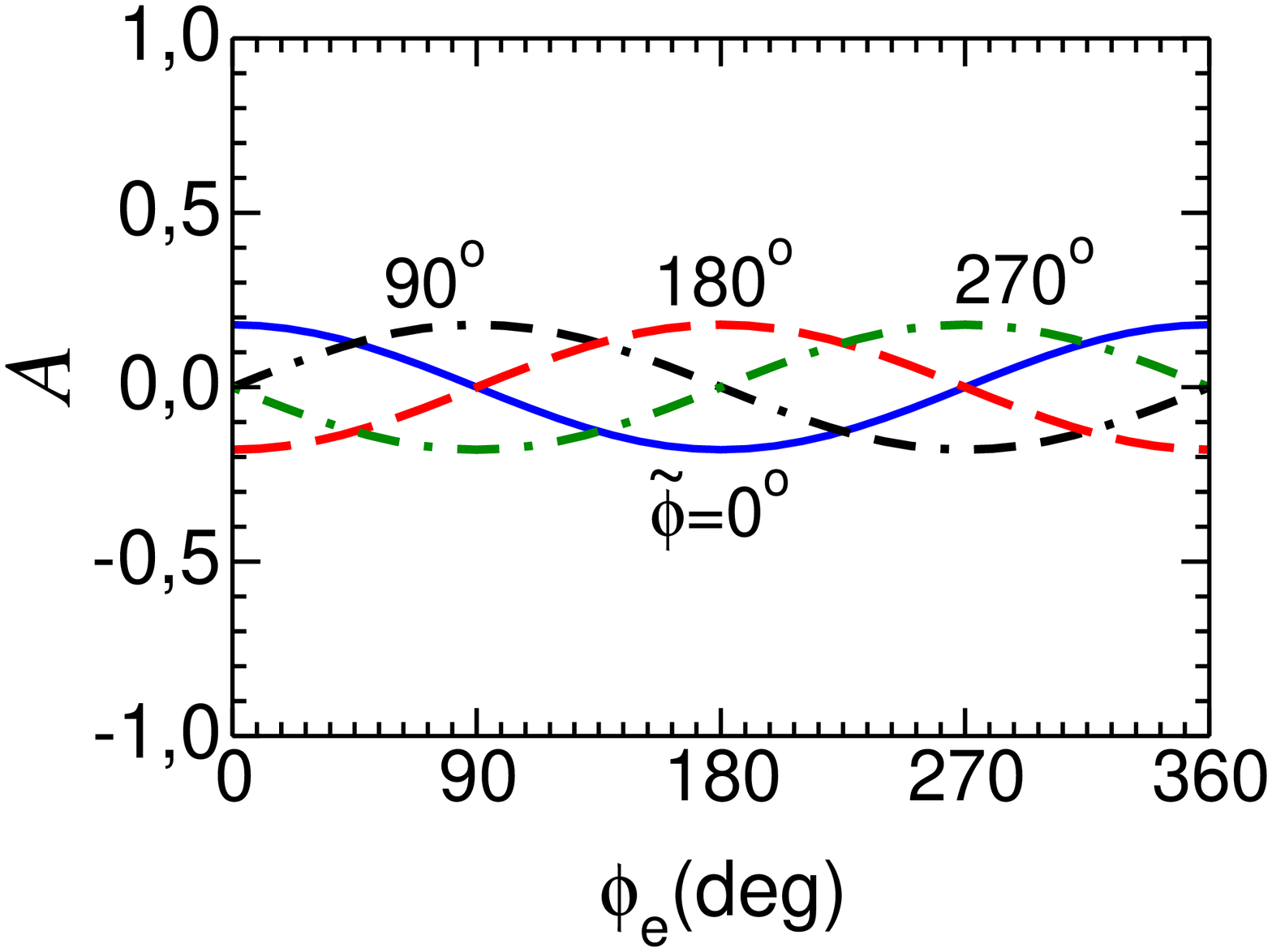}
\end{center}
\caption{\small{ The same as in Fig.~\ref{Fig:14A} but
for short pulse with $N=2$.
 \label{Fig:15A}}}
\end{figure}

The effect of the carrier phase decreases
when the duration of pulse increases.
Thus, when the number of oscillations in a pulse
is $N\geq2$, the inequality
of two terms in (\ref{UU81}) does not hold, instead they
have the same order of magnitude and the alignment of the
differential distributions with respect to
$\phi_e\simeq\tilde\phi$ becomes very weak. The corresponding results
are exhibited in Fig.~{\ref{Fig:15A}}. The probabilities (rates)
$2\pi dW/d\phi_e$ as a function of
$\phi_e$ for the short pulse with $N=2$ and $\zeta=4$ and $\xi^2=0.1$
for different $\tilde\phi$ are shown in the left panel.
One can see a very weak dependence of the rates on
$\phi_e$ and $\tilde\phi$. The rates are concentrated near the
 value $\sim 10^{-3}$~eV. Although, a small enhancement
 in the vicinity $\phi_e\simeq\tilde\phi$ still exists.
 This also is manifest in the
 anisotropy shown in the right panel. The anisotropy is finite,
 but its absolute value is less than 0.2.

\begin{figure}[ht]
\begin{center}
\includegraphics[width=0.35\columnwidth]{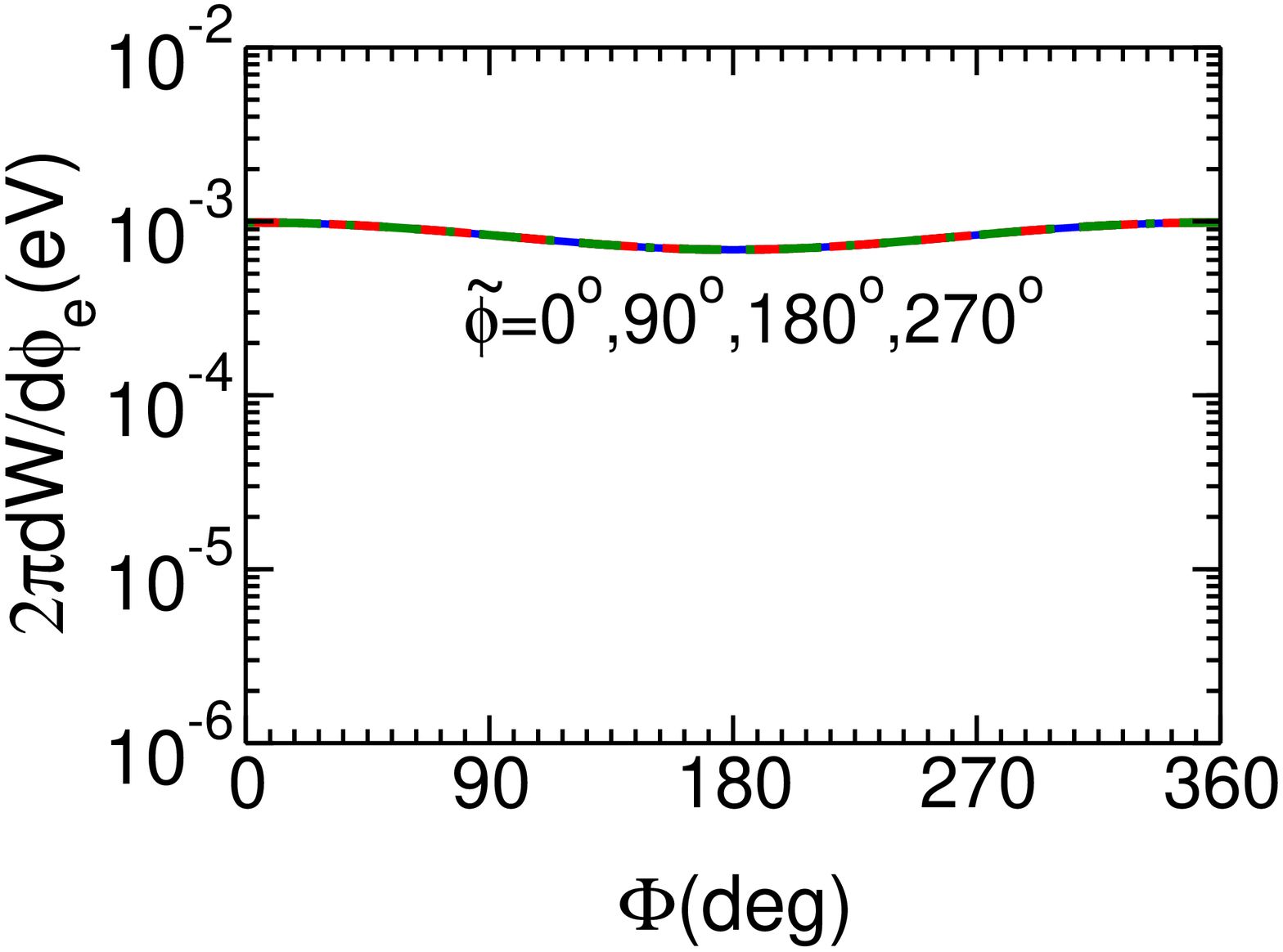}\qquad
\includegraphics[width=0.35\columnwidth]{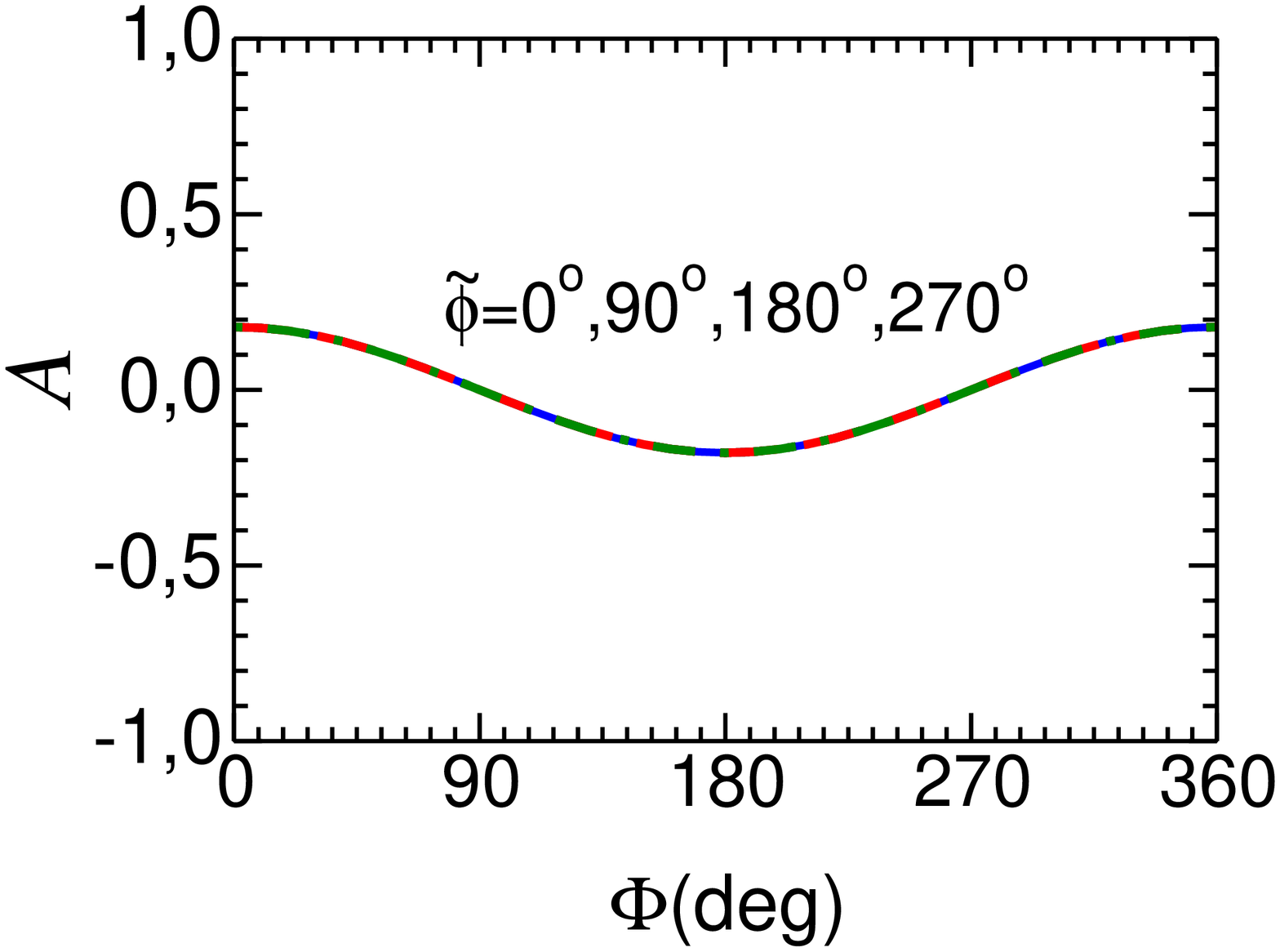}
\end{center}
\caption{\small{The same as in Fig.~\ref{Fig:15A} but as a function
 of the scale variable $\Phi=\phi_e - \tilde\phi$.
 \label{Fig:17A}}}
\end{figure}

 In order to stress the alignment
 of the differential azimuthal-angle distributions
 along $\phi_e\simeq\tilde\phi$ one can plot
 differential distributions and anisotropies
 as a function of the "scale"\ variable
 $\Phi=\phi_e - \tilde\phi$. In this case,
 all curves shown, for example in the left and right panels
 in Fig.~\ref{Fig:14A},
 are merged into a single carrier phase independent curve.
 \begin{figure}[ht]
 \begin{center}
\includegraphics[width=0.35\columnwidth]{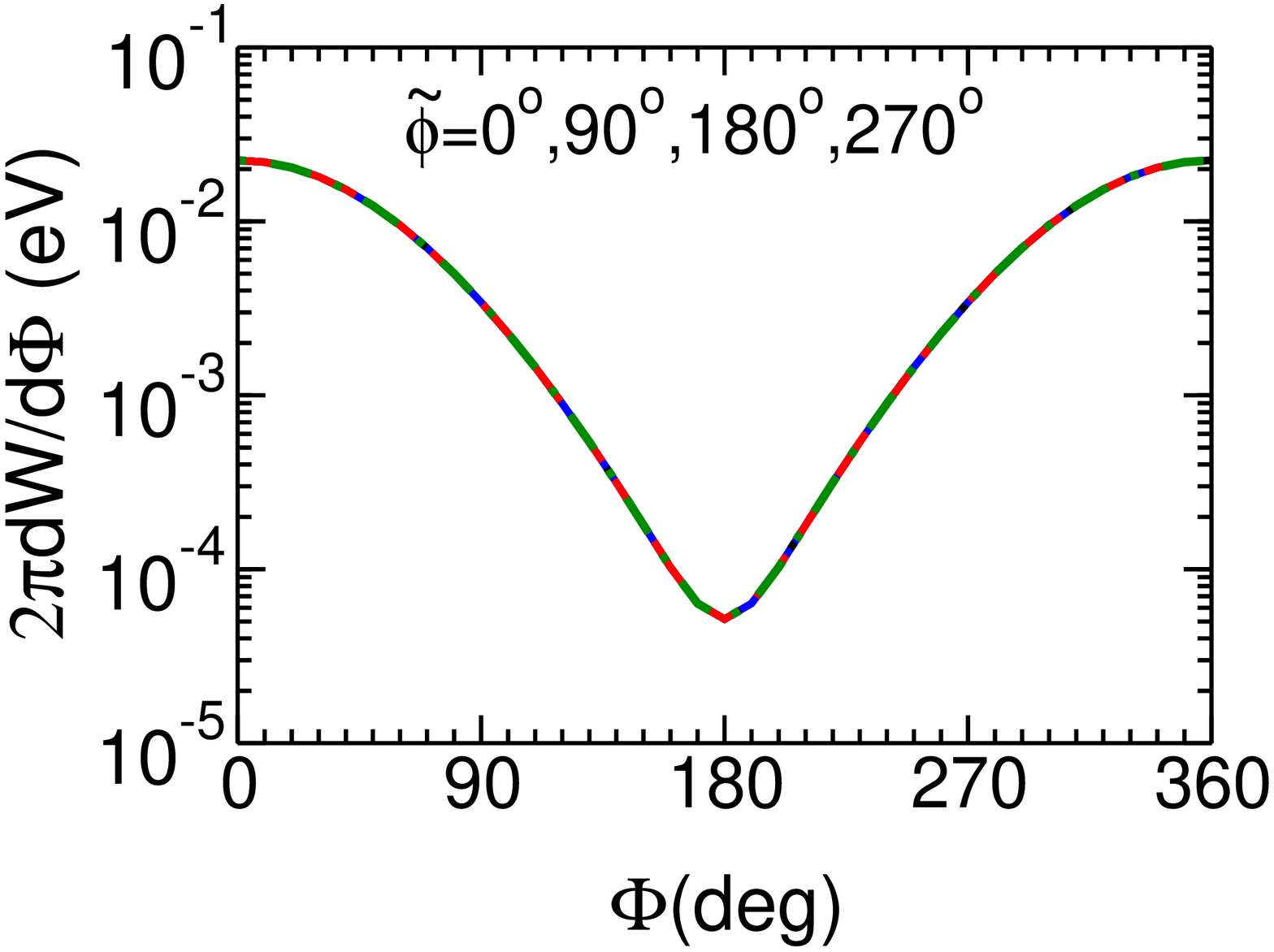}\qquad
\includegraphics[width=0.35\columnwidth]{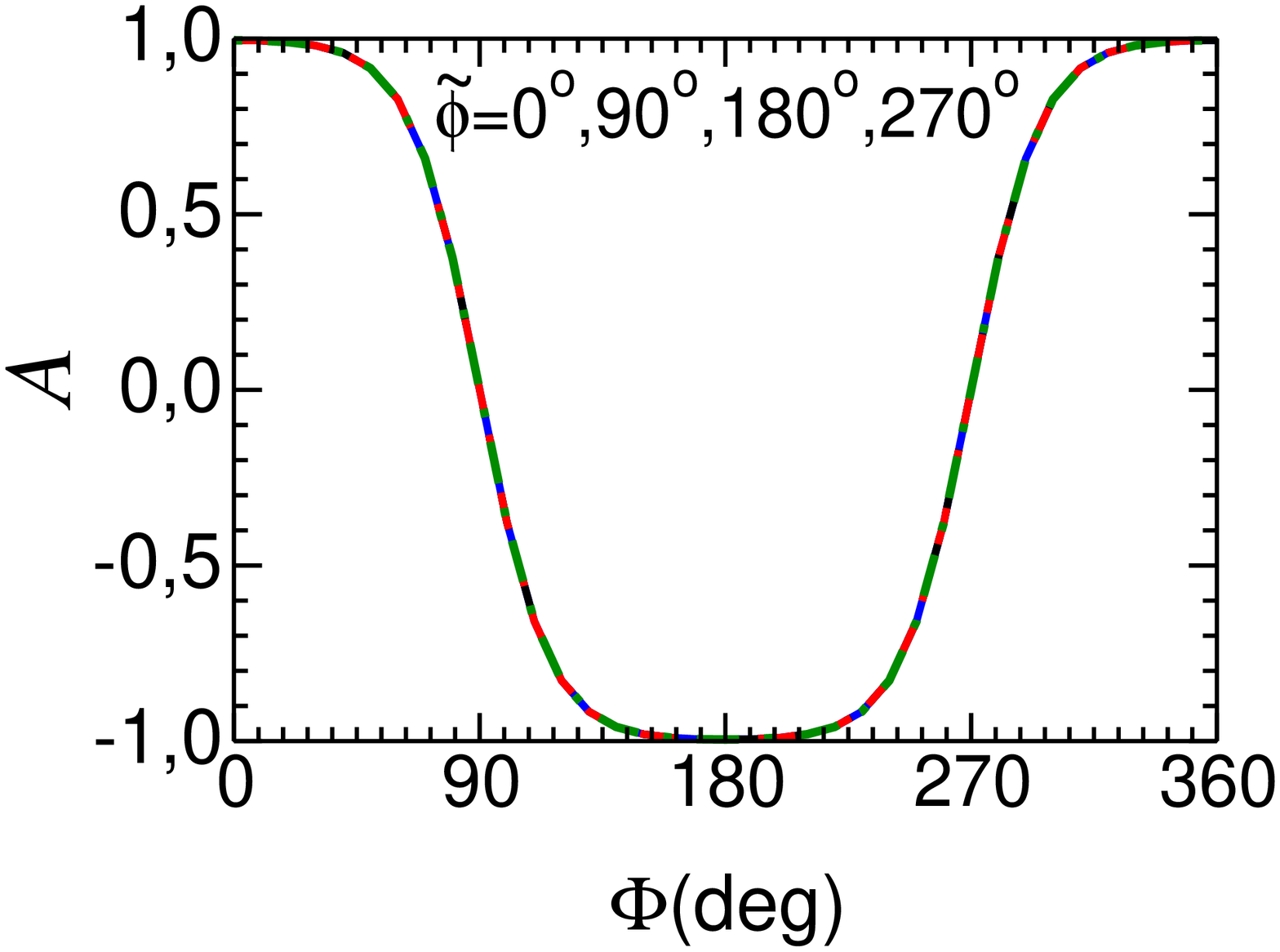}
\end{center}
\caption{\small{The same as in Fig.~\ref{Fig:14A} but as a function
 of the scale variable $\Phi=\phi_e - \tilde\phi$.
 \label{Fig:16A}}}
\end{figure}
 The corresponding  result is exhibited in Fig.~\ref{Fig:16A},
 where one can see a carrier phase independence of the
 differential distributions and anisotropies shown in the
 left and right panels, respectively.
 Similarly, a carrier phase independent result is obtained
 for the short pulse with $N=2$ shown in Fig.~\ref{Fig:17A}.

Formally, this follows from the fact that the
carrier phase is included in the expressions for
the basic functions (\ref{CP3}) and (\ref{CP4})
in the combination $\phi_e-\tilde\phi$. Therefore,
the differential distributions are a function
of $\Phi=\phi_e-\tilde\phi$ rather than of $\phi_e$
(for finite $\tilde\phi$).
From the physical point of view this means that
at finite $\tilde\phi$ the differential azimuthal distributions
is convenient to study in the coordinates $x',y'$ rotated
relative to the initial coordinates $x,y$
by an angle equal to the carrier phase $\tilde\phi$.

\section{Compton scattering in short laser pulse}

\subsection{General formalism}

The Compton scattering process is considered here as a spontaneous emission
of one photon off an electron in an external e.m.\ field. Similarly to the
Breit-Wheeler process, we employ
the four-potential of a circularly polarized laser field in
form of Eq.~(\ref{III1})
with  the envelope function $f(\phi)$ discussed in Sect.~II.
Here, we also use one-parameter hyperbolic secant (hs) envelope and
two-parameter symmetrized Fermi (sF) shape with $b/\Delta=0.15$.
All details and notations are given in Sect.~II.
Using the same arguments as before, we start our consideration
assuming $\tilde\phi=0$ and discuss the impact of the finite
carrier phase later.

Utilizing the e.m.\ potential (\ref{III1}) and the Volkov solution for
the electron wave function in this background
field one finds to the following
expression for the $S$ matrix element
\begin{eqnarray}
S =-i e \int\limits_{-\infty}^\infty dl \, M(l) \frac{(2\pi)^4
\delta^4(p+ lk-p'-k')}{\sqrt{2E \, 2E' \, 2\omega'}}~,
 \label{ES}
\end{eqnarray}
 where $k$, $k' = (\omega', \mathbf{k}')$, $p = (E, \mathbf{p})$ and $p' = (E', \mathbf{p}')$
 refer to the four-momenta of the
 background (laser) field (\ref{III1}), scattered photon,
 as well as asymptotic incoming (in-state)
 and outgoing (out-state) electrons.
 All quantities are considered in the laboratory system.
 Similarly to the Breit-Wheeler process the
 transition matrix $M(l)$ consists of four terms
 (cf.~Eq.~(\ref{III4})),
 \begin{eqnarray}
 M(l)=\sum\limits_{i=0}^3  M^{(i)}\,C^{(i)}(l)~,
 \label{EM}
 \end{eqnarray}
 where the transition operators have now the form
 $M^{(i)}=\bar u_{p'}\,\hat M^{(i)}\,u_p$ 
 with
\begin{eqnarray}
 \hat M^{(0)}&=&\fs\varepsilon'~,\quad
 \hat M^{(1)}=
  \frac{ e^2a^2 \,
 (\varepsilon'\cdot k)\,\fs k}
 {2(k\cdot p)(k\cdot p')}~,\nonumber\\
 \hat M^{(2,3)}&=&\frac{e\fs a_{(1,2)}\fs k\fs
 \varepsilon'}{2(k\cdot p')} + \frac{e\fs \varepsilon'\fs k\fs
 a_{(1,2)}}{2(k\cdot p)}~.
 \label{B2C}
\end{eqnarray}
 Here, $u_p$ and $\bar u_{p'}$ are free Dirac spinors depending on
 the momenta $p$ and $p'$;
 and $\varepsilon'$ denotes the polarization four vector
 of the scattered photon. Since the Compton scattering is crossing channel
 of the Breit-Wheeler processes, the identity
 $\hat M^{(i)}_{\rm Compton}(p,p',k,k')=
 \hat M^{(i)}_{\rm BW}(-p,p',k,-k')$ is realized.
Utilizing the prescription of Sect.~3.~1 one can
express the coefficients $C^{(i)}(l)$ through
basic functions $Y_l(z)$ and $X_l(z)$ (cf. Eqs.~(\ref{III24}) and (\ref{III25}))
with
\begin{eqnarray}
 {\cal P(\phi)} = z\int_{-\infty}^{\phi}\,d\phi'\,
 \cos(\phi'-\phi_0)f(\phi')
 -\xi^2\frac{u}{u_0} \int_{-\infty}^\phi\,d\phi'\,f^2(\phi')
 \label{PC}
 \end{eqnarray}
 and
\begin{eqnarray}
 z=2l\xi\sqrt{\frac{u}{u_l}\left(1-\frac{u}{u_l}\right)},\,\,
 u\equiv(k'\cdot k)/(k\cdot p'),\,\, u_l=l/\,u_0~,
 \label{zC}
\end{eqnarray}
 where $u_0=2{k\cdot p}/m^2$.
 Now, the phase $\phi_0$ is equal to the azimuthal angle of the
 direction of flight of the outgoing electron, $\phi_0 = \phi_{e'}$,
 and it is related to
 the azimuthal angle of the momentum of the outgoing photon as
 $\phi_{\gamma'}=\phi_0 + \pi$.

 This representation of functions $C^{(i)}(l)$ allows to define a
 partial differential cross section
 \begin{eqnarray}
 \frac{d\sigma(l)}{d\omega'\,d\phi_{e'}}
 =\frac{2\alpha^2}{N_0\,\xi^2\,(s-m^2)\,|p - l\omega|}\,w(l)
  \label{S1}
 \end{eqnarray}
with
\begin{eqnarray}
 w(l)&=&
 -2 \widetilde Y^2_l(z)+\xi^2(1 +\frac{u^2}{2(1+u)})
 \nonumber\\
 &\times& \left(Y^2_{l-1}(z)+ Y^2_{l+1}(z)
 -2\widetilde Y_l(z)X^*_l(z)\right)~.
 \label{S2}
\end{eqnarray}
 Equation (\ref{S2}) resembles the corresponding expression for the partial
 probability of photon emission in the case
 of IPA~\cite{LL4} with the substitutions $l\to n = 1, 2, \cdots$ and
 ${\widetilde Y^2_l(z)},\,Y_l^2(z),\,\widetilde Y_l(z)X^*_l(z)\to J_n^2(z')$,
 namely
 \begin{eqnarray}
 w_n&=&
 -2 J^2_n(z')+\xi^2(1 +\frac{u^2}{2(1+u)})
 \nonumber\\
 &\times& \left(J^2_{n-1}(z')+ J^2_{n+1}(z')
 -2 J_n^2(z')\right)~,
 \label{S2INF}
\end{eqnarray}
 where $J_n(z')$ denotes Bessel functions with
  $z'=\frac{2n\xi}{\sqrt{1+\xi^2}}\sqrt{\frac{u}{u_n}\left(1-\frac{u}{u_n}\right)}$
  and  $u_n=\frac{2n(k\cdot p)}{m^2(1+\xi^2)}$.
 Similarly to IPA, the phase $\phi_0$ can be determined
 through invariants $\alpha_{1,2}$ as $\cos\phi_0=\alpha_1/z$,
 $\sin\phi_0=\alpha_2/z$ with $\alpha_{1,2}=e\left(a_{1,2}\cdot
 p/k\cdot p-a_{1,2}\cdot p'/k\cdot p'\right)$.
 The dimensionless field intensity $\xi^2$ is described by
 Eqs.~(\ref{S22})~-~(\ref{S25}).

 The frequency $\omega'$ of the emitted photon is related to the
 auxiliary variable $l$  and the polar angle $\theta'$
 of the direction of the momentum $\mathbf{k}'$ via
\begin{eqnarray}
 \omega'=\frac{l\,\omega (E+|\mathbf {p}|)}{E + |\mathbf {p}| \cos\theta'
 +l \omega(1-\cos\theta') }
 \label{S3}
\end{eqnarray}
 and increases with $l$ at fixed $\theta'$ since $\omega'$ is
 a function of $l$ at fixed $\theta'$. For convenience,
 we also present a similar expression for IPA,
 where the fermions are dressed and the integer
 quantity $n$, together with the field intensity $\xi^2$, appear:
 \begin{eqnarray}
 \omega'=\frac{n\,\omega (E+|\mathbf {p}|)}{E + |\mathbf {p}| \cos\theta'
 +\omega(n + \frac{m^2\xi^2}{2(k\cdot p)} )(1-\cos\theta')}~.
 \label{S3_}
\end{eqnarray}

The differential cross section of the one-photon production is
eventually
\begin{eqnarray}
 \frac{d\sigma}{d\omega'}=\int\limits_{\eta}
 dl\,\int\limits_{0}^{2\pi}d\phi_{e'}
 \frac{d\sigma (l)}{d\omega' d \phi_{e'}}
 \delta \left(l-l(\omega') \right)~.
 \label{S33}
\end{eqnarray}
 The lower integration limit $\eta >0$
 is defined by kinematics, i.e.\ by the minimum value of
 the considered $\omega'$, in accordance with Eq.~(\ref{S3}). In
 the IPA case, the variable $n = 1, 2,\, \cdots$ refers to
 the contribution of the individual harmonics
 ($n = 1$ with $\xi^2\ll 1$ recovers the Klein-Nishina cross
 section, cf.~\cite{Ritus-79}). The value $n\omega$ is
related to the energy
 of the background field involved in Compton scattering.
 Obviously, this value is a multiple of $\omega$.
 In FPA, the internal quantity $l$ is a continuous
 variable, implying a continuous distribution of the differential
 cross section over the $\omega' - \theta'$ plane.
 The quantity $l\omega$ can be considered as energy
 of the laser beam involved in the Compton process,
 which is not a multiple $\omega$.
 Mindful of this fact, without loss of generality, we
 denote the processes with $l>1$ as a multi-photon
 generalized Compton scattering, remembering
 that $l$ is a continuous quantity.

 The multi-photon effects become most clearly evident in the
 partially energy-integrated cross section
 \begin{eqnarray}
 {\tilde\sigma_{}(\omega')} = \int\limits_{\omega'}^{\infty}
 d\bar\omega' \frac{d\sigma (\bar\omega')}{d\bar\omega'}
 =\int\limits_{l'}^{\infty} dl
 \frac{d\sigma(l)}{dl}~,
 \label{S6}
\end{eqnarray}
 where
 $d\sigma(l) / dl =
 ( d\sigma(\omega') / d\omega')
 (d\omega'(l) / dl)$,
 and the minimum value of $l'$ is
 \begin{eqnarray}
 l'=\frac{\omega'}{\omega}\,
 \frac{E+|\mathbf{p}|\cos\theta'}{E+|\mathbf{p}|-\omega'(1-\cos\theta')}~.
 \label{S66}
 \end{eqnarray}
 The cross section (\ref{S6}) has the meaning of a cumulative distribution.
 In this case, the subthreshold, multi-photon events correspond to
 frequencies $\omega'$ of the outgoing photon which
 exceed the corresponding
 threshold value $\omega_1'=\omega'(l=1)$ (cf. Eq.~(\ref{S3})),
 and ratio $\kappa=\omega'/\omega'_1>1$ represents the sub-threshold
 parameter.


\subsection{The differential cross section}

In IPA \cite{Ritus-79,LL4}, the cross section of the multi-photon
Compton scattering increases with $\theta'$ towards $180^o$.
For instance, it peaks at about $170^o$ for the chosen
electron energy of 4 MeV (all quantities are considered in the
laboratory frame) and rapidly drops to zero when $\theta'$
approaches $180^o$ for the harmonics $n > 1$ yielding thus the
blind spot for back-scattering. Therefore, in our subsequent
analysis we choose the near-backward photon production at $\theta'
= 170^o$ and an optical laser with $\omega=1.55$~eV. Defining
one-photon events by $n = 1$, this kinematics leads via
Eq.~(\ref{S3_}) to $\omega_1'\equiv \omega'(n=1,\xi^2\ll1,\,
\theta' = 170^o) \simeq 0.133$~keV which we refer to as a
threshold value. Accordingly, $\omega' > \omega'_1$ is enabled by
non-linear effects, which in turn may be related loosely to
multi-photon dynamics with $n>1$ in IPA or $l>1$ in FPA where,
 we remind again, the internal
 variable $l$ can not be interpreted strictly as number of laser
 photons involved (cf.~\cite{DSeipt-2014}).
 Note that all calculations for IPA
 are performed in a standard way~\cite{Ritus-79,LL4}.
 The energy of the outgoing photon in IPA is calculated using
 Eq.~(\ref{S3_}), where dressing of electrons in the
 background field is taken into account.

\begin{figure}[ht]
\begin{center}
\includegraphics[width=0.35\columnwidth]{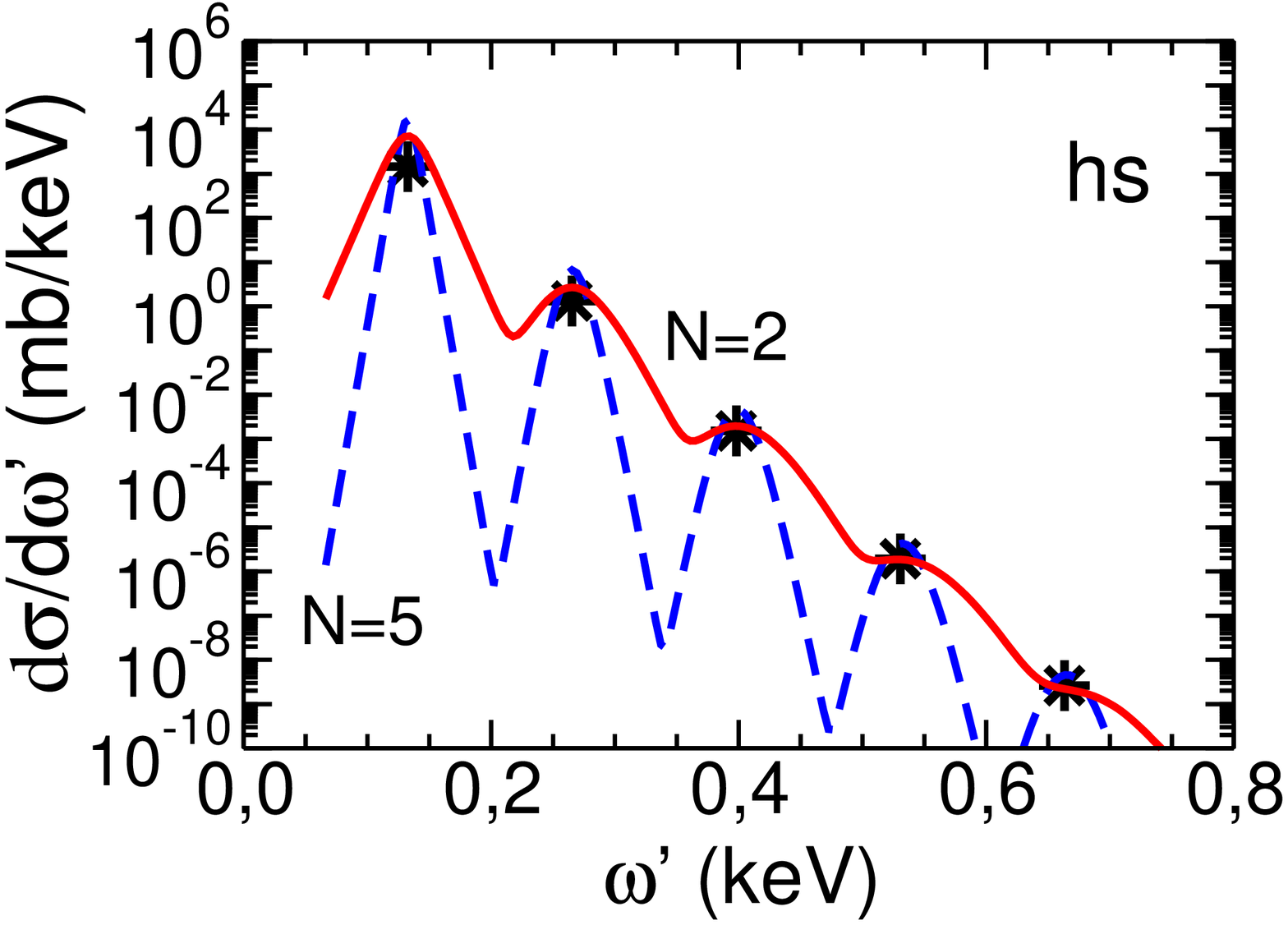}\qquad
\includegraphics[width=0.35\columnwidth]{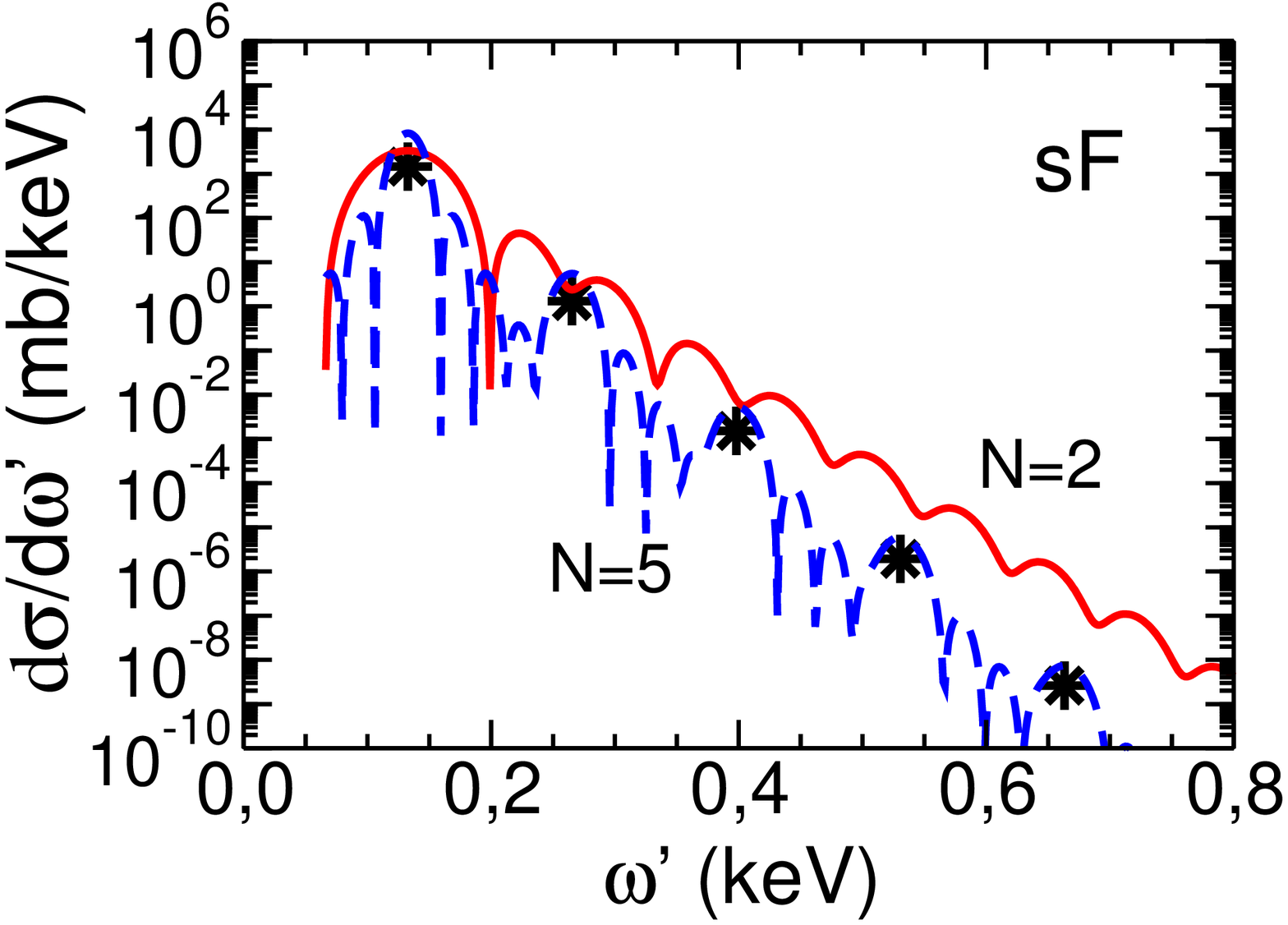}
\end{center}
 \caption{\small{Differential cross section $d \sigma / d\omega' \,
 \vert_{\theta' = 170^o}$ of the Compton scattering for
 $\xi^2=10^{-3}$.
 The solid and dashed curves are for $N = 2$ and  5, respectively.
 The stars depict the IPA results for lowest
 harmonics. Left and right panels
 correspond to hyperbolic secant (hs) and symmetrized Fermi (sF)
 shapes of the envelopes, respectively.
 \label{gFig:4}}}
\end{figure}

 Let us consider first an example of short pulses
 with moderate intensity, $\xi^2 = 10^{-3}$,
 similar to a recent experiment
 of Compton backscattering~\cite{Jochmann}.
 Results for the hs and sF
 shapes are exhibited in Fig.~\ref{gFig:4}.
 The solid and the dashed curves
 correspond to pulses with
 $N=2$ and 5, respectively.
 The stars depict the IPA results, i.e., the harmonics at fixed
 scattering angle $\theta'$.
 Their positions correspond to integer values of
 $n=1, \,2, \cdots$ in accordance with Eq.~(\ref{S3_}).
 i.e.\ the distribution of scattered photon energies is
 a discrete function of $\omega'$.
 We stress that the cross section at $\omega'>\omega_1'$
 is essentially "sub-threshold"\ , i.e.\ outside the kinematically
 allowed region of the Klein-Nishina process due to multi-photon effects.

 In the FPA case, the energy distribution
 becomes a continues function of $\omega'$. The actual shape
 is determined by both the pulse duration and the envelope form.
 Consider first the case of the hs shape
 (cf.~Fig.~\ref{gFig:4}, left panel).
 The cross section displays sharp bumps with peak positions
 corresponding to integer values of $l=n$ (as in IPA).
 In the vicinity of the bumps, at $l=n \pm \epsilon$,
 $\epsilon \ll 1$,
 the cross section is rapidly decreasing.
 Such a behavior reflects the properties of the functions
 $Y_{l=n+\epsilon}(z)$ (cf. Eq.~(\ref{B6}))
 which is proportional to the
 Fourier transform of the $(n+1)$-th degree of the
 envelope function $F^{(n+1)}(\epsilon)$.
 At $\xi^2 \ll 1$, the contribution of terms $\propto
 X_l$ is negligible.

 The behavior of the cross section in the
 vicinity of the first bump is proportional to $F^2_{\rm hs}(\epsilon)$
 with $F_{\rm hs}(\epsilon)$ given in Eq.~(\ref{U5}),
 or
 $ F_{\rm hs}(x)\simeq \Delta \exp[-\pi\Delta x/2]$.
 Thus,  the cross section becomes steeper
 with increasing pulse duration $\Delta$. This result
 qualitatively agrees with that of Ref.~\cite{Krajewska-2012}.

 In the case of the
 sF shape, the dependence $F_{\rm sF}(\epsilon)$ is more complicated
 (cf. Eq.~(\ref{U5})).
 Together with the overall decrease of the cross section
 proportional to $\exp[-2\pi b\,l(\omega')]$ it also
 indicates fast oscillations
 with a frequency $\propto \Delta$. Such oscillations show up
 in the cross section as some secondary bumpy structures. These properties
 are manifest in Fig.~\ref{gFig:4}~(right panel):
 the overall decrease of the cross section decreases with
 decreasing pulse duration, and the number of the secondary
 bumps in the region of $\omega'$, corresponding to the nearest
 integer values of $l$,  increases with pulse duration.

 \begin{figure}[ht]
\begin{center}
\includegraphics[width=0.35\columnwidth]{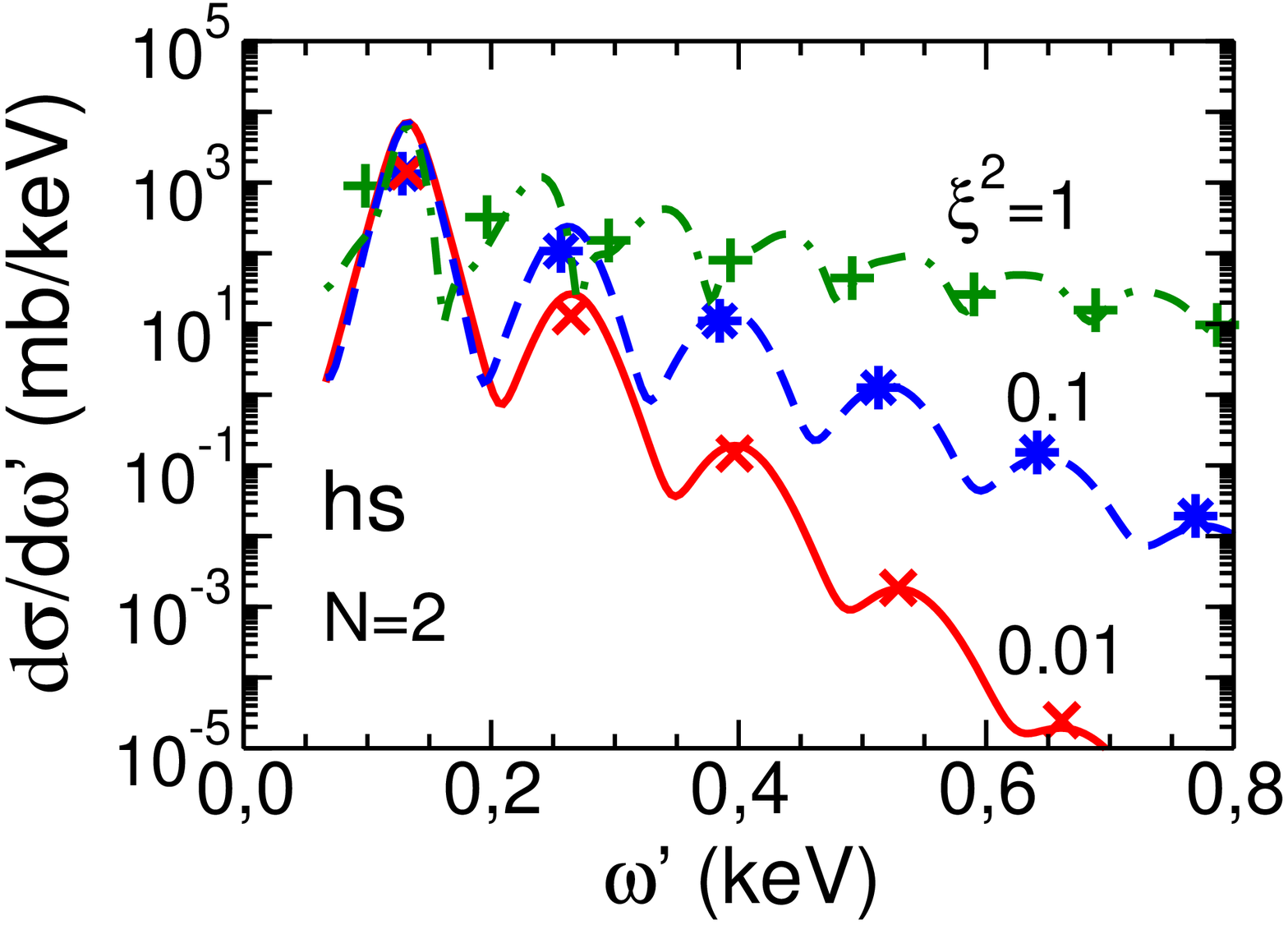}\qquad
\includegraphics[width=0.35\columnwidth]{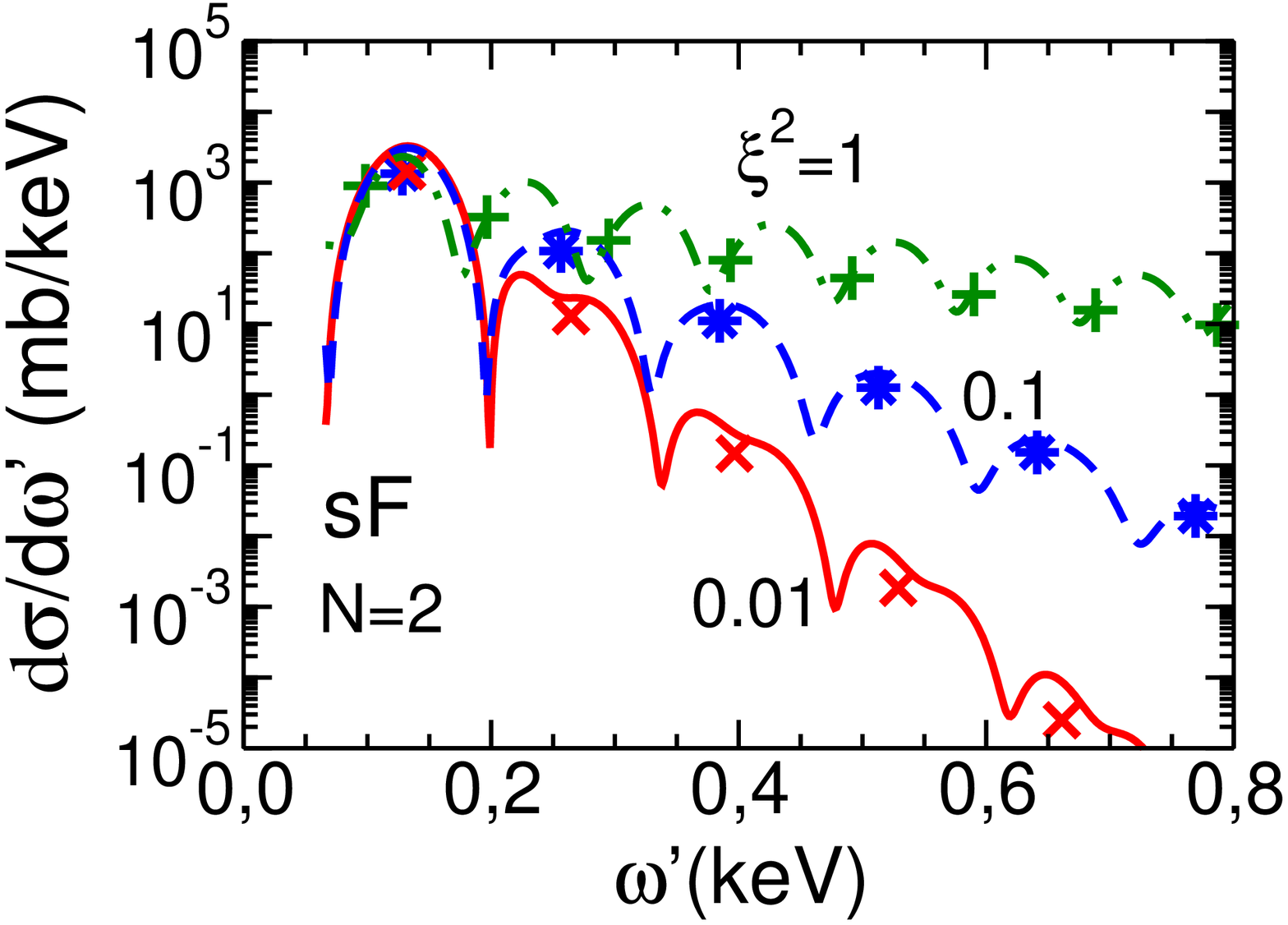}
\end{center}
 \caption{\small{Differential cross section $d \sigma / d\omega' \,
 \vert_{\theta' = 170^o}$ of the Compton scattering for
 $\xi^2=0.01,\,0.1$ and 1, shown
 by solid, dashed, and dot-dashed curves,
 respectively, for $N=2$.
 The symbols "x"\ , stars and pluses  depict the IPA results
 for the lowest harmonics for $\xi^2= 0.01,\,0.1,$ and 1, respectively.
 Left and right panels
 correspond to hyperbolic secant (hs) and symmetrized Fermi (sF)
 shapes of the envelopes.
 \label{gFig:5}}}
\end{figure}
 In Fig.~\ref{gFig:5} we present the
 differential cross sections
 for different field intensities $\xi^2=0.01,\,0.1$ and 1,
 depicted by solid, dashed,
 and dot-dashed curves, respectively.
 The duration of the pulse corresponds to $N=2$. The bump positions
 for FPA in Fig.~\ref{gFig:5} are shifted relative to the discrete
 positions of contributions from the individual harmonics in IPA,
 shown by corresponding symbols.
 These shifts are a consequence of the electron dressing in IPA
 which depends on $\xi^2$.

\begin{figure}[ht]
\begin{center}
\includegraphics[width=0.35\columnwidth]{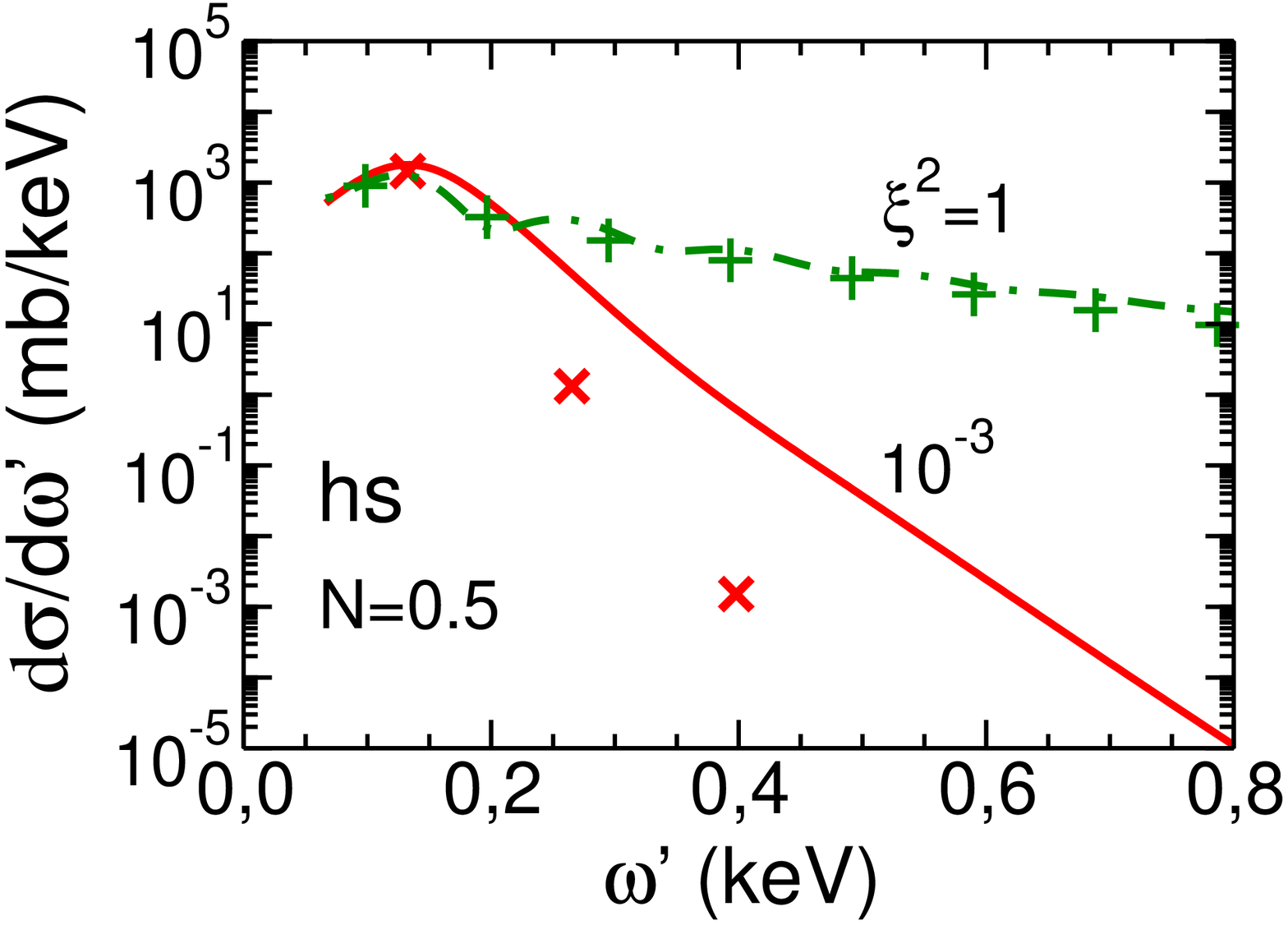}\qquad
\includegraphics[width=0.35\columnwidth]{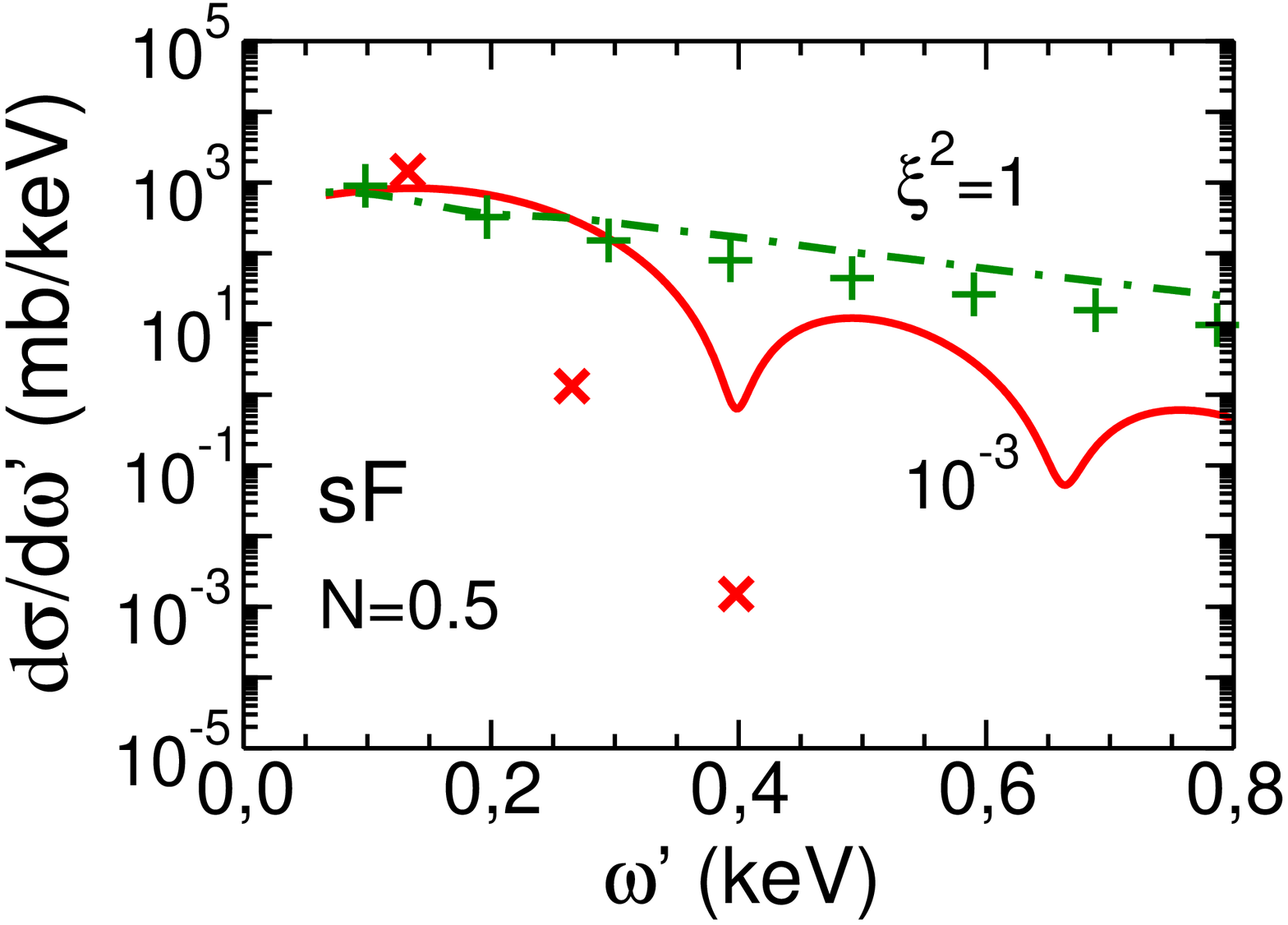}
\end{center}
 \caption{\small{Differential cross section
 $d \sigma / d\omega' \,
 \vert_{\theta' = 170^o}$ of Compton scattering for
 $\xi^2=10^{-3}$ and 1 shown by solid
 and  dot-dashed curves, respectively, for $N=0.5$.
 Crosses and pluses depict the discrete IPA results
 for lowest harmonics
 for $\xi^2=10^{-3}$ and 1, respectively.
 Left and right panels
 correspond to hyperbolic secant (hs) and symmetrized Fermi (sF)
 envelope shapes.
 \label{gFig:6}}}
\end{figure}

 For completeness, in Fig.~\ref{gFig:6} we exhibit
 the differential cross sections for a sub-cycle pulse with $N=0.5$
 for $\xi^2=10^{-3}$ and 1, shown by solid
 and  dot-dashed curves, respectively,
 for the hs (left) and sF (right) envelope shapes.
 Crosses and pluses depict the IPA results
 for $\xi^2=10^{-3}$, and 1.
For the hs shape, the cross sections decrease almost monotonically,
with a large enhancement of the FPA result compared to IPA for
small field intensities ($\xi^2\ll 1$). In the case of the flat-top
envelope the cross section exhibits some oscillations which
point to more complicated spectral properties of the flat-top
envelope shape.

 To summarize this part we can conclude that
 the results for fully differential cross sections
 for IPA and FPA are quite different. In IPA, the cross section
 represents the discrete spectrum where the frequencies of the
 outgoing photons $\omega'$ are fixed according to Eq.~(\ref{S3_}).
 In FPA, the differential cross sections are continuous
 functions of $\omega'$. Some similarities of IPA and FPA
 can be seen in the case of small field intensities $\xi^2\ll1$ and
the smooth one-parameter envelope shape with $N=2\dots10$. Here,
 the differential cross sections
 have a bump structure, where the position of bumps
 and bump heights are close to that of IPA.
 The situation changes drastically for more complicated
 (and probably more realistic) flat-top envelope shapes. In this case
 one can see a lot of additional bumps which reflect the more
 complicated spectral properties of the flat-top  shape;
it is difficult to find a relation not only between IPA and
 FPA, but also within FPA for different pulse durations.
 Experimentally, studying multi-photon effects using rapidly
 oscillating fully differential cross sections seems
 to be rather complicated. An
 analysis of integral observables helps to overcome this problem.
 In particular, the partially integrated cross sections have a
 distinct advantage: they are smooth functions of $\omega'$ and
 allow to study directly the multi-photon dynamics.

\subsection{Partially integrated cross sections}

 The non-linear dynamics becomes most transparent  in the
 partially energy-integrated cross section defined in Eq.~(\ref{S6}).
 In this case, the sub-threshold multi-photon events are filtered
 when the lower limit of integration $\omega'$ exceeds the
 threshold value $\omega_1'=\omega'(n=1,\xi^2)$ (with $\xi^2\ll1$
 for the pure Klein-Nishina process). Thus, events with
 $\omega'(l)\gg\omega'_1$ and $l\gg1$ correspond essentially to
 multi-photon process, where the energy $l\omega\gg\omega$ is
 absorbed from the pulse.
 Experimentally, this can be realized by an absorptive medium
 which is transparent for frequencies above a certain threshold
 $\omega'$. Otherwise, such a partially integrated spectrum can be
 synthesized from a completely measured spectrum. Admittedly, the
 considered range of energies with a spectral distribution
 uncovering many decades is experimentally challenging.

 \begin{figure}[ht]
\begin{center}
\includegraphics[width=0.35\columnwidth]{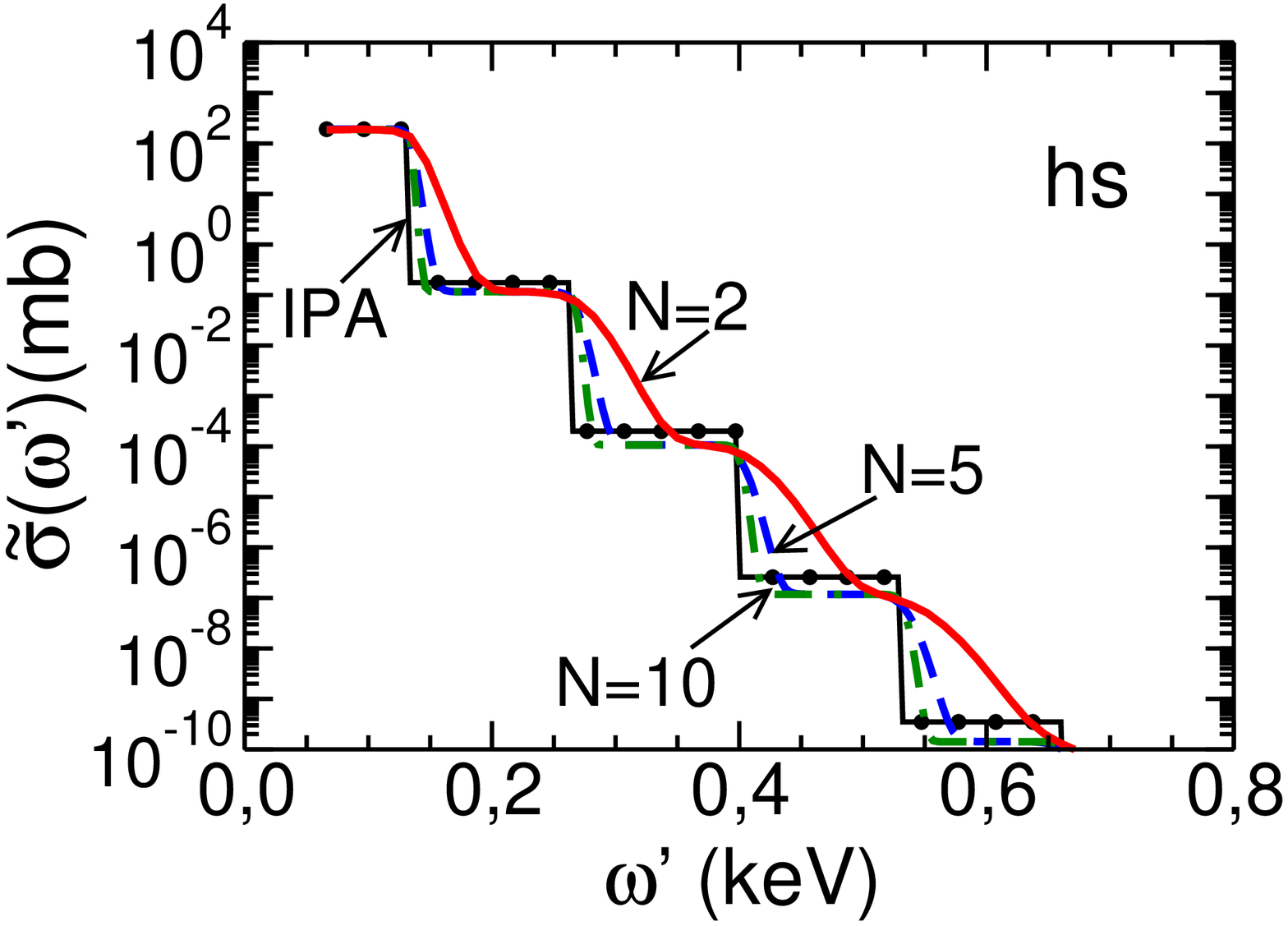}\qquad
\includegraphics[width=0.35\columnwidth]{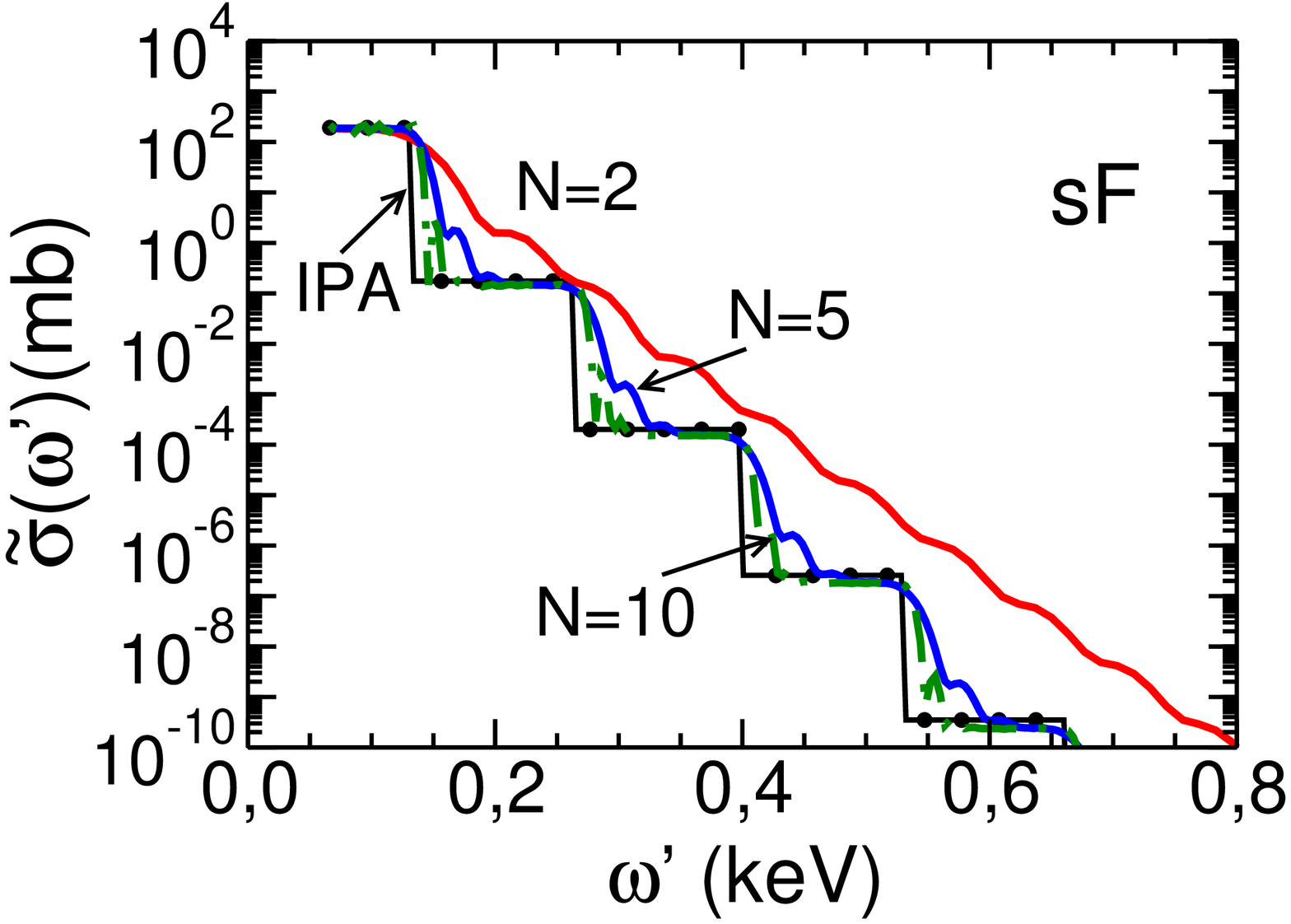}
\end{center}
 \caption{\small{The partially
 integrated cross section (\ref{S6})
 for $\xi^2=10^{-3}$.
 The thin solid curve marked by dots depicts the IPA result.
 The solid, dashed, and dot-dashed curves
 correspond to $N =2$, 5 and 10, respectively. Left and right
 panels are for hyperbolic secant (hs) and symmetrized Fermi (sF) envelopes.
\label{gFig:7} }}
\end{figure}
  The partially integrated cross sections of Eq.~(\ref{S6})
  are presented in Fig.~\ref{gFig:7}.
  The thin solid curve (marked by dots) depicts IPA results given by
\begin{eqnarray}
\tilde\sigma^{IPA} (\omega')
 =\int\limits_{l'(\omega')}^{\infty}dl\sum\limits_{n=1}^{\infty}
 \frac{d\sigma^{IPA}_n}{d\omega'_n}
 \frac{d\omega'_n}{d n}\theta(n-l)~,
 \label{S6_IPA}
\end{eqnarray}
 where $\omega'(n)$ is defined by Eq.~(\ref{S3_}).
 That is, the partially integrated cross section becomes a step-like
 function, where  each new step corresponds to the contribution of
 a new (higher) harmonic $n$,
 which can be interpreted  as $n$-laser photon process. Results for
 the finite pulse
 exhibited by solid, dashed,
 and dot-dashed curves
 correspond to $N=2, \,5$ and
 10, respectively. In the above-threshold region with
 $\omega'\leq \omega'_1$, the cross sections do not depend
 on the widths and shapes of the envelopes, and the results of IPA and FPA
 coincide. The situation changes significantly
 in the deep sub-threshold region, where $\omega'>\omega_1'$ $(l\gg1),\, n\gg1$.
 For  short pulses with $N\simeq 2$,
 the FPA results exceed that of IPA considerably,
 and the excess may reach
 several orders of magnitude, especially for the flat-top envelope
 shown by the solid curve in Fig.~\ref{gFig:7}~(right panel).
 However, when the number of oscillation in a pulse
 increases ($N\gtrsim 10$)
 there is  a qualitative convergence
 of FPA and IPA results, independently of the pulse shape.
 Thus, at $N=10$ and $\omega'=0.6$~keV the difference between predictions
 for hs and sF shapes is a factor of two, as compared with
 the difference of the few orders of
 magnitude at $N=2$ for the same value of $\omega'$.

\begin{figure}[ht]
\begin{center}
 \includegraphics[width=0.35\columnwidth]{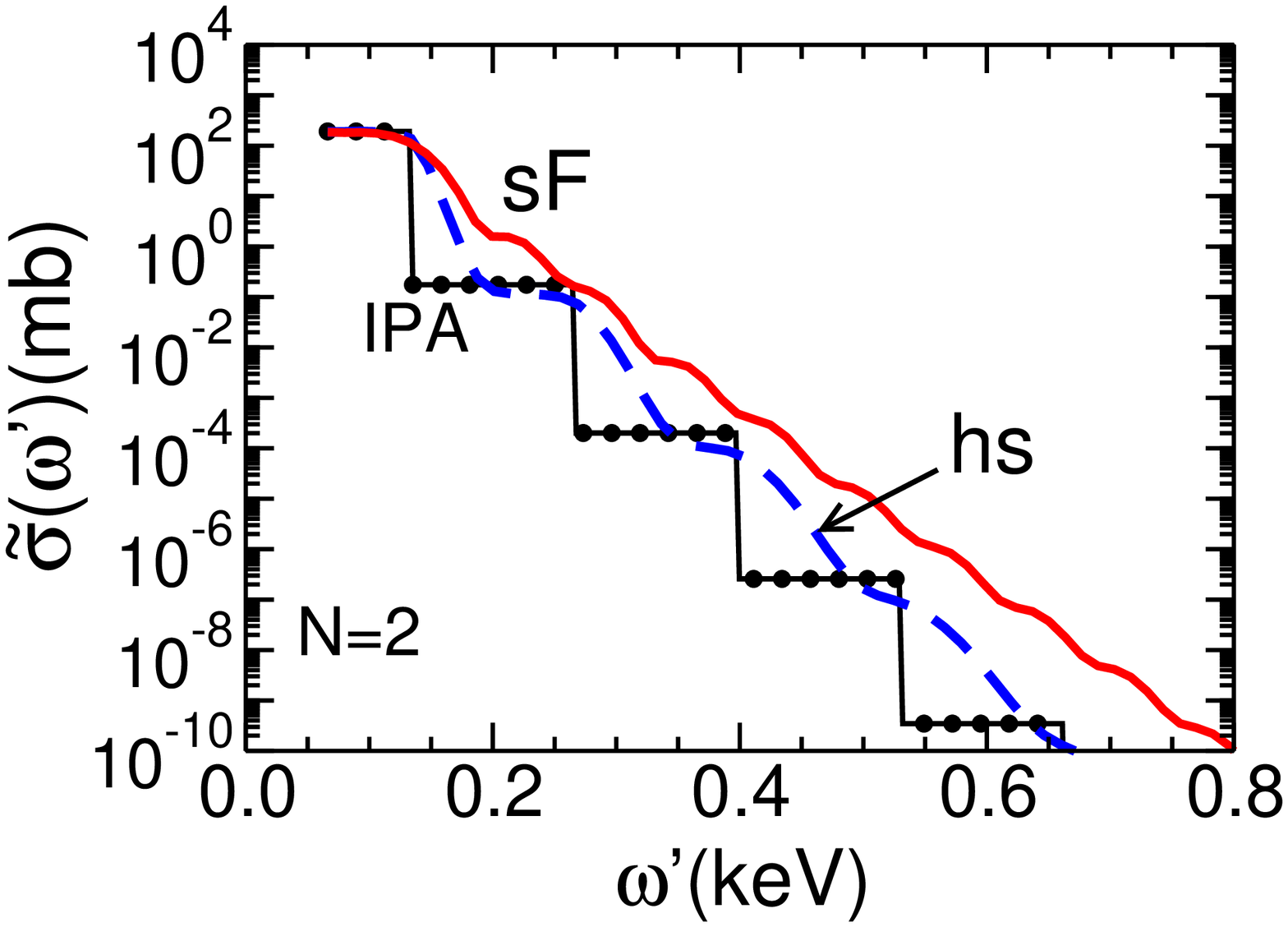}\qquad
 \includegraphics[width=0.35\columnwidth]{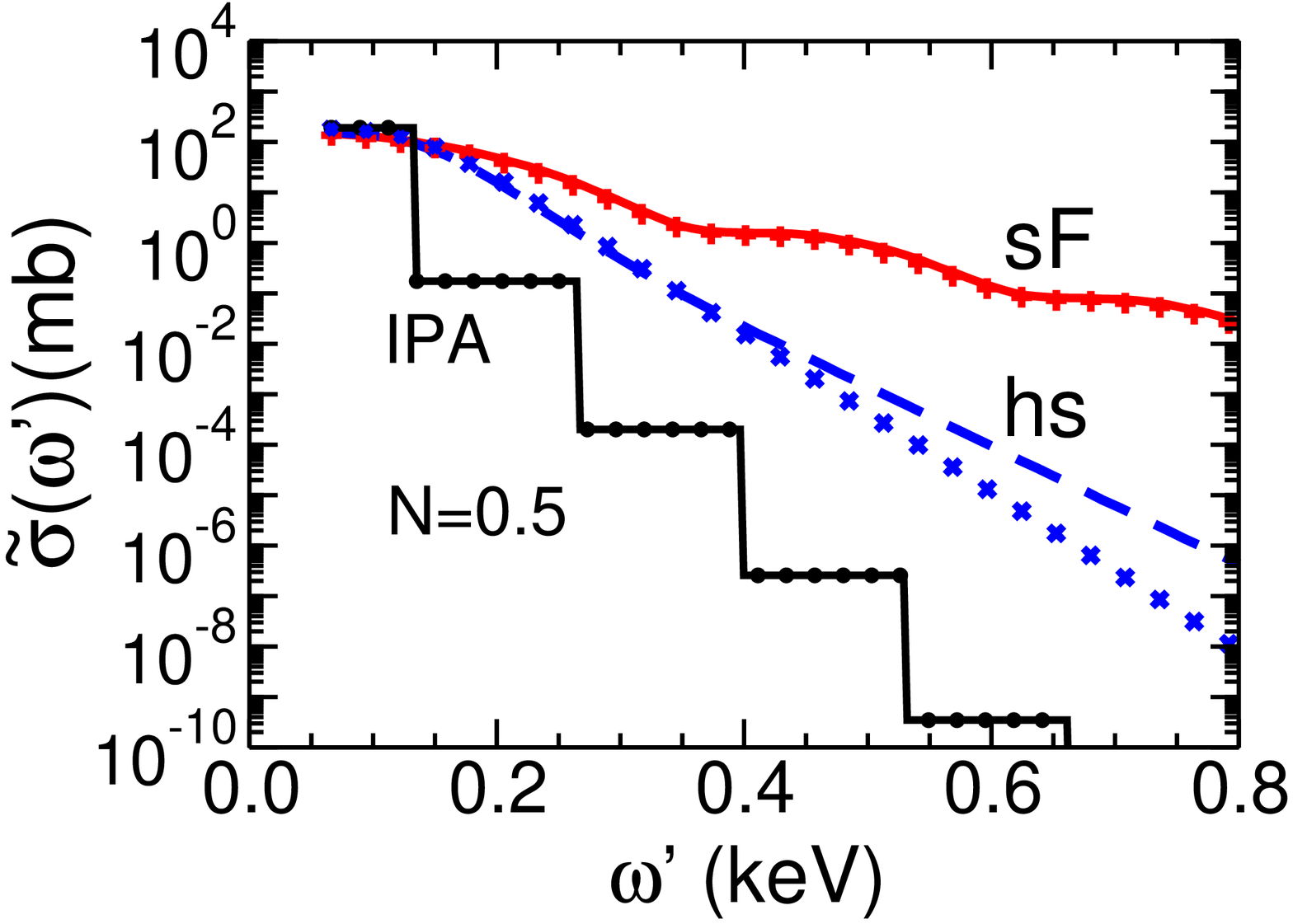}
 \end{center}
 \caption{\small{
 The partially integrated cross section (\ref{S6}) for $\xi^2=10^{-3}$.
 Left  panel: $N=2$, for the hyperbolic secant (hs, dashed curve)
 and symmetrized Fermi (sF, solid curve) shapes.
 Right panel: The same as in left panel, but for a sub-cycle pulse
 with $N=0.5$.
 The crosses and pluses correspond to the asymptotic solutions
 for hs and sF shapes, respectively, described in
 the text.
 \label{gFig:8} }}
\end{figure}

 To highlight the difference of the hs and
 sF (flat-top) shapes for short pulse we exhibit in Fig.~\ref{gFig:8}
 (left panel) results for $N=2$.
 At $\omega' \gtrsim 0.6$~keV,
 the difference between them is more than two orders of magnitude.

 Consider now the case of sub-cycle pulses
 with $N<1$. Our result for $N=0.5$ 
 is exhibited in Fig.~\ref{gFig:8}~(left panel).
 One can see a large enhancement
 of the cross section with respect to the IPA case
 for the sub-cycle pulse in the sub-threshold region.
 The enhancement for the sF shape is much greater
 pointing to a sensible dependence on the actual pulse shape.
 For a qualitative estimate of such a behavior we can drop
 the $\phi_{e'}$ dependence by taking $\phi_{e'}=0$.
 This choice is quite reasonable for the flat-top
 sF envelope shape and may serve as an upper limit
 for the cross sections in the case of the smooth
 hs envelope shape.
 Under the considered
 conditions the basic function $Y_l$ in Eq.~(\ref{III24})
 can be approximated as
 \begin{eqnarray}
  Y_l&\simeq& \frac{1}{2\pi}{\rm e}^{-il\phi_0}\int dq\,F(q)\int d\phi
  {\rm e}^{i(l-q)\phi -i{\cal P}(\phi)}\nonumber\\
  &\simeq&\frac{1}{2\pi}\int dq\,F(q)\int d\phi
  {\rm e}^{i(l-q-l\beta\xi)\phi -i\delta}
  ={\rm e}^{-i\delta} F(\tilde l)~,
 \label{S_FF}
 \end{eqnarray}
 where $F(l)$ is the Fourier transform of the envelope
 function,
 $\tilde l=l(1-\beta\xi)$ with
 $\beta=2\sqrt{\frac{u}{u_l}(1-\frac{u}{u_l})}<1$
 and $\delta=z\int_{-\infty}^0 d\phi \cos\phi\,f(\phi)-l\phi_0$.
 As a result, the cross section is almost completely defined
 by the square of the Fourier transforms
 (cf.\  Eqs.~(\ref{U5})), i.e.
 $\tilde\sigma(\omega')\simeq
 g(l(\omega'))\,F^2(\tilde l(\omega')-1 )$,
 where $g(\omega')$ is a smooth function of $l=l(\omega')$
 (cf. Eq.~(\ref{ASY4})).
 The Fourier transform for the sF shape decreases slower with
 increasing $l$.
 Such a dependence is evident in Fig.~\ref{gFig:7} (right panel).
 For an illustration,
 the crosses depict the result of a calculation where
 the basic functions $Y_l$ and $X_l$
 in the partial probability $\omega' (l)$ in Eq.~(\ref{S2})
 are replaced by their asymptotic values $F^{(1)}(\tilde l-1)$
 and $F^{(2)}(\tilde l-1)$, respectively. A more detail discussion of
 the asymptotic result is presented below (cf.\ Eq.~(\ref{ASY4})).

\begin{figure}[ht]
 \begin{center}
 \includegraphics[width=0.35\columnwidth]{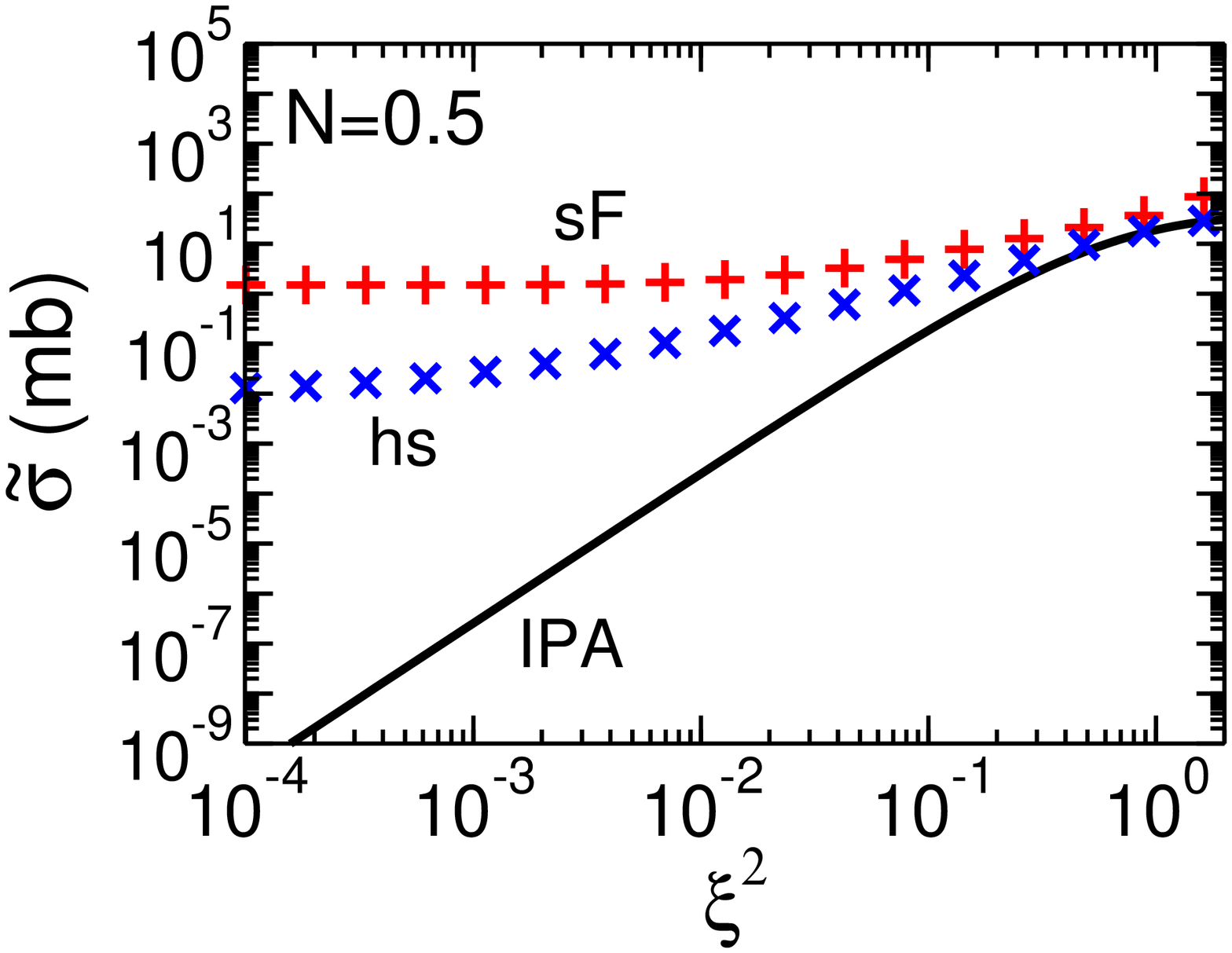}\qquad
 \includegraphics[width=0.35\columnwidth]{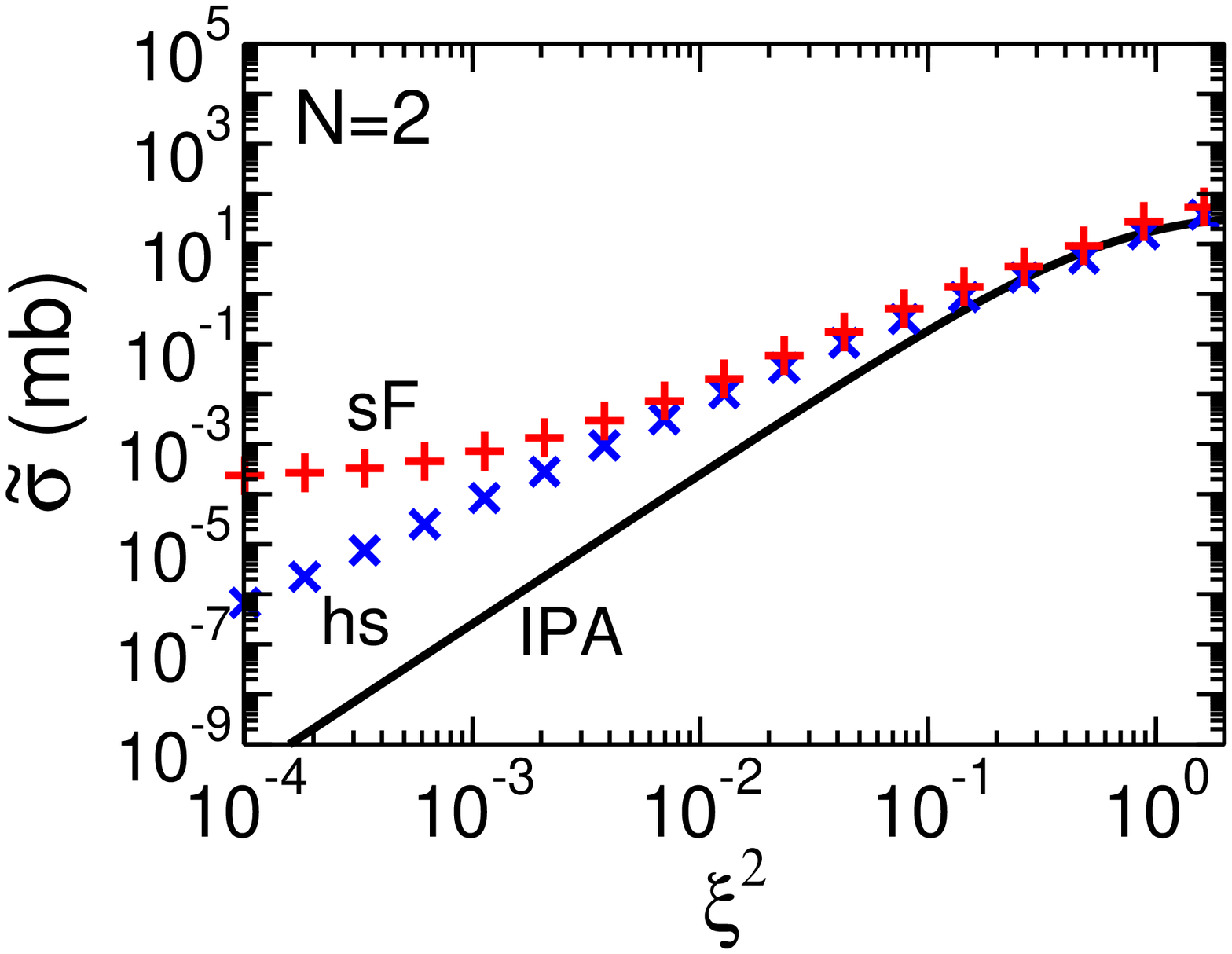}
 \end{center}
 \caption{\small{The partially integrated cross
 section as a function of $\xi^2$ at $\kappa=\omega'/\omega'_1=3$
 for short pulses with $N=0.5$ (left) and 2 (right).
 The solid curve and symbols correspond to IPA and FPA
 (hs and sF envelope functions), respectively.
\label{gFig:9}}}
 \end{figure}

  The dependence of the partially integrated cross
  section as a function of $\xi^2$ at fixed ratio $\kappa\equiv\omega'/\omega'_1=3$
  for short pulses with $N=0.5$ and 2 is exhibited in Fig.~\ref{gFig:9}
  in left and right panels, respectively.
  Note that the minimum value of $l'(\omega')$
  is related to $\kappa$ as
  \begin{eqnarray}
 l'(\omega')&=&\kappa\frac{E+|\mathbf{p}|\cos\theta'}
 {E+|\mathbf{p}|\cos\theta'+\omega(1-\kappa)(1-\cos\theta')}~,
 \label{S_Lim_l}
 \end{eqnarray}
 meaning $l' < \kappa$. Similarly, for $n_{\rm min}$ one has
 $n_{\rm min}=x$, for $I(x)=x$ and  $n_{\rm min}=x+1$
 for $I(x)<x$ with
\begin{eqnarray}
 x =\frac{E+|\mathbf{p}|\cos\theta'
 +\frac{\omega m^2\xi^2}{2(k\cdot p)}(1-\cos\theta')}
 {E+|\mathbf{p}|\cos\theta'+\omega(1-\kappa
 + \frac{m^2\xi^2}{2(k\cdot p)})(1-\cos\theta')} .
 \label{S_Lim}
 \end{eqnarray}
 The solid curves and symbols correspond to IPA and FPA,
  respectively, with different pulse shapes.
 One can see that the main difference of
 IPA and FPA, as well as the pulse shape dependence,
 appears at small field intensities
 $\xi^2\ll1$, where the dependence of the cross section on
 the pulse shape and duration is essential.

 To explain this result we use
 the asymptotic solution for $\tilde \sigma$ which is obtained
 by keeping leading terms in $\xi^2$ in Eqs.~(\ref{S2}) and (\ref{S2INF})
 and taking into account
 that the dominant contribution to the integrals of
 Eqs.~(\ref{S6}) and (\ref{S6_IPA}) stems
 from $l\sim l'$ and $n\sim I(l')+1$, respectively.
 Consider first the partially integrated cross section in IPA.
 Using the asymptotic expression for the Bessel
 functions
 \begin{eqnarray}
 J_k(z)\simeq \left(\frac{z}{2}\right)^k\frac{1}{k!} \quad
{\rm for}\,\,\,z\ll1~,
 \label{ASY1}
 \end{eqnarray}
 and keeping the leading terms in Eq.~(\ref{S2INF}) with
 $J_{n-1}^2(z)$ and $n=I(l')+1$, one obtains
 \begin{eqnarray}
 \tilde\sigma^{IPA}\simeq
 \frac{2\pi\alpha^2}{(E+|\mathbf{p}|\cos\theta')|\mathbf{p}|}\xi^{2k}
 \Phi(k)~,
 \label{ASY2}
 \end{eqnarray}
 where $k=I(l')\simeq I(\kappa)$ and
  \begin{eqnarray}
 \Phi(k)&=&\frac{(k+1)^{2(k+1)}}{(k+1)!^2}
 \left(t_k(1-t_k)\right)^{2k}\nonumber\\
 &\times&\left(1+\frac{u}{2(1+u)}-2t_k(1-t_k)\right)
 \label{ASY22}
 \end{eqnarray}
 with $t_k=u/u_k$, where
 $u=\omega'(1-\cos\theta')/(E+|\mathbf{p}|-\omega'(1- \cos\theta') )$
 and $u_k=2k\omega(E+|\mathbf{p}|)/m^2$. Within the considered kinematics,
 $t_k$ does not depend on $k$ and can be approximated by
 $t_k\simeq m^2(1-\cos\theta')/(2(E+|p|\cos\theta')(E+|p|))\simeq
 0.35$.

\begin{figure}[ht]
\begin{center}
 \includegraphics[width=0.4\columnwidth]{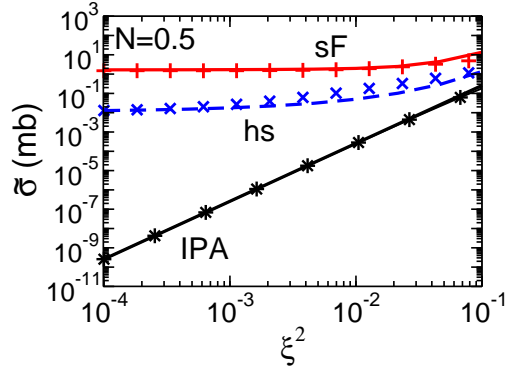}
\end{center}
 \caption{\small{
 The partially integrated cross
 sections as a function of $\xi^2\ll1$ for a sub-cycle pulse
 with $N=0.5$.
 The stars are for the full IPA result. The thick solid curve corresponds to the
 asymptotic solution of Eq.~(\ref{ASY2}).
 The pluses and crosses  are for full calculations
 for sF and hs shapes, respectively,
 while the dashed and dot-dashed curves
 are the corresponding asymptotic results
 of Eq.~(\ref{ASY4}).
 \label{gFig:10}}}
 \end{figure}

 The result for the asymptotic solution for IPA of (\ref{ASY2})
 is shown by
 the solid black curve in Fig.~\ref{gFig:10} together with a full
 calculation depicted by stars. One can see an excellent agreement
 of these two results.

 For FPA, in the case of sub-cycle pulse with $N=0.5$,
 we use the asymptotic
 representation for the basic functions $Y_l$ in the form of
 Eq.~(\ref{S_FF}) which allows to express
 the partially integrated cross section as
  \begin{eqnarray}
  &&\tilde\sigma\simeq
  \frac{2\pi\alpha^2}{N_0(E+|\mathbf{p}|\cos\theta')|\mathbf{p}|}\nonumber\\
  &\times& \left(1+\frac{u}{2(1+u)}-2t_{l'}(1-t_{l'})\right)
  \int\limits_{l'}^{l'+1} dl\,F^2(\tilde l-1)~,
  \nonumber\\
  \label{ASY4}
  \end{eqnarray}
 where $F(x)$ is the Fourier transform of the envelope
 function (cf. Eq.~(\ref{U5})).
 Results for the sub-cycle pulse with $N=0.5$ are
 presented in Fig.~\ref{gFig:10}, where
 the pluses and crosses  are for full calculations
 for the sF and hs shapes, respectively.
 The dashed and dot-dashed curves are the asymptotic
 solution of Eq.~(\ref{ASY4})
 for sF and hs shapes, respectively.
  \begin{figure}[ht]
  \begin{center}
 \includegraphics[width=0.4\columnwidth] {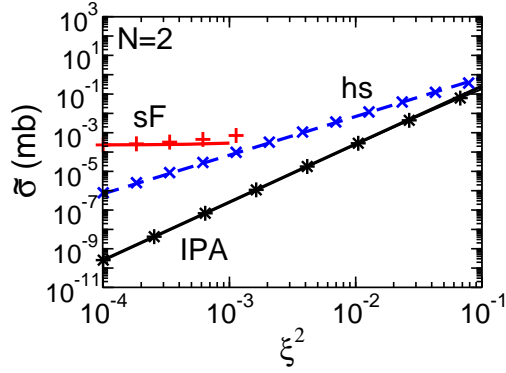}
\end{center}
 \caption{\small{
 Results for a short pulse with $N=2$.
 The symbols stars, pluses and crosses are full calculations,
 for IPA and FPA for sF and hs shapes, respectively;
 the solid black, dashed, and dot-dashed curves
 are the corresponding asymptotic results
 of Eqs.~(\ref{ASY2}), (\ref{ASY4})
 and (\ref{ASY7}), respectively.
 \label{gFig:11}}}
 \end{figure}

  We would like to note that, at $\xi^2\ll1$, our asymptotic solution
  for sub-cycle pulse weakly depends on $\xi$ only through
  the weak $l(1-\beta\xi)$ dependence in the Fourier transform.
  The leading $\xi^2$
  dependence of the partial harmonics $w_l$ in (\ref{S2}) is
  compensated by the $\xi^2$ dependence of the flux factor in the
  denominator of Eq.~(\ref{S1}). Nevertheless, such a weak $\xi$ dependence
  is in qualitative agreement of full and
  asymptotic solutions, both for sF and hs envelope shapes.
  Thus, we can conclude that the partially integrated
  cross section for the sub-cycle pulse at $\xi^2\ll1$ is almost completely
  determined by the square of the Fourier transform of the envelope function
  which is a measure of high momentum frequencies
  generated by the pulse shape.

  In the case of a short pulse with $N=2$ and $\xi^2\leq0.1$,
  we use for the asymptotic solution
  asymptotic expression of the basic functions of Eq.~(\ref{B6}).
  Note that such an expression is valid only for the smooth
  one-parameter envelope shapes, where the function  ${\cal P}(\phi)$,
  defined in Eq.~(\ref{III24}),
  takes a simple form ${\cal P}(\phi)=z\sin(\phi-\phi_0)f(\phi)+{\cal
  O}(\xi^2)$~(cf. Eq.~(\ref{III21})).
  One can see that, if the argument obeys $l' > I(l')$,
  then the main contribution to the cross section comes from
  the two terms with
 \begin{eqnarray}
 Y_{k,\varepsilon_1}(z)\,\,\,\,{\rm and}\,\,\,\,
 Y_{k+1,\varepsilon_2}(z)~,
 \label{ASY6}
 \end{eqnarray}
 where $k=I(l')$, $\varepsilon_1=l'-I(l')\equiv \varepsilon > 0$, and
 $\varepsilon_2=\varepsilon-1 < 0$.
 Then, keeping the leading terms in $\xi^2$ in (\ref{S2})
 one can get an approximate expression for the partially
 integrated cross section in the form
 \begin{eqnarray}
 && \tilde\sigma\simeq
 \frac{2\pi\alpha^2}{N_0(E+|\mathbf{p}|\cos\theta')|\mathbf{p}|}\xi^{2(k-1)}
 \nonumber\\
 &\times&\left(
 \Phi(k-1)\int\limits_{\varepsilon}^1\,d\epsilon \,(F^{(k)}(\epsilon))^2
 \right.
 \nonumber\\
 &&\left.\qquad\qquad +\xi^2\,\Phi(k)\int\limits_{\varepsilon-1}^1\,d\epsilon
  (F^{(k+1)}(\epsilon))^2
  \right)~,
 \label{ASY7}
 \end{eqnarray}
  where $F^{(m)}$ is the Fourier transform of $m$-th power
  of the envelope function $f(\phi)$.
  The full and approximate results for $\tilde\sigma$ are shown in
  Fig.~\ref{gFig:11} by crosses and the
  dot-dashed thick curve, respectively. One can see a
  fairly good agreement of approximate and full results
  up to $\xi^2=0.1$.

  In the case of the flat-top envelope, the integrand of $\tilde\sigma$
  has a more complicated structure with a large number of bumps.
  The asymptotic solution for the basic functions of Eq.~(\ref{B6})
  does not apply here. However, as a first approximation
  one can use the asymptotic solution of Eq.~(\ref{S_FF}). Then, the
  cross section $\tilde\sigma$ is determined by Eq.~(\ref{ASY4}).
  The full and approximate results for $\tilde\sigma$ are shown in
  Fig.~\ref{gFig:11} by pluses  and the
  dashed  curve, respectively. One can see an agreement of
  full and approximated results,
  however, in a very limited
  range of $\xi^2\ll1$.

  To summarize this part we note that, in the case of
  short pulses  and small field intensities, the partially
  integrated cross section is determined by
  the interplay of pulse shape and multi-photon dynamics.
  For both considered shapes, the cross sections are
  described by the simple asymptotic expressions
  which can be used in practical research.

  At large values $\xi^2\gg 1$, our analysis shows
  that the dependence on the envelope shape
  disappears because, similar to
  the Breit-Wheeler process, only
  the central part of the envelope becomes important.
  Formally, under a change of the variable
  $l\to l_{\rm eff}=l + m^2\xi^2u/2(k\cdot p)$,
  the basic functions $Y_l(z)$ with $l\gg1, \,z\gg1$
  become similar to the asymptotic
  form of the Bessel functions $J_l(z)$ and, as a consequence,
  one can get the total production probability
  (or the total cross section) in the form of IPA\cite{Ritus-79}
  with a slightly modified pre-exponential factor.

\subsection{Effect of the finite carrier phase}

 The generalization of our approach to the case
 of the finite carrier phase $\tilde \phi$ in
 e.m. potential~(\ref{III1}) is carried out by
 the same method as in the case of $\ee$ pair production
 described in Sect.~3.~7. The functions $C^{(i)}(l)$
 in transition matrix~(\ref{EM}) are transformed according
 to Eq.~(\ref{CP1}) with
 \begin{eqnarray}
 {\cal P(\phi)} = z\int_{-\infty}^{\phi}\,d\phi'\,
 \cos(\phi'-\phi_0+\tilde\phi)f(\phi')
 -\xi^2\frac{u}{u_0}
 \int_{-\infty}^\phi\,d\phi'\,f^2(\phi')~,
 \label{PCP}
 \end{eqnarray}
 where the variables $z,\,u$ and $u_0$ are defined in (\ref{zC}).
 Then, using the basic functions $Y_l(z)$ and $X_l(z)$ in the form
 of Eq.~(\ref{CP3}) and utilizing Eq.~(\ref{CP4}) one can
 obtain the partial differential cross section
 $d\sigma(l)/d\omega'd\phi_{e'}$ in the form of Eq.~(\ref{S1})
 with $w(l)$ given by Eq.~(\ref{S2}), but with new basic functions
 $Y_l$ and $X_l$ which now depend on carrier phase $\tilde\phi$
 according to  Eq.~(\ref{CP3}) with (\ref{PCP}).
 Recall, that $\phi_0=\phi_{e'}$ is
 the azimuthal angle of the outgoing electron momentum.
 The differential partially integrated cross section
 reads
  \begin{eqnarray}
 \frac{d\tilde\sigma_{}(\omega')}{d\phi_{e'}} = \int\limits_{\omega'}^{\infty}
 d\bar\omega' \frac{d\sigma (\bar\omega')}{d\bar\omega'\,d\phi_{e'}}~.
 \label{CPS6}
\end{eqnarray}

 It is natural to expect that the  effect of the finite carrier
 phase essentially appears in the differential cross section of
 the generalized Compton scattering as a function of the
 azimuthal angle of the outgoing electron momentum
 because the carrier phase is included in the expressions for the basic
 functions (\ref{CP3}) in the combination $\phi_{e'}-\tilde\phi$.

\begin{figure}[ht]
\begin{center}
\includegraphics[width=0.35\columnwidth]{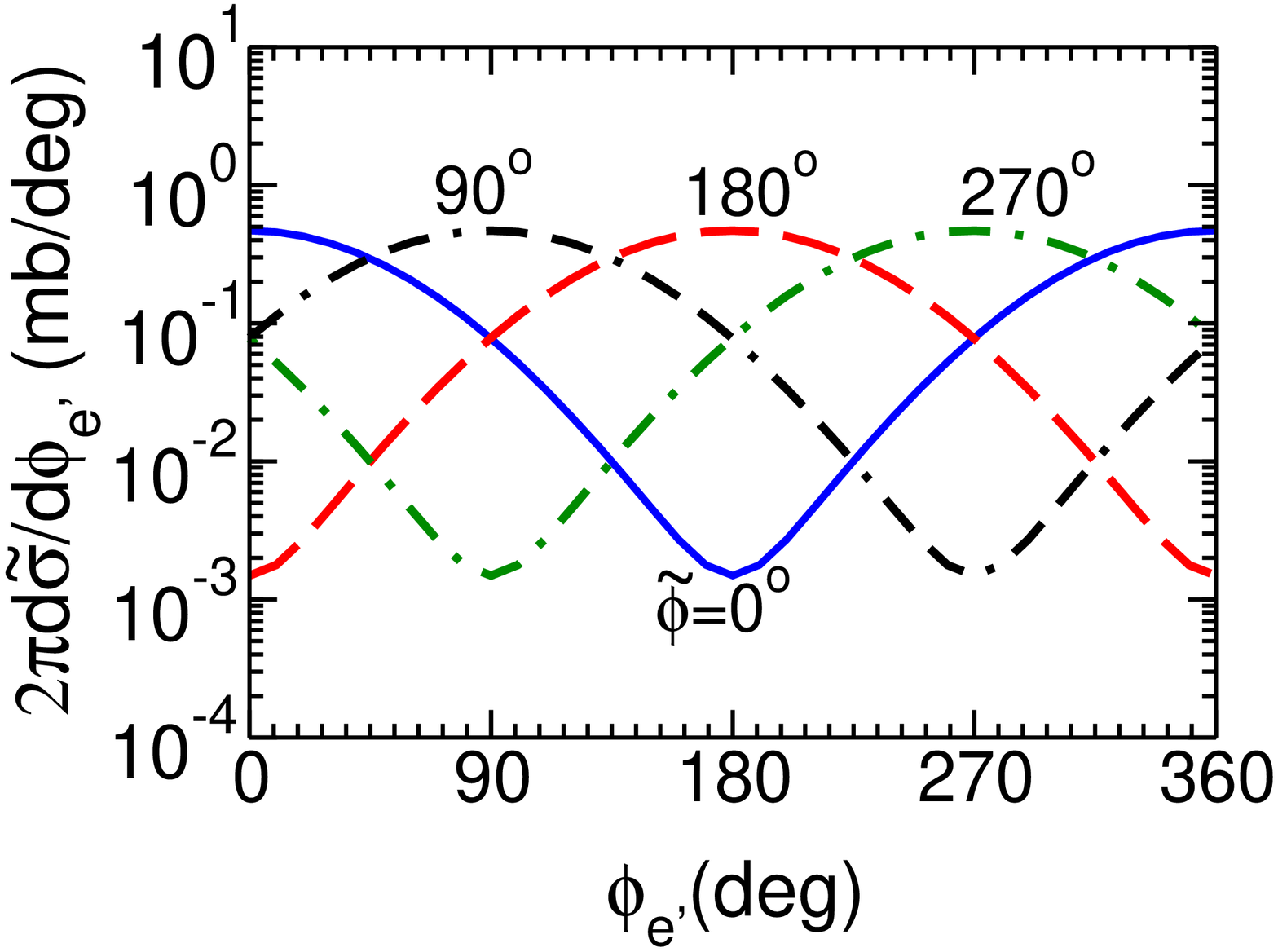}\qquad
\includegraphics[width=0.35\columnwidth]{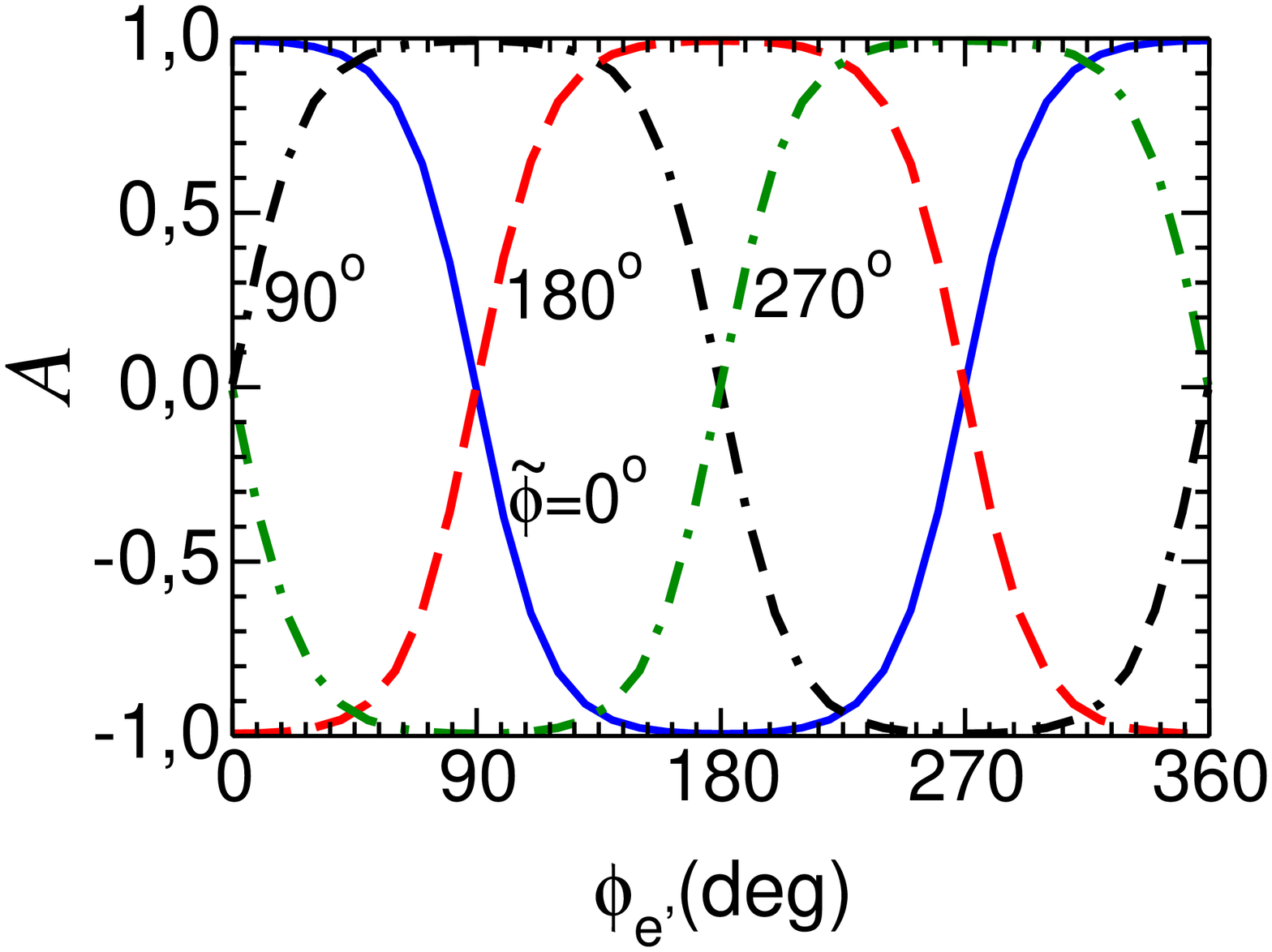}
\end{center}
\caption{\small{
 Left panel: The differential cross section (\ref{CPS6}) as
 a function of the azimuthal
 angle of the outgoing electron momentum
 $\phi_{e'}$ for different values of the
 carrier phase $\tilde\phi$.
 The solid, dash-dash-dotted, dashed and dash-dotted
 curves correspond to the carrier phase equal 0,
 90, 180 and 270 degrees, respectively.
 Right panel: The anisotropy~(\ref{U9}) as a function
 of $\phi_{e'}$ for different $\tilde\phi$.
 For the hyperbolic secant shape with $N=0.5$;
 $\xi^2=0.1$ and $ \kappa=\omega'/\omega'_1=4$.
 \label{Fig:14AC}}}
\end{figure}
 As an example, in Fig.~\ref{Fig:14AC} (left panel) we show the differential
 cross section (\ref{CPS6}) as a  function of the azimuthal angle
 $\phi_{e'}$ for different values of the carrier phase $\tilde\phi$
 for the sub-cycle pulse with $N=0.5$ for the hyperbolic secant shape
 with $\kappa=\omega'/\omega'_1=4$ and $\xi^2=0.1$.
 One can see a clear bump-like structure of the distribution,
 where the bump position coincides with the corresponding value of
 the carrier phase.
 The reason of such behaviour is the same as an alignment
 of the probability along $\phi_{e}=\tilde\phi$,
 described in Sect.~3.~7.
  Corresponding anisotropies defined  as
  \begin{eqnarray}
  {\cal A}=\frac{d\tilde\sigma(\phi_{e'}) - d\tilde\sigma(\phi_{e'}+\pi)}
  {d\tilde\sigma(\phi_{e'}) + d\tilde\sigma(\phi_{e'}+\pi)}
  \label{ASYCP1}
  \end{eqnarray}
   are exhibited in
  Fig.~\ref{Fig:14AC} (right panel). One can see a strong dependence
  of the anisotropy on the carrier phase which follows to the
  bump-like behavior of the differential probabilities shown in the
  left panel. Similar to the Breit-Wheeler process, the
  anisotropy takes a maximum value ${\cal A}\simeq 1$
  at $\phi_{e'}=\tilde\phi$ and  $|{\cal A}|<1$ at $\phi_{e'}\neq\tilde\phi$.
  It takes a minimum value ${\cal A}\simeq -1$
  at  $\phi_e-\tilde\phi=\pm\pi$.

\begin{figure}[ht]
 \begin{center}
\includegraphics[width=0.35\columnwidth]{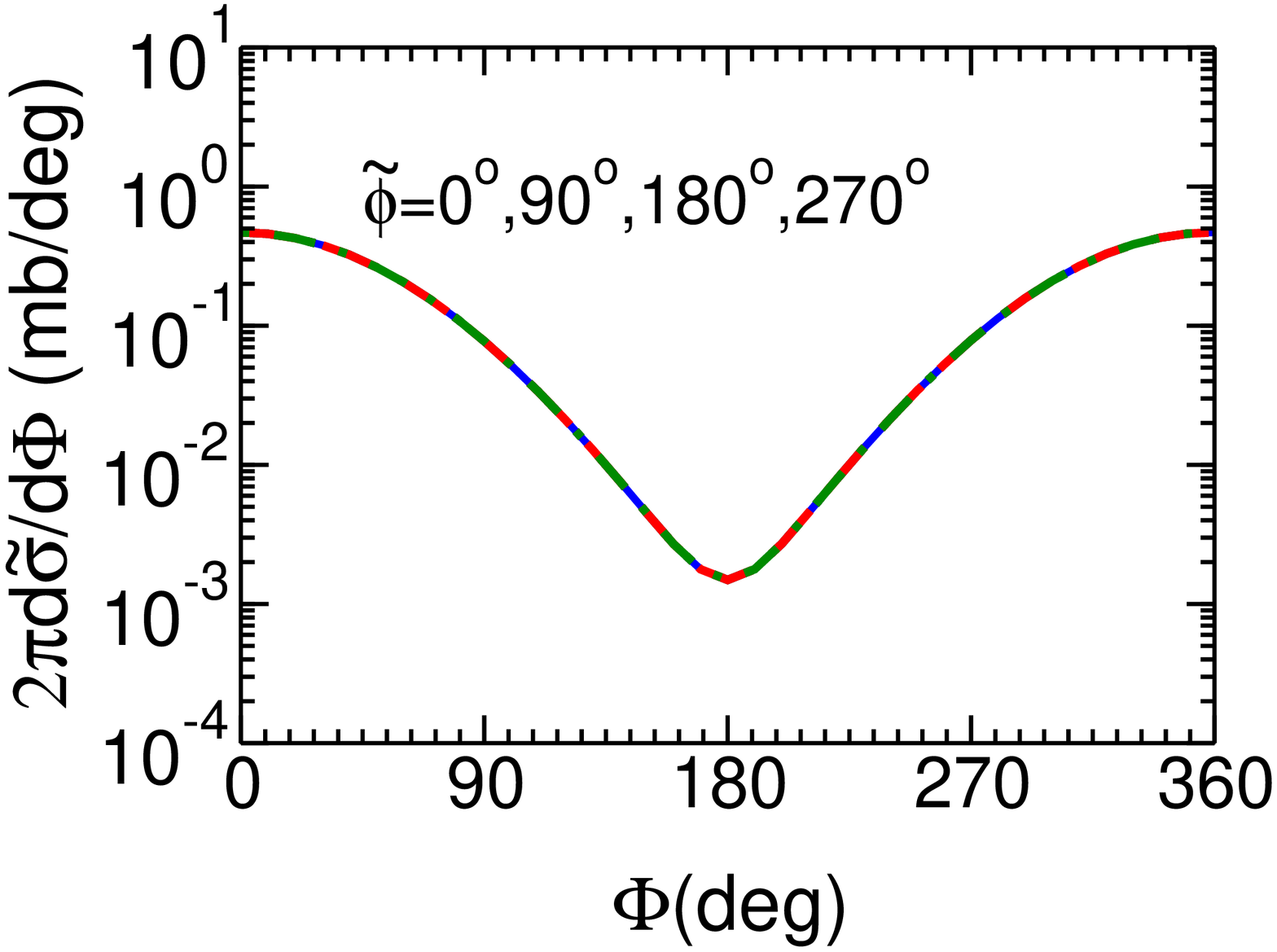}\qquad
\includegraphics[width=0.35\columnwidth]{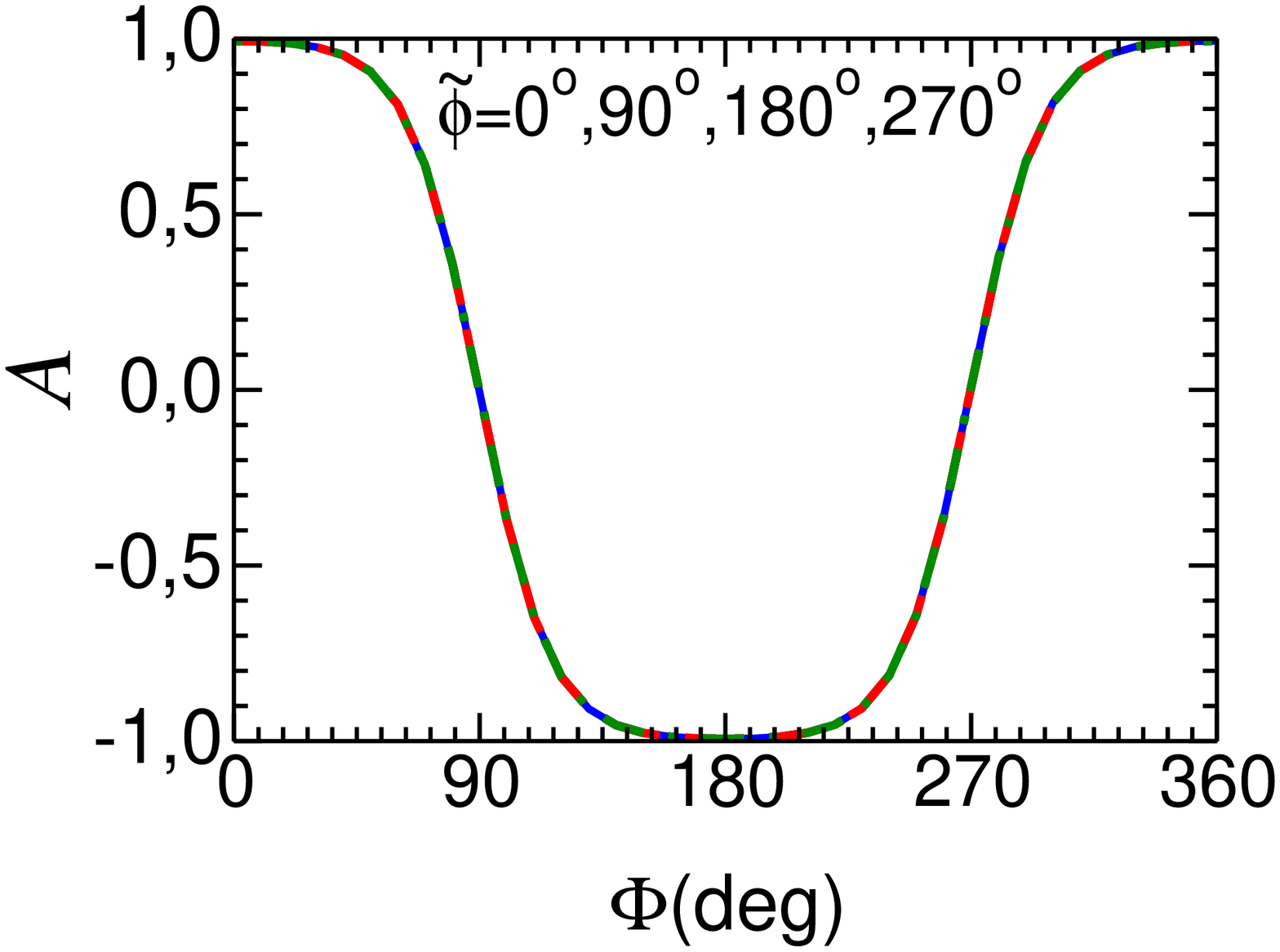}
\end{center}
\caption{\small{The same as in Fig.~\ref{Fig:14AC} but as a function
 of the scale variable $\Phi=\phi_{e'} - \tilde\phi$.
 \label{Fig:15AC}}}
\end{figure}

 The differential cross sections and anisotropies
 as  functions of  the "scale"\ variable
 $\Phi=\phi_{e'} - \tilde\phi$ at fixed values of
 $\tilde\phi$ are exhibited in Fig.~\ref{Fig:15AC}
 in the left and right panels, respectively.
 All curves shown, in the left and right panels
 in Fig.~\ref{Fig:15AC},
 are merged into a single carrier phase independent curve.
 Similar to the Breit-Wheeler process,
 such a carrier phase independence of the
 differential cross sections and anisotropies
 is a consequence of the $\phi_{0}-\tilde\phi=\phi_{e'}-\tilde\phi$
 dependence of the basic functions in
 Eqs.~(\ref{CP3}) and (\ref{CP4}).

 The effect of the carrier phase
 decreases with increasing pulse duration. Taking into account
 the similarity between Breit-Wheeler and Compton scattering processes
 we do not show here the result for the Compton scattering
 (for $N\geq2$), limiting
 to the most striking example of sub-cycle pulse, exhibited in
 Figs.~\ref{Fig:14AC} and \ref{Fig:15AC}.


\section{Summary}

 In summary we have
 considered two elementary quantum processes occurring
 in a short and intense electromagnetic
 (laser) pulses. They are the $\ee$ pair production
 (generalized Breit-Wheeler process)
  and the crossed process i.e. emission of single photon off an electron
 (generalized Compton scattering).
 We emphasized the very significant impact
 of the temporal pulse structure.
 Still, the pulses are approximated by plane waves, meaning
 that curved wave fronts deserve in future also
 dedicated investigations.

 The pair production in the sub-threshold region
 with $\zeta>1$ is currently a subject of great interest.
 We have shown that the production probability is determined
 by a non-trivial interplay of two dynamic effects. The first one is
 related to the shape and duration of the pulse.
 The second one is the non-linear dynamics of charged particles
 in the strong
 electromagnetic field itself, independently of the pulse
 geometry.
 These two effects play quite different roles in two limiting cases.
The pulse shape effects are manifested clearly at small values
of product the $\xi \zeta$, where $\xi$ characterizes the laser
intensity and $\zeta$ refers to the threshold kinematics. The
rapid variation of the e.m. field in a very short pulse amplifies
the multi-photon events, and moreover, the probability of
multi-photon events in the finite pulse approximation
(FPA) can exceed the prediction of the infinite pulse approximation
(IPA) by orders of magnitude.
Thus, for example in the case of an ultra-short (sub
cycle) pulse with the "number of oscillations"\ $N$ in the pulse less
than one, the production probability as a function of $\zeta$ is
almost completely determined by the square of the Fourier
transform of the pulse envelope function. High-$l$ components,
where $l$ is the Fourier conjugate to the invariant phase variable
$\phi$, lead to the enhancement of the production probability.
Among the considered envelope shapes, the flat-top shape with fast
ramping and deramping intervals is most promising to obtain the highest
probability. We also find that the different envelope shapes lead
to anisotropies of the electron (positron) emission which can be
studied experimentally. For short pulses with $N<10$, the effects
of the pulse shape are also important and the final yield differs
significantly from the IPA prediction. This difference depends on
the envelope shapes and the pulse duration.

Contrary to that, the non-linear multi-photon
dynamics of $\ee$ production in a strong
electromagnetic field plays a crutial role at
large field intensities, $\xi^2\gg1$.
Here, the effects of the pulse shape and duration
disappear since the dominant contribution comes from the central
part of the envelope function. As a result, the probabilities
in FPA and IPA coincide.

In the transition region of intermediate intensities
$\xi^2\sim 1$,  the probability
is determined
by the complex interplay of the both effects, and they must be taken
into account simultaneously by a direct numerical
evaluation of the multi-dimensional
integrals with rapidly oscillating integrands.

The effect of the carrier phase manifests itself most clearly
in ultra short (sub-cycle) pulses in  azimuthal distributions
of direction of flight of the outgoing electron (positron).
The production probability has a bump-like structure where the bump position
coincides with the value of carrier phase. This leads
to a definite alignment of the differential
cross section and anisotropy in the $x-y$ plane along the
angle equal to the carrier phase.
The impact of the carrier phase decreases with increasing pulse duration.


  The considered generalized
  nonlinear (multi-photon) Compton scattering in
  short and ultra-short (sub-cycle)
  laser pulses is a crossing channel to the Breit-Wheeler process
  and, therefore, reflects the main features of the latter one.
  We have shown that the fully differential
  cross section as a function of the frequency of the
  outgoing photon at fixed production angle is a rapidly
  oscillating function for short pulses with the duration
  determined by the number of oscillations $N=2\cdots 10$,
  especially for the flat-top envelope shapes. An
  experimental study of multi-photon effects in the case of rapidly
  oscillating cross sections seems to be rather challenging.
  To overcome the problem of
  such a staggering we suggest to utilize the
  partially integrated  cross section
  which seems to be a powerful tool for studying
  the non-linear (multi-photon) dynamics in the sub-threshold region.
  We find that these cross sections at selected
  pulse properties (field intensity, pulse duration)
  are very sensitive to the pulse shape.
  In the case of small e.m.\ field intensities, the cross section may be
  enhanced by several orders of magnitude as compared to
  an infinitely long pulse. Such an enhancement is more important
  for flat-top envelope shapes which generate intensive
  high-frequency harmonics and play a role of a power
  amplifier.
  In the above-threshold region, the partially integrated cross section
  manifest some "universality" \ , i.e. an independence of the pulse
  shape structure, where results for FPA and IPA are close to each other.
  Note that such a "universality" \ does not appear in fully
  differential cross section, where one can find
  rapidly oscillating cross section as a function of $\omega'$,
  especially for the flat-top envelope shape.
  At high field intensity, the central part of envelopes
  becomes dominant and the integrated cross sections
  coincide with that for infinitely long pulses.
  It provides a rationale for the use of simple analytical
  expressions of IPA for Monte Carlo transport approaches.
  Finally, we have shown that the effect of the carrier phase
  is important and might be seen clearly in sub-cycle pulses.
  Similarly to the Breit-Wheeler process, we predict
  a definite alignment of the differential
  cross section and anisotropy in the $x-y$ plane along the
  angle equal to the carrier phase.

  Our considerations are focused on circularly polarized
  photon beams. However, we expect that qualitatively,
  in the case of a linearly polarized  pulse, our main results,
  i.e.\ the sensitivity of the production probability of pair production
  and partially integrated cross section of the Compton scattering
  to the sub-threshold multi-photon
  interactions and to the pulse structure, would be similar.
  The main difference is expected for the anisotropies since
  the  momentum of the outgoing electron will be correlated with the direction
  of pulse polarization.

  Our considerations are devoted essentially to the elementary processes in
  optical laser beams. With the availability of X ray beams
  (XFELs cf.[LCLS, SACLA, European XFEL, Swiss XFEL, .... ])
  already now or in near future a further field of interesting
  phenomena is entered, where the here presented theory also applies.

\vspace{5mm}

\centerline{\bf Acknowledgments}

The authors acknowledge fruitful discussions with
D.~Seipt, T.~Nousch, T.~Shibata, R.~Sauerbrey
and T.~E.~Cowan.

\vspace{5mm}

\centerline{\bf Appendix}
\begin{center}
{\bf{Production probability at
large values of $\xi$}}
\end{center}
The total probability $W$ in the limit of large $\xi$ and
and small $\xi/\zeta$,
was evaluated by Narozhny and Ritus~\cite{NikishovRitus}
and summarized by Ritus~\cite{Ritus-79} in compact form.
Below, for completeness and easy reference, we recall
some details of Ritus's evaluation making an expansion for an
arbitrary value of $\xi/\zeta$, applying it for the case of
the finite pulse (cf. Sect.~3.~5).

In IPA, the
total probability is represented as an infinite sum
of partial harmonics~\cite{Ritus-79}
\begin{eqnarray}
W&=&\frac14 {\alpha M_e\zeta}\,\sum\limits_{n=n_{0}}^{\infty}
\int\limits_1^{u_n} \frac{du}{u^{3/2}\sqrt{u-1}}
 \{2J^2_n(z)\nonumber\\
&+&\xi^2(2u-1)\left(
J^2_{n+1}(z) + J^2_{n-1}(z)-2J^2_n(z)
\right)\}~,
\label{II8-1}
\end{eqnarray}
where  $n_0\equiv n_{\rm min}=\zeta (1+\xi^2)$,
$u_n=n/n_0$, and $J_n(z)$ is the Bessel function
of the first kind  (cylindrical harmonics).
Using the identities
\begin{eqnarray}
2\,\frac{n}{z}\, J_n(z)=   J_{n-1}(z)  + J_{n+1}(z) , \,\,\,
2\,{J'}_n(z) =    J_{n-1}(z)  - J_{n+1}(z) ~,
\label{II111}
\end{eqnarray}
the total probability takes the following form
\begin{eqnarray}
W&=&\frac12 {\alpha M_e\zeta^{1/2}  }   \sum\limits_{n_{0}}^{\infty}
\int\limits_1^{u_n}\frac{du}{u^{3/2}\sqrt{u-1}}
\left(
J^2_n(z)\right.\nonumber\\
&+&\xi^2(2u-1)\left.\left(
(\frac{n^2}{z^2} -1  ) {J}^2_{n}(z) + { J'}^2_{n}(z)
\right)\right)~.
\label{II112}
\end{eqnarray}
At large $\xi\gg1$, $\zeta\gg1$, $n,\,z\gg1$ and $n>z$
one can replace the sum over integer $n$ by an integral over $dn$,
replacing, for convenience, integer $n$ to continues $l$ with
$l_{\rm min}\equiv l_0=\zeta(1+\xi^2)$.
Using Watson's asymptotic
expression for the Bessel functions one finds
\begin{eqnarray}
J_l\left(\frac{l}{\cosh\alpha}\right)
=\frac{1}{\sqrt{2\pi l\tanh\alpha}}
{\rm }e^{-l(\alpha   - \tanh\alpha)} +{\cal O}\left(\frac{1}{\xi}\right)
\label{II113}
\end{eqnarray}
with   $\cosh\alpha=l/z$.
If $l$ is large the first term represents a good approximation
irrespectively whether $\xi/\zeta$
is small or large~\cite{WatsonBook}.
The corresponding derivative reads
\begin{eqnarray}
J'_l(z)\simeq\sinh\alpha\,J_l(z)\,
\left(1+\frac{1}{2l\sinh^2\alpha\tanh\alpha} \right)~.
\label{II14}
\end{eqnarray}
Consider first the case of small  $\xi/\zeta\ll1$,
when the second term in (\ref{II14}) can be neglected.
Then, the total probability becomes
\begin{eqnarray}
W=\frac{e^2 M_e\zeta^{1/2} }{8\pi^2  }
 \int\limits_{l_0}^{\infty}dl\,
\int\limits_1^{u_l}\frac{du}{u^{3/2}\sqrt{u-1}}
\frac{1+2\xi^2(2u-1)\sinh^2\alpha }
{l\,\tanh\alpha}
\,{\rm e}^{f(u,l)}~,
\label{II15}
\end{eqnarray}
where $u_l=l/l_0$ and  $\hat f(u,l)=-2l(\alpha -\tanh(\alpha))$
with
\begin{eqnarray}
\tanh^2(\alpha)=\frac{1+\xi^2\left(1-   \frac{2u}{u_l}\right)^2  }{1+\xi^2}~.
\label{C2}
\end{eqnarray}
To avoid a notational confusion
with respect to the standard variable $\alpha$,
we replace below the fine structure
constant by $e^2/4\pi$.

The two-dimensional integral is evaluated using the
saddle point approximation since the function $\hat f(u,l)$ has a sharp
minimum at the point $u=\bar u$ defined by
the equation $\hat f'_u(u=\bar u)=0$.
That allows  (i) to expand it into a Taylor series
\begin{eqnarray}
f(u,l)\simeq \hat f(\bar u, l) +\frac12 \hat f_u{''}(\bar u,l)(u-\bar u)^2~,
\label{C3}
\end{eqnarray}
and (ii) to take the rest (smooth) part of the integrand in Eq.~(\ref{II15})
at the point $u=\bar u$ yielding
\begin{eqnarray}
W=\frac{e^2 M_e\zeta^{1/2} }{16\pi^2}
 \int\limits_{l_0}^{\infty}dl\,
 {\cal A}_0(\bar u, l){\rm e}^{\hat f(\bar u, l)}
 \int\limits_1^{u_l}\frac{du}{\sqrt{u-1}}
\,\,{\rm e} ^{\frac12 \hat f^{''}(\bar u,l)(u-\bar u)^2 }~,
\label{C33}
\end{eqnarray}
with
\begin{eqnarray}
{\cal A}_0(u,l) =
\frac{1+2\xi^2(2u-1)\sinh^2\alpha}
{u^{3/2}l\tanh\alpha}~.
\label{C34}
\end{eqnarray}
The explicit  expression
\begin{eqnarray}
\hat f'_u(u,l)= \frac{4l_0\sinh^2\alpha}{\tanh\alpha}
\frac{\xi^2}{1+\xi^2}
\left(1-\frac{2u}{u_l} \right)
\label{C4}
\end{eqnarray}
leads to the solution
\begin{eqnarray}
\bar u= \frac{u_l}{2}=\frac{l}{2l_0}~,
\label{C5}
\end{eqnarray}
which results in the following equalities
\begin{eqnarray}
&&\tanh\bar\alpha\equiv \tanh\alpha(\bar u)=\frac{2}{\sqrt{1+\xi^2}} , \,\,\sinh\bar\alpha=\frac{1}{\xi},
\,\,\,\hat f{''}_u(\bar u,l)=-\frac{8l_0^2}{l\sqrt{1+\xi^2}}\nonumber\\
&&{\cal A}_0=\frac{1+2(2\bar u-1)}{{\bar u}^{3/2} l}\sqrt{1+\xi^2},
\,\,\,\hat f(\bar u, l)=-2l(\bar \alpha -\tanh\bar\alpha)~.
\label{C6}
\end{eqnarray}
Using the substitutions $u=t+1$,  $a=2(\bar \alpha -\tanh\bar\alpha)$, and
 $A=-\frac12\hat f^{''}(\bar u,l)$ one can rewrite Eq.~(\ref{C33}) as
 \begin{eqnarray}
W= \frac{e^2 M_e\zeta^{1/2} }{16\pi^2 }
 \int\limits_{l_0}^{\infty}dl\,
 {\cal A}_0(\bar u, l){\rm e}^{-al -  A(1-\bar u)^2 }
 \int\limits_{0}^{\infty}dt\,  t^{\nu-1}{\rm e}^{-\beta t^2 -\gamma t} ~,
\label{C66}
\end{eqnarray}
with $\nu=1/2$, $\beta=A$, and  $\gamma=2A(1-\bar u)$.
The integral over $dt$ is expressed via the parabolic cylinder function $D_{-\nu}$
\begin{eqnarray}
\int\limits_{0}^{\infty}dt\,  t^{\nu-1}{\rm e}^{-\beta t^2 -\gamma t}
=\left( \frac{1}{2\beta}\right)^{\nu/2}\,\Gamma(\nu)\,
{\rm  exp} [\frac{\gamma^2}{8\beta}]\,
\,D_{-\nu}\left(\frac{\gamma}{\sqrt{2\beta}} \right)~,
\label{C9}
\end{eqnarray}
which results in
\begin{eqnarray}
W= \frac{e^2 M_e\zeta^{1/2} }{16\pi^{3/2} }
 \int\limits_{l_0}^{\infty}dl\,
 \left(\frac{1}{2A} \right)^{\frac14}
 {\cal A}_0(\bar u, l){\rm e}^{-al -  \frac{A}{2}(1-\bar u)^2 }
 D_{-\frac12}(y)
\label{C99}
\end{eqnarray}
with $y= \sqrt{2A}(1-\bar u)$.
The main contribution to this integral comes from
the region $\bar u\sim 1$ ($l\sim \bar l=2l_0$) and, therefore,
one can use
the  substitution
\begin{eqnarray}
\int\limits_{l_0}^\infty dl =-\frac{2l_0}{\sqrt{2A}}
\int\limits_{\sqrt{A/2}}^{-\infty}dy\approx
\frac{2l_0}{\sqrt{2A}}
\int\limits_{-\infty}^{\infty}dy~,
\label{C99_}
\end{eqnarray}
which results in
\begin{eqnarray}
W= \frac{e^2 M_e \zeta^{1/2} }{16\pi^{3/2} }
\left( \frac{1}{2A}  \right)^{\frac14}\frac{2l_0}{\sqrt{2A}}
{\cal A}_0(\bar u, \bar l){\rm e}^{-2l_0a }
 \int\limits_{-\infty}^{\infty}dy\,
 {\rm e}^{Zy -  y^2/4 }
 D_{-\frac12}(y)
\label{C999}
\end{eqnarray}
with $Z=2l_0a/\sqrt{2A}$. Using the identity
\begin{eqnarray}
 \int\limits_{-\infty}^{\infty}dy\,
 {\rm e}^{Zy -  y^2/4 }
 D_{-\frac12}(y)=\sqrt{\frac{2\pi}{Z}}\,{\rm e}^{Z^2/2}~,
\label{C999_}
\end{eqnarray}
one can rewrite the production probability as
\begin{eqnarray}
W= \frac{e^2 M_e\zeta^{1/2} }{16\pi  }
\sqrt{\frac{2l_0}{aA}}\,
{\cal A}_0(\bar u, \bar l)\,
{\exp}[ -2l_0a +\frac{l_0^2a^2}{A}]~.
\label{C9999}
\end{eqnarray}

In order to reproduce the Ritus result~\cite{Ritus-79} in terms of
the kinematic factor $\zeta$ and the field
intensity $\xi$ one has to use the identity $l_0=\zeta(1+\xi^2)$
and to represent $a(\bar \alpha)$ as a series for small values $1/\xi$
utilizing the expansions
\begin{eqnarray}
\bar\alpha &=&{\rm arsinh}\frac{1}{\xi}
\simeq\frac{1}{\xi} - \frac{1}{6\xi^3} + \frac{3}{40\xi^5}~,\nonumber\\
\tanh\bar\alpha &=&\frac{1}{\sqrt{1+\xi^2}}\simeq
\frac{1}{\xi} -\frac{1}{2\xi^3} +\frac{3}{8\xi^5},\,
{\cal A}_0=\frac{3}{2\zeta\xi}~,
\label{C13}
\end{eqnarray}
which leads to (\ref{H7}) with $d=1$.
Inclusion of the second term in~(\ref{II14}) modifies eventually
${\cal A}_0$
as
\begin{eqnarray}
{\cal A}_0=
\frac{3}{2\zeta\xi}
\left(1+ \frac{\xi}{6\zeta}\left(1+\frac{\xi}{8\zeta} \right)\right)
\label{C14}
\end{eqnarray}
yielding the result displayed in~(\ref{H7})
which generalizes the Ritus result for arbitrary values of $\xi/\zeta$.
We emphasize that, in the strong field regime, IPA is representative
(with taking into account the pre-exponential factor $d(\xi/\zeta)$ in~(\ref{H7}))
since, as stressed above, pulse shape and pulse duration effects are sub leading.


\end{document}